\pdfoutput=1
\documentclass[english]{IEEEtran}
\usepackage[T1]{fontenc}
\usepackage[latin9]{inputenc}
\usepackage{color}
\usepackage{amsmath}
\usepackage{amsthm}
\usepackage{amssymb}
\usepackage{graphicx}

\makeatletter
\let\oldforeign@language\foreign@language
\DeclareRobustCommand{\foreign@language}[1]{%
  \lowercase{\oldforeign@language{#1}}}
\theoremstyle{plain}
\newtheorem{thm}{\protect\theoremname}
\theoremstyle{definition}
\newtheorem{defn}[thm]{\protect\definitionname}
\theoremstyle{plain}
\newtheorem{prop}[thm]{\protect\propositionname}
\theoremstyle{remark}
\newtheorem{rem}[thm]{\protect\remarkname}
\theoremstyle{plain}
\newtheorem{cor}[thm]{\protect\corollaryname}

\ifCLASSOPTIONcompsoc
  \usepackage[nocompress]{cite}
\else
  \usepackage{cite}
\fi
\ifCLASSINFOpdf
\else
\fi

\usepackage{setspace}
 \usepackage{xargs}

\usepackage{amsxtra,amssymb,amsmath}
\usepackage{graphicx}
\usepackage{hyperref}
\usepackage{epstopdf}
 
\author{Viet Hung Tran\thanks{%
V. H. Tran was with Telecom Paris institute, 75013 Paris, France. He is now with Univeristy of Surrey, GU27XH Surrey, U.K. (e-mails: v.tran@surrey.ac.uk, tranviethung@hcmut.edu.vn).}}

\makeatother

\usepackage{babel}
\providecommand{\corollaryname}{Corollary}
\providecommand{\definitionname}{Definition}
\providecommand{\propositionname}{Proposition}
\providecommand{\remarkname}{Remark}
\providecommand{\theoremname}{Theorem}

\begin{document}

\title{Copula Variational Bayes inference \\
via information geometry}

\markboth{IEEE Transactions on Information theory 2018 (preprint)}{}

\maketitle

\global\long\def\argmin{\operatornamewithlimits{arg\,min}}

\global\long\def\argmax{\operatornamewithlimits{arg\,max}}


\global\long\def\calN{\mathcal{N}}

\global\long\def\Mul{Mu}

\global\long\def\iW{i\mathcal{W}}


\global\long\def\transpose{T}

\global\long\def\TRIANGLEQ{\triangleq}

\global\long\def\REAL{\mathbb{R}}

\global\long\def\NATURAL{\mathbb{N}}

\global\long\def\COMPLEX{\mathbb{C}}

\global\long\def\Re{\text{Re}}

\global\long\def\Ran{\text{Range}}

\global\long\def\card#1{\mbox{card\ensuremath{\left(#1\right)}}}

\global\long\def\bigO{O}


\global\long\def\ZERO{\mathbf{0}}

\global\long\def\ONE{\mathbf{1}}

\global\long\def\IDEN{\mathbf{I}}

\global\long\def\UNIT{\mathbb{I}}

\global\long\def\Diag{\text{Diag}}

\global\long\def\diag{\text{diag}}

\global\long\def\Trace{\text{Tr}}


\global\long\def\Lzero{\ell_{0}}

\global\long\def\Lone{\ell_{1}}

\global\long\def\Ltwo{\ell_{2}}

\global\long\def\Linfty{\ell_{\infty}}

\global\long\def\lnorm{\ell}

\global\long\def\lp{\ell_{p}}

\global\long\def\Lnorm{\mathcal{L}}

\global\long\def\Lp{\mathcal{L}_{p}}


\global\long\def\ntime{N}

\global\long\def\nstate{M}

\global\long\def\ndim{K}

\global\long\def\itime{i}

\global\long\def\ipick{j}

\global\long\def\istate{m}

\global\long\def\idim{k}

\global\long\def\iton{\itime=1,\ldots,\ntime}

\global\long\def\iinn{\itime\in\{1,\ldots,\ntime\}}

\global\long\def\jtom{\istate=1,\ldots,\nstate}

\global\long\def\kinK{\idim\in\{1,\ldots,\ndim\}}

\global\long\def\ktom{\istate=1,\ldots,\nstate}

\global\long\def\seti#1#2{#1\in\{1,2,\ldots,#2\}}

\global\long\def\setk#1#2{#1{}_{1},#1{}_{2},\ldots,#1_{#2}}

\global\long\def\setp#1#2{(#1{}_{1},#1{}_{2},\ldots,#1_{#2})}

\global\long\def\setv#1#2{[#1{}_{1},#1{}_{2},\ldots,#1_{#2}]}

\global\long\def\setu#1#2#3{[#1{}_{1}^{#3},#1{}_{2}^{#3},\ldots,#1_{#2}^{#3}]}

\global\long\def\setd#1#2{\{#1{}_{1},#1{}_{2},\ldots,#1_{#2}\}}

\global\long\def\sete#1#2#3{[#1{}_{1,#3},#1{}_{2,#3},\ldots,#1_{#2,#3}]}

\global\long\def\setf#1#2#3#4{[#1{}_{1,#3}^{#4},#1{}_{2,#3}^{#4},\ldots,#1_{#2,#3}^{#4}]}

\global\long\def\setx#1#2#3{[#1,\ldots,#2,\ldots,#3]}


\global\long\def\EXPECTATION{\mathbb{E}}

\global\long\def\VARIANCE{\text{Var}}

\global\long\def\UNIT{\mathbb{I}}

\global\long\def\Entropy{H}

\global\long\def\Mutual{I}


\global\long\def\BX{\mathcal{X}}

\global\long\def\Bx{x}

\global\long\def\Ba{\alpha}

\global\long\def\Bb{\beta}

\global\long\def\Bc{\gamma}

\global\long\def\bBx{\boldsymbol{x}}

\global\long\def\bBa{\boldsymbol{\alpha}}

\global\long\def\bBb{\boldsymbol{\beta}}

\global\long\def\bBc{\boldsymbol{\gamma}}

\global\long\def\Bfunc{\phi}

\global\long\def\BD{\mathcal{D}}

\global\long\def\BH{\mathcal{H}}

\global\long\def\Bp{p}

\global\long\def\Bmixture{f_{\Bp}}


\global\long\def\Copula{C}

\global\long\def\copula{c}

\global\long\def\ctilde{\widetilde{\copula}}

\global\long\def\pdf{f}

\global\long\def\fbold{\boldsymbol{\pdf}}

\global\long\def\ftilde{\widetilde{\pdf}}

\global\long\def\fdelta{\pdf_{\delta}}

\global\long\def\cmf{F}

\global\long\def\cmftilde{\widetilde{\cmf}}

\global\long\def\pseudo{\cmf^{\leftarrow}}

\global\long\def\tpseudo{\widetilde{\cmf}^{\leftarrow}}

\global\long\def\inverse{\cmf^{-1}}

\global\long\def\tinverse{\widetilde{\cmf}^{-1}}

\global\long\def\Cu{u}

\global\long\def\bCu{\boldsymbol{u}}

\global\long\def\Cutilde{\widetilde{u}}

\global\long\def\bCutilde{\widetilde{\bCu}}


\global\long\def\para{\theta}

\global\long\def\PARA{\boldsymbol{\Theta}}

\global\long\def\hPARA{\widehat{\PARA}}

\global\long\def\bpara{\boldsymbol{\theta}}

\global\long\def\tbpara{\widetilde{\bpara}}

\global\long\def\tpara{\widetilde{\para}}

\global\long\def\hpara{\widehat{\para}}

\global\long\def\hyperpara{\xi}

\global\long\def\bhpara{\boldsymbol{\xi}}

\global\long\def\thyper{\widetilde{\hyperpara}}

\global\long\def\dtilde{\widetilde{\delta}}

\global\long\def\data{x}

\global\long\def\xbold{\boldsymbol{x}}

\global\long\def\xBar{\overline{\xbold}}

\global\long\def\xVar{\boldsymbol{s}}

\global\long\def\Data{\boldsymbol{X}}


\global\long\def\KL{\text{KL}}

\global\long\def\KLff{\KL{}_{\ftilde||f}}

\global\long\def\VB{\text{VB}}

\global\long\def\CVB{\text{CVB}}

\global\long\def\CEF{\text{CEF}}

\global\long\def\MR{\text{MR}}

\global\long\def\EM{\text{EM}}

\global\long\def\ICM{\text{ICM}}

\global\long\def\ELBO{\text{ELBO}}

\global\long\def\Converged{\text{[c]}}

\global\long\def\VBzeta{\zeta}

\global\long\def\Normalizing{Z}

\global\long\def\fzeta{\zeta}

\global\long\def\iVB{\nu}

\global\long\def\nVB{\iVB_{c}}


\global\long\def\weight{p}

\global\long\def\tweight{\widetilde{p}}

\global\long\def\bweight{\boldsymbol{\weight}}

\global\long\def\tbweight{\widetilde{\boldsymbol{\weight}}}

\global\long\def\cweight{\bar{\weight}}

\global\long\def\cbweight{\bar{\bweight}}

\global\long\def\tbWeight{\widetilde{\boldsymbol{P}}}

\global\long\def\qweight{q}

\global\long\def\tqweight{\widetilde{\qweight}}

\global\long\def\bqweight{\boldsymbol{\qweight}}

\global\long\def\tbqweight{\widetilde{\bqweight}}

\global\long\def\uweight{\gamma}

\global\long\def\tuweight{\widetilde{\uweight}}

\global\long\def\tukappa{\widetilde{\kappa}}

\global\long\def\transition{w}

\global\long\def\ttransition{\widetilde{\transition}}

\global\long\def\btransition{\boldsymbol{\transition}}

\global\long\def\bttransition{\widetilde{\boldsymbol{\transition}}}

\global\long\def\Transition{\boldsymbol{W}}

\global\long\def\tTransition{\widetilde{\Transition}}

\global\long\def\Label{l}

\global\long\def\bLabel{\boldsymbol{l}}

\global\long\def\bhLabel{\widehat{\boldsymbol{l}}}

\global\long\def\tLabel{\widetilde{\Label}}

\global\long\def\hLabel{\widehat{\Label}}

\global\long\def\tbLabel{\widetilde{\bLabel}}

\global\long\def\hbLabel{\widehat{\bLabel}}

\global\long\def\LABEL{\boldsymbol{L}}

\global\long\def\tLABEL{\widetilde{\LABEL}}

\global\long\def\hLABEL{\widehat{\LABEL}}

\global\long\def\element{\boldsymbol{\epsilon}}


\global\long\def\Radius{R}

\global\long\def\mean{\mu}

\global\long\def\bmean{\boldsymbol{\mean}}

\global\long\def\Mean{\boldsymbol{\Upsilon}}

\global\long\def\tmean{\widetilde{\mean}}

\global\long\def\tbmean{\widetilde{\boldsymbol{\mean}}}

\global\long\def\hbmean{\widehat{\boldsymbol{\mean}}}

\global\long\def\tMean{\widetilde{\boldsymbol{\Upsilon}}}

\global\long\def\bcMean{\bar{\Mean}}

\global\long\def\hMean{\widehat{\Mean}}

\global\long\def\bcmean{\overline{\boldsymbol{\mean}}}

\global\long\def\bhmean{\widehat{\boldsymbol{\mean}}}

\global\long\def\std{\sigma}

\global\long\def\tstd{\widetilde{\std}}

\global\long\def\covmatrix{\boldsymbol{\Sigma}}

\global\long\def\tcovmatrix{\widetilde{\boldsymbol{\Sigma}}}

\global\long\def\correff{\rho}

\global\long\def\tcorreff{\widetilde{\rho}}

\global\long\def\rcorr{\beta}

\global\long\def\trcorr{\tilde{\rcorr}}

\begin{abstract}
Variational Bayes (VB), also known as independent mean-field approximation,
has become a popular method for Bayesian network inference in recent
years. Its application is vast, e.g. in neural network, compressed
sensing, clustering, etc. to name just a few. In this paper, the independence
constraint in VB will be relaxed to a conditional constraint class,
called copula in statistics. Since a joint probability distribution
always belongs to a copula class, the novel copula VB (CVB) approximation
is a generalized form of VB. Via information geometry, we will see
that CVB algorithm iteratively projects the original joint distribution
to a copula constraint space until it reaches a local minimum Kullback-Leibler
(KL) divergence. By this way, all mean-field approximations, e.g.
iterative VB, Expectation-Maximization (EM), Iterated Conditional
Mode (ICM) and k-means algorithms, are special cases of CVB approximation. 

For a generic Bayesian network, an augmented hierarchy form of CVB
will also be designed. While mean-field algorithms can only return
a locally optimal approximation for a correlated network, the augmented
CVB network, which is an optimally weighted average of a mixture of
simpler network structures, can potentially achieve the globally optimal
approximation for the first time. Via simulations of Gaussian mixture
clustering, the classification's accuracy of CVB will be shown to
be far superior to that of state-of-the-art VB, EM and k-means algorithms.
\end{abstract}


\section{Introduction}

Originally, the idea of mean-field theory is to approximate an interacting
system by a non-interacting system, such that the mean values of system's
nodes are kept unchanged \cite{VH:PhDThesis:14}. Variational Bayes
(VB) is a redefined method of mean-field theory, in which the joint
probability distribution $f_{\bpara}$ of a system is approximated
by a free-form independent distribution $\ftilde_{\bpara}=\prod_{\idim=1}^{\ndim}\ftilde_{\para_{\idim}}$,
such that the Kullback-Leibler (KL) divergence $\KL_{\ftilde_{\bpara}||\pdf_{\bpara}}$
is minimized \cite{AQ::Book:06}, $\bpara\TRIANGLEQ\setd{\para}{\ndim}$.
The term ``variational'' in VB originates from ``calculus of variations''
in differential mathematics, which is used to find the derivative
of KL divergence over distribution space \cite{VB:proof:Lagrange:VEM:03,VB:proof:Lagrange:book:15}. 

The VB approximation is particularly useful for estimating unknown
parameters in a complicated system. If the true value of parameters
$\bpara$ is unknown, we assume they follow a probabilistic model
a-priori. We then apply Bayesian inference, also called inverse probability
in the past \cite{ch2:origin:Bayes:Bayes1763,ch2:BK:statistics:History86},
to minimizing the expected loss function between true value $\bpara$
and posterior estimate $\widehat{\bpara}(\xbold)$. In practice, the
computational complexity of posterior estimate $\widehat{\bpara}(\xbold)$
often grows exponentially with arriving data $\xbold$ and, hence,
yields the curse of dimensionality \cite{Bayes:CurseOfDimension:97}.
For tractable computation, as shown in this paper, the VB algorithm
iteratively projects the originally complex distribution into simpler
independent class of each unknown parameter $\para_{\idim}$, one
by one, until the KL divergence converges to a local minimum. For
this reason, the VB algorithm has been used extensively in many fields
requiring tractable parameter's inference, e.g. in neural networks
\cite{VBcite:neural:NIPS:11}, compressed sensing \cite{VBcite:Compress:10},
data clustering \cite{VBcite:clustering:14}, etc. to name just a
few. 

Nonetheless, the independent class is too strict in practice, particularly
in case of highly correlated model \cite{AQ::Book:06}. In order to
capture the dependence in a probabilistic model, a popular method
in statistics is to consider a copula class. The key idea is to separate
the dependence structure, namely copula, of a joint distribution from
its marginal distributions. In this way, the copula concept is similar
to nonnegative compatibility functions over cliques in factor graphs
\cite{VB:proof:MichealJordan:08,VB:FreeEnergy:Yedidia:05}, although
the compatibility functions are not probability distributions like
copula. Indeed, a copula $\copula_{\bpara}$ is a joint distribution
whose marginals are uniform, as originally proposed in \cite{copula:Sklar:59}.
For example, the copula of a bivariate discrete distribution is a
bi-stochastic matrix, whose sum of any row or any column is equal
to one \cite{copula:BOOK:Carlo:15,copula:discrete:bimatrix:06}. More
generally, by Sklar's theorem \cite{copula:Sklar:59,copula:Sklar:96},
any joint distribution $f_{\bpara}$ can always be written in copula
form $\pdf_{\bpara}=\copula_{\bpara}\prod_{\idim=1}^{\ndim}\pdf_{\para_{\idim}}$,
in which $\copula_{\bpara}$ fully describes the inter-dependence
of variables in a joint distribution. For independent class, the copula
density $\copula_{\bpara}$ is merely a constant and equal to one
everywhere \cite{copula:BOOK:Carlo:15}. 

In this paper, the novel copula VB (CVB) approximation $\ftilde_{\bpara}=\ctilde_{\bpara}\prod_{\idim=1}^{\ndim}\ftilde_{\para_{\idim}}$
will extend the independent constraint in VB to a copula class of
dependent distributions. After fixing the distributional form of $\ctilde_{\bpara}$,
the CVB iteratively updates the free-form marginals $\ftilde_{\para_{\idim}}$
one by one, similarly to traditional VB, until KL divergence $\KL_{\ftilde_{\bpara}||\pdf_{\bpara}}$
converges to a local minimum. The CVB approximation will become exact
if the form of $\ctilde_{\bpara}$ is the same as that of original
copula $\copula_{\bpara}$. The study of copula form $\copula_{\bpara}$
is still an active field in probability theory and statistics \cite{copula:BOOK:Carlo:15},
owing to its flexibility to modeling the dependence of any joint distribution
$\pdf_{\bpara}$. Also, because the mutual information $f_{\bpara}$
is equal to entropy of its copula $\copula_{\bpara}$ \cite{copula:Entropy:11},
the copula is currently an interesting topic for information criterions
\cite{copula:cite:info:14a,copula:cite:info:14b}.

In information geometry, the KL divergence is a special case of the
Bregman divergence, which, in turn, is a generalized concept of distance
in Euclidean space \cite{Shun-ichiAmari:BOOK:16}. By reinterpreting
the KL minimization in VB as the Bregman projection, we will see that
CVB, and its special case VB, iteratively projects the original distribution
to a fixed copula constraint space until convergence.\textcolor{green}{{}
}Then, similar to the fact that the mean is the point of minimum total
distance to data, an augmented CVB approximation will also be designed
as a distribution of minimum total Bregman divergence to the original
distribution in this paper.

Three popular special cases of VB will also be revisited in this paper,
namely Expectation-Maximization (EM) \cite{EM:majorization:04,EM:majorization:17},
Iterated Conditional Mode (ICM) \cite{ch4:art:ICM:Besag86,ch4:art:ICM:localMAP_2006}
and k-means algorithms \cite{kmean:LLoyd:82,kmean:cite:15}. In literature,
the well-known EM algorithm was shown to be a special case of VB \cite{VH:PhDThesis:14,AQ::Book:06},
in which one of VB's marginal is restricted to a point estimate via
Dirac delta function. In this paper, the EM algorithm will be shown
that it does not only reach a local minimum KL divergence, but it
may also return a local maximum-a-posteriori (MAP) point estimate
of the true marginal distribution. This justifies the superiority
of EM algorithm to VB in some cases of MAP estimation, since the peaks
in VB marginals might not be the same as those of true marginals. 

If all VB marginals are restricted to Dirac delta space, the iterative
VB algorithm will become ICM algorithm, which returns a locally joint
MAP estimate of the original distribution. Also, for standard Normal
mixture clustering, the ICM algorithm is equivalent to the well-known
k-means algorithm, as shown in this paper. The k-means algorithm is
also equivalent to the Lloyd-Max algorithm \cite{kmean:LLoyd:82},
which has been widely used in quantization context \cite{VH:CVA}.

For illustration, the CVB and its special cases mentioned above will
be applied to two canonical models in this paper, namely bivariate
Gaussian distribution and Gaussian mixture clustering. By tuning the
correlation in these two models, the performance of CVB will be shown
to be superior to that of state-of-the-art mean-field methods like
VB, EM and k-means algorithm. An augmented CVB form for a generic
Bayesian network will also be studied and applied to this Gaussian
mixture model.

\subsection{Related works}

Although some generalized forms of VB have been proposed in literature,
most of them are merely variants of mean-field approximations and,
hence, still confined within independent class. For example, in \cite{name:CVB:01,name:CVB:06},
the so-called Conditionally Variational algorithm is an application
of traditional VB to a joint conditional distribution $\ftilde(\bpara|\hyperpara)=\prod_{\idim=1}^{\ndim}\ftilde_{\idim}(\para_{\idim}|\hyperpara)$,
given a latent variable $\hyperpara$. Hence, different to CVB above,
the approximated marginal $\ftilde_{\hyperpara}$ was not updated
in their scheme. In \cite{VBcite:Generalized:Jordan:03}, the so-called
generalized mean-field algorithm is merely to apply the traditional
VB method to the independent class of a set of variables, i.e. each
$\para_{\idim}$ consists of a set of variables. In \cite{VBcite:structureVB:05},
the so-called Structured Variational inference is the same as the
generalized mean-field, except that the dependent structure inside
the set $\para_{\idim}$ is also specified. In summary, they are different
ways of applying traditional VB, without changing the VB's updating
formula. In contrast, the CVB in this paper involves new tractable
formulae and broader copula constraint class.

The closest form to the CVB of this paper is the so-called Copula
Variational inference in \cite{CVB:Blei:gradient:15}, which fixes
the form of approximated distribution $\ftilde_{\bpara|\hyperpara}=\ctilde_{\bpara|\hyperpara}\prod_{\itime=1}^{\ntime}\ftilde_{\para_{\itime}|\hyperpara}$
and applies gradient decent method upon the latent variable $\hyperpara$
in order to find a local minimum of KL divergence. In contrast, the
CVB in this paper is a free-form approximation, i.e. it does not impose
any particular form initially, and provides higher-order moment's
estimates than a mere point estimate. Hence, the fixed-form constraint
class in their Copula Variational inference is much more restricted
than the free-form copula constraint class of CVB in this paper. Also,
the iterative computation for CVB will be given in closed form with
low complexity, rather than relying point estimates of gradient decent
methods. 

\subsection{Contributions and organization}

The contributions of this paper are summarized as follows:
\begin{itemize}
\item A novel copula VB (CVB) algorithm, which extends the independent constraint
class of traditional VB to a copula constraint class, will be given.
The convergence of CVB will be proved via three methods: Lagrange
multiplier method in calculus of variation, Jensen's inequality and
the Bregman projection in information geometry. The two former methods
have been used in literature for proof of convergence of traditional
VB, while the third method is new and provides a unified scheme for
the former two methods.
\item The EM, ICM and k-means algorithms will be shown to be special cases
of the traditional VB, i.e. they all locally minimize the KL divergence
under a fixed-form independent constraint class.
\item An augmented form of CVB, namely hierarchical CVB approximation, with
linear computational complexity for a generic Bayesian network will
also be provided.
\item In simulations, the CVB algorithm for Gaussian mixture clustering
will be illustrated. The classification's performance of CVB will
be shown to be far superior to that of VB, EM and k-means algorithms
for this model.
\end{itemize}
The paper is organized as follows: since the Bregman projection in
information geometry is insightful and plays central role to VB method,
it will be presented first in section \ref{sec:Information-geometry}.
The definition and property of copula will then be introduced in section
\ref{sec:Copula-theory}. The novel copula VB (CVB) method and its
special cases will be presented in section \ref{sec:Copula-Variational-Bayes}.
The computational flow of CVB for a Bayesian network is studied in
section \ref{sec:Bayesian-network} and will be applied to simulations
in section \ref{sec:Case-study}. The paper is then concluded in section
\ref{sec:Conclusion}. 

Note that, for notational simplicity, the notion of probability density
function (p.d.f.) for continuous random variable (r.v.) in this paper
will be implicitly understood as the probability mass function (p.m.f)
in the case of discrete r.v., when the context is clear.

\section{Information geometry \label{sec:Information-geometry}}

In this section, we will revisit a geometric interpretation of one
of fundamental measures in information theory, namely Kullback-Leibler
(KL) divergence, which is also the central part of VB (i.e. mean-field)
approximation. For this purpose, the Bregman divergence, which is
a generalization of both Euclidean distance and KL divergence, will
be defined first. Two important theorems, namely Bregman pythagorean
theorem and Bregman variance theorem, will then be presented. These
two theorems generalize the concept of Euclidean projection and variance
theorem to the probabilistic functional space, respectively. The Bregman
divergence is also a key concept in the field of information geometry
in literature \cite{Bregman:BayesFunction:08,Shun-ichiAmari:BOOK:16}.

\subsection{Bregman divergence for vector space}

For simplicity, in this subsection, we will define Bregman divergence
for real vector space first, which helps us visualize the Bregman
pythagorean theorem later.

\begin{figure}
\centering{}\includegraphics[width=0.7\linewidth]{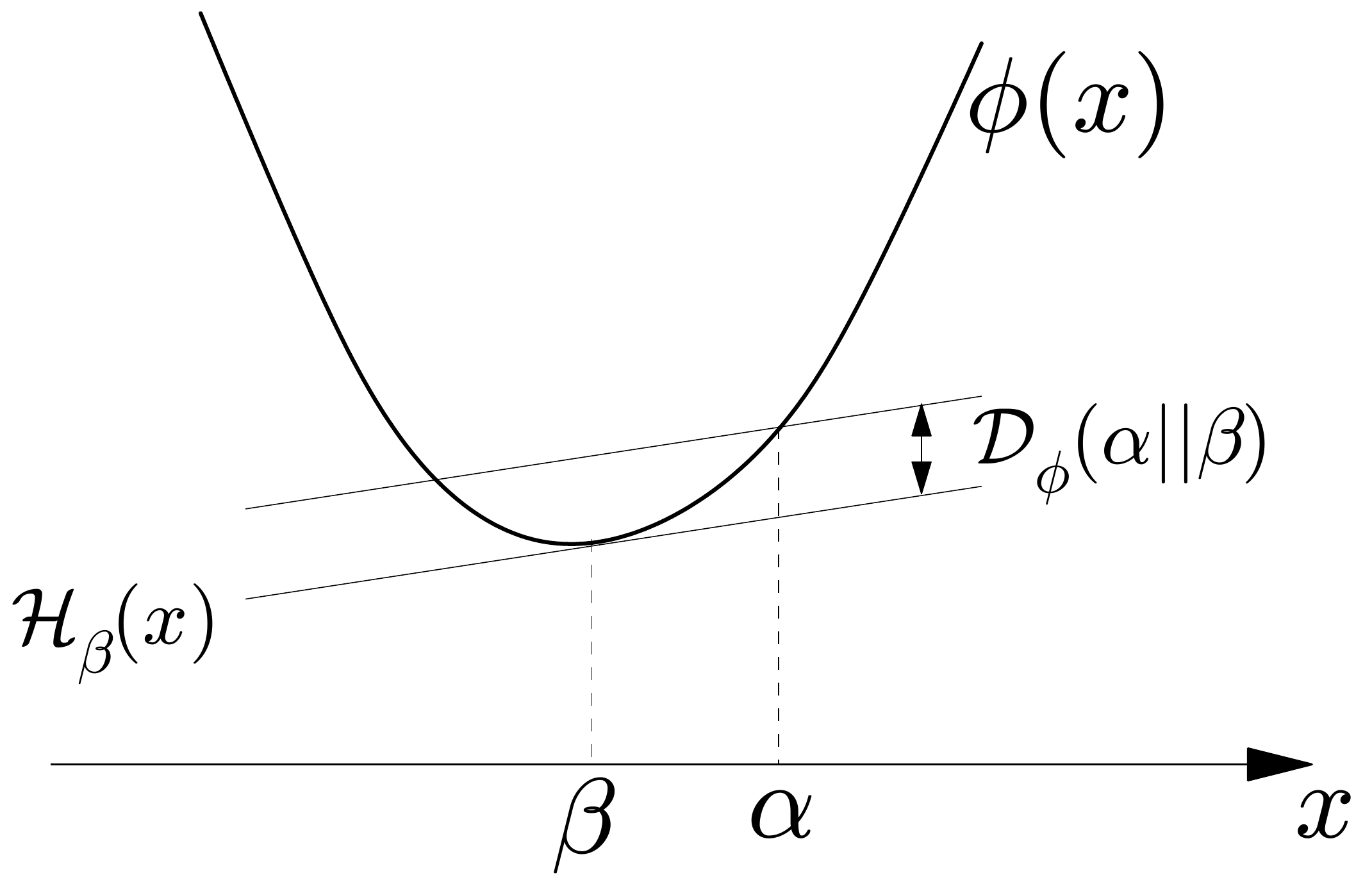}\caption{\label{fig:Bregman}Illustration of Bregman divergence $\protect\BD$
for convex function $\protect\Bfunc$. The hyperplane $\protect\BH_{\protect\bBb}(\protect\bBa)\protect\TRIANGLEQ\protect\Bfunc(\protect\bBb)+\left\langle \protect\bBa-\protect\bBb,\nabla\protect\Bfunc(\protect\bBb)\right\rangle $
is tangent to $\protect\Bfunc$ at point $\protect\bBb$. Note that,
if $\protect\Bfunc(\protect\bBa)$ is equal to the continuous entropy
function $H_{\protect\bBa}(\protect\bBa)$, the hyperplane $\protect\BH_{\protect\bBb}(\protect\bBa)$
is equal to the cross entropy $H_{\protect\bBb}(\protect\bBa)$ from
$\protect\bBa$ to $\protect\bBb$ and $\protect\BD(\protect\bBa||\protect\bBb)=H_{\protect\bBa}(\protect\bBa)-H_{\protect\bBb}(\protect\bBa)$
is equal to Kullback-Leibler (KL) divergence (c.f. section \ref{subsec:Kullback-Leibler}).}
\end{figure}

\begin{figure}
\centering{}\includegraphics[width=0.7\linewidth]{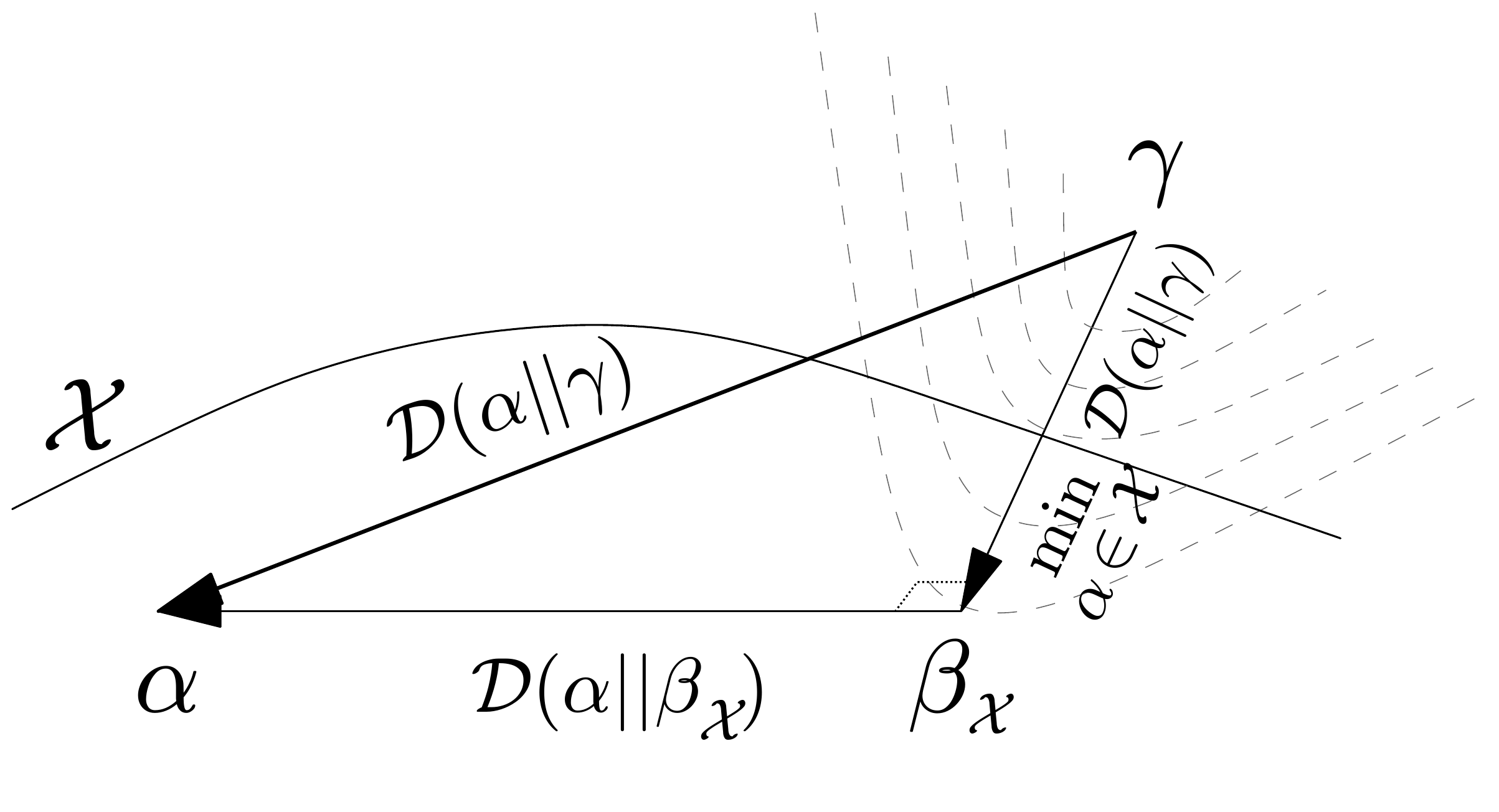}\caption{\label{fig:Pythagore}Illustration of Bregman pythagorean inequality
over closed convex set $\protect\bBa\in\protect\BX$. The point $\protect\bBb_{\protect\BX}\in\protect\BX$
is called the Bregman projection of $\protect\bBc\in\protect\REAL^{\protect\ndim}$
onto $\protect\BX\subset\protect\REAL^{\protect\ndim}$. The dashed
contours represent the convexity of $\protect\BD(\protect\bBb||\protect\bBc)$
over arbitrary point $\protect\bBb\in\protect\REAL^{\protect\ndim}$
in general.}
\end{figure}

\begin{defn}
(Bregman divergence) \label{def:(Bregman-divergence)}\\
Let $\Bfunc:\REAL^{\ndim}\rightarrow\REAL$ be a strictly convex and
differentiable function. Given two points $\bBa,\bBb\in\REAL^{\ndim}$,
with $\bBa\TRIANGLEQ\setv{\Ba}{\ndim}^{\transpose}$ and $\bBb\TRIANGLEQ\setv{\Bb}{\ndim}^{\transpose}$,
the Bregman divergence $\BD:\REAL^{\ndim}\times\REAL^{\ndim}\rightarrow\REAL^{+}$,
with $\REAL^{+}\TRIANGLEQ[0,+\infty)$ , is defined as follows:
\begin{align}
\BD(\bBa||\bBb) & \TRIANGLEQ\BH_{\bBa}(\bBa)-\BH_{\bBb}(\bBa)\label{eq:BREGMAN}\\
 & =\Bfunc(\bBa)-\Bfunc(\bBb)-\left\langle \bBa-\bBb,\nabla\Bfunc(\bBb)\right\rangle ,\nonumber 
\end{align}
where $\nabla$ is gradient operator, $\left\langle \cdot,\cdot\right\rangle $
denotes inner product and $\BH_{\bBb}(\bBa)\TRIANGLEQ\Bfunc(\bBb)+\left\langle \bBa-\bBb,\nabla\Bfunc(\bBb)\right\rangle $
is hyperplane tangent to $\Bfunc$ at point $\bBb$, as illustrated
in Fig. \ref{fig:Bregman}.
\end{defn}
For simplicity, the notations $\BD$ and $\BD_{\Bfunc}$ are used
interchangeably in this paper when the context is clear. Some well-known
properties of Bregman divergence (\ref{eq:BREGMAN}) are summarized
below:
\begin{prop}
(Bregman divergence's properties)\label{prop:Bregman-properties}
\end{prop}
\begin{enumerate}
\item Non-negativity: $\BD(\bBa||\bBb)\geq0$.
\item Equality:$\BD(\bBa||\bBb)=0\Leftrightarrow\bBa=\bBb$. 
\item Asymmetry: $\BD(\bBa||\bBb)\neq\BD(\bBb||\bBa)$ in general. 
\item Convexity: $\BD(\bBa||\bBb)$ is convex over $\bBa$, but not over
$\bBb$ in general.
\item Gradient: $\nabla_{\bBa}\BD(\bBa||\bBb)=\nabla\Bfunc(\bBa)-\nabla\Bfunc(\bBb)$
and $\nabla_{\bBb}\BD(\bBa||\bBb)=\nabla^{2}\Bfunc(\bBb)[\bBb-\bBa]$.
\item Affine equivalence class: $\BD_{\Bfunc}(\bBa||\bBb)=\BD_{\widetilde{\Bfunc}}(\bBa||\bBb)$
if $\widetilde{\Bfunc}(\bBx)=\Bfunc(\bBx)+\left\langle \bBc,\bBx\right\rangle +c$,
e.g. $\widetilde{\Bfunc}(\bBx)=\BD_{\Bfunc}(\bBx||\bBb)$. 
\item Three-point property: 
\end{enumerate}
\begin{align}
\BD(\bBa||\bBb)+\BD(\bBb||\bBc)-\BD(\bBa||\bBc) & =\left\langle \bBb-\bBa,\underset{\nabla_{\bBb}\BD(\bBb||\bBc)}{\underbrace{\nabla\phi(\bBb)-\nabla\phi(\bBc)}}\right\rangle \label{eq:3POINTs}
\end{align}
The points $\{\bBa,\bBb,\bBc\}$ in (\ref{eq:3POINTs}) are called
\textit{Bregman orthogonal} at point $\bBb$ if $\left\langle \bBb-\bBa,\nabla\phi(\bBb)-\nabla\phi(\bBc)\right\rangle =0$. 
\begin{IEEEproof}
All properties~1-7 are direct consequence of Bregman definition (\ref{eq:BREGMAN}).
The derivation of well-known properties~1-4 and 6-7 can be found
in \cite{Shun-ichiAmari:BOOK:16,Bregman:cluster:Voronoi:10} and \cite{Bregman:cluster:affine:05,Shun-ichiAmari:Bregman:09},
respectively, for any $\bBx,\bBc\in\REAL^{\ndim}$, $c\in\REAL$.
 In property~6, since $\BD_{\Bfunc}(\bBx||\bBb)$ is both convex
and affine over $\bBx$, as defined in (\ref{eq:BREGMAN}), we can
assign $\widetilde{\Bfunc}(\bBx)=\BD_{\Bfunc}(\bBx||\bBb)$. In property~7,
the $\nabla_{\bBb}$ form is a consequence of gradient property. The
gradient property, i.e. the property~5, can be derived from definition
(\ref{eq:BREGMAN}) as follows: $\nabla_{\bBa}\BD_{\Bfunc}(\bBa||\bBb)=\nabla_{\bBa}\Bfunc(\bBa)-\nabla_{\bBa}\left\langle \bBa-\bBb,\nabla\Bfunc(\bBb)\right\rangle =\nabla_{\bBa}\Bfunc(\bBa)-\nabla\Bfunc(\bBb)$.
Similarly, from (\ref{eq:BREGMAN}), we have $\nabla_{\bBb}\BD_{\Bfunc}(\bBa||\bBb)=-\nabla_{\bBb}\Bfunc(\bBb)-\nabla_{\bBb}\left\langle \bBa-\bBb,\nabla\Bfunc(\bBb)\right\rangle =-\nabla_{\bBb}\Bfunc(\bBb)-\left(-\nabla\Bfunc(\bBb)+\nabla^{2}\Bfunc(\bBb)[\bBa-\bBb]\right)=\nabla^{2}\Bfunc(\bBb)[\bBb-\bBa]$,
in which $\nabla^{2}$ denotes Hessian matrix operator.
\end{IEEEproof}
\begin{rem}
The gradient property gives us some insight on Bregman divergence.
For example, from gradient property, we can see that $\bBa=\bBb$
is the stationary and minimum point of $\BD(\bBa||\bBb)$. Also, $\BD(\bBa||\bBb)$
is convex over $\bBa$ but not over $\bBb$ since $\Bfunc(\cdot)$
is a convex function, as shown intuitively in the form of $\nabla_{\bBa}\BD(\bBa||\bBb)$
and $\nabla_{\bBb}\BD(\bBa||\bBb)$, respectively. The gradient form
$\nabla_{\bBb}\BD(\bBb||\bBc)$ in (\ref{eq:3POINTs}) represents
the changing value of $\BD(\bBb||\bBc)$ over $\bBb$ and, hence,
explains the three-point property intuitively, as illustrated in Fig.
\ref{fig:Pythagore}.
\end{rem}
Let us now consider the most important property of Bregman divergence
in this paper, namely Bregman pythagorean inequality, which defines
the Bregman projection over a closed convex subset $\BX\subset\REAL^{\ndim}$. 

\begin{thm}
(Bregman pythagorean inequality)\label{thm:Bregman-Pythagorean}\\
Let $\BX$ be a closed convex subset in $\REAL^{\ndim}$. For any
points $\bBa\in\BX$ and $\bBc\in\REAL^{\ndim}$, we have: 
\begin{equation}
\BD(\bBa||\bBb_{\BX})+\BD(\bBb_{\BX}||\bBc)\leq\BD(\bBa||\bBc),\label{eq:PYTHAGORE}
\end{equation}
where the unique point $\bBb_{\BX}$ is called the Bayesian projection
of $\bBc$ onto $\BX$ and defined as follows: 
\begin{equation}
\bBb_{\BX}\TRIANGLEQ\argmin_{\bBa\in\BX}\BD(\bBa||\bBc).\label{eq:Projection_point}
\end{equation}
From three-point property (\ref{eq:3POINTs}), we can see that the
Bregman pythagorean inequality in (\ref{eq:PYTHAGORE}) becomes equality
for all $\bBa\in\BX$ if and only if $\BX$ is an affine set (i.e.
the triple points $\{\bBa,\bBb_{\BX},\bBc\}$ are Bregman orthogonal
at $\bBb_{\BX}$, $\forall\bBa\in\BX$).
\end{thm}
\begin{IEEEproof}
Note that $\bBb_{\BX}$, as defined in (\ref{eq:Projection_point}),
is not necessarily unique if $\BX$ is not convex \cite{Shun-ichiAmari:BOOK:16}.
The uniqueness of $\bBb_{\BX}$ (\ref{eq:Projection_point}) for convex
set $\BX$ can be proved either via contradiction \cite{Bregman:cluster:Voronoi:10}
or via convexity of $\BX$ in three-point property (\ref{eq:3POINTs}),
c.f. \cite{Shun-ichiAmari:BOOK:16,Bregman:projection:14}. Substituting
$\bBb_{\BX}$ in (\ref{eq:Projection_point}) to three-point property
(\ref{eq:3POINTs}) yields the Bregman pythagorean inequality (\ref{eq:PYTHAGORE}). 
\end{IEEEproof}
\begin{figure}
\centering{}\includegraphics[width=1\linewidth]{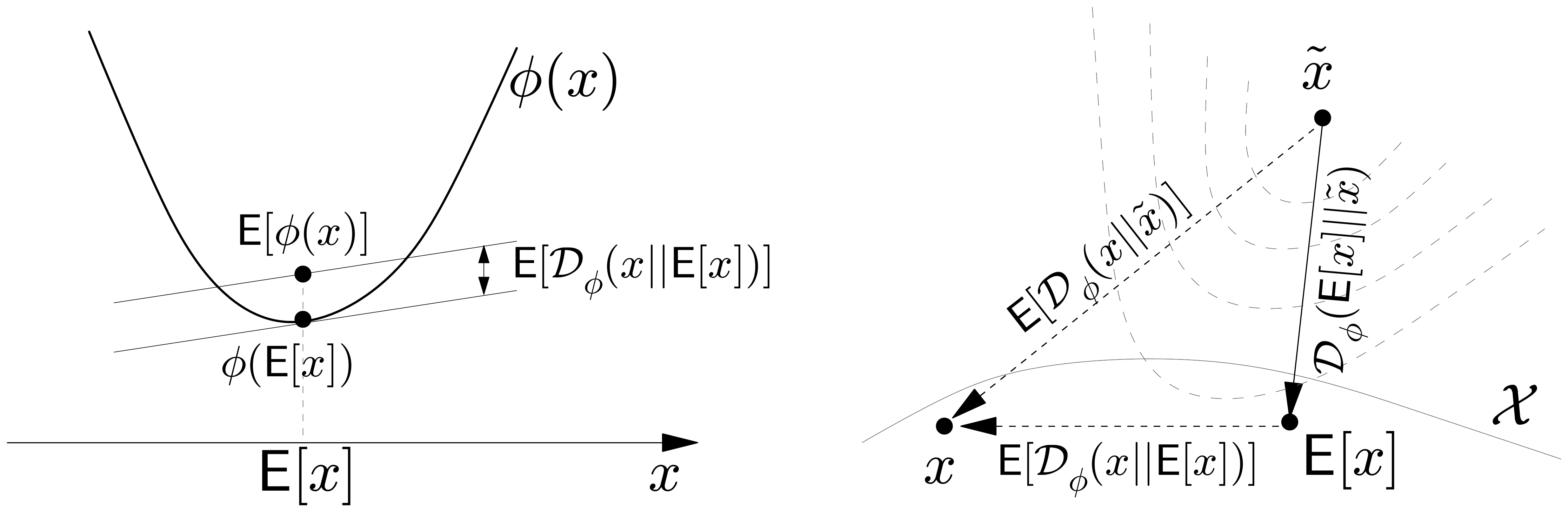}\caption{\label{fig:Jensen}Illustration of equivalence between Jensen's inequality
(left) and Bregman variance theorem (right). Similar to Fig. \ref{fig:Bregman}
and Fig. \ref{fig:Pythagore}, the dashed contours on the right represent
the convexity of $\protect\BD_{\phi}(\protect\bBx||\widetilde{\protect\bBx})$
over $\protect\bBx$, which, in turn, can be regarded as another convex
function $\widetilde{\protect\Bfunc}$ for Jensen's inequality on
the left.}
\end{figure}

Owing to Bregman divergence, we also have a geometrical interpretation
of probabilistic variance, as shown in the following theorem on Jensen's
inequality:
\begin{thm}
(Bregman variance theorem - Jensen's inequality) \label{thm:Bregman-variance-Jensen}\\
Let $\bBx\in\REAL^{\ndim}$ be a r.v. with mean $\EXPECTATION[\bBx]$
and variance $\VARIANCE[\bBx]$. The Bregman variance $\text{\ensuremath{\VARIANCE}}_{\Bfunc}[\bBx]$
is defined as follows: 
\begin{equation}
\text{\ensuremath{\VARIANCE}}_{\Bfunc}[\bBx]\text{\ensuremath{\TRIANGLEQ}}\EXPECTATION[\BD_{\phi}(\bBx||\EXPECTATION[\bBx])]=\EXPECTATION[\phi(\bBx)]-\phi(\EXPECTATION[\bBx])\geq0.\label{eq:Jensen inequality}
\end{equation}
Equivalently, we have: 
\begin{equation}
\text{\ensuremath{\VARIANCE}}_{\Bfunc}[\bBx]\text{\ensuremath{\TRIANGLEQ}}\EXPECTATION[\BD_{\phi}(\bBx||\EXPECTATION[\bBx])]=\EXPECTATION[D_{\phi}(\bBx||\widetilde{\bBx})]-\BD_{\phi}(\EXPECTATION[\bBx]||\widetilde{\bBx})\geq0\label{eq:Bregman_var}
\end{equation}
for any fixed point $\widetilde{\bBx}\in\REAL^{\ndim}$. The right
hand side (r.h.s.) of (\ref{eq:Jensen inequality}) is called Jensen's
inequality in literature, i.e. $\EXPECTATION(\phi(\bBx))\geq\phi(\EXPECTATION(\bBx))$,
for any convex function $\phi$ \cite{Jensen:cite:16}. Also, from
(\ref{eq:Bregman_var}), we have:
\begin{equation}
\bBx_{0}\TRIANGLEQ\EXPECTATION[\bBx]=\argmin_{\widetilde{\bBx}}\EXPECTATION[D(\bBx||\widetilde{\bBx})],\label{eq:Bregman_mean}
\end{equation}
as illustrated in Fig. \ref{fig:Jensen}.
\end{thm}
\begin{IEEEproof}
Let us show the proof in reverse way. Firstly, the mean property (\ref{eq:Bregman_mean})
is a consequence of (\ref{eq:Bregman_var}), i.e. we have: $\EXPECTATION[D(\bBx||\widetilde{\bBx})]=\EXPECTATION[\BD(\bBx||\EXPECTATION[\bBx])]+\BD(\EXPECTATION[\bBx]||\widetilde{\bBx})$
and $\BD(\EXPECTATION[\bBx]||\widetilde{\bBx})=0\Leftrightarrow\widetilde{\bBx}=\EXPECTATION[\bBx]$.
Secondly, by replacing $\phi(\bBx)$ in (\ref{eq:Jensen inequality})
with $\widetilde{\Bfunc}(\bBx)=D_{\Bfunc}(\bBx||\widetilde{\bBx})$,
the form (\ref{eq:Bregman_var}) is equivalent to (\ref{eq:Jensen inequality}),
owing to the affine equivalence property in Proposition \ref{prop:Bregman-properties}.
Lastly, the form (\ref{eq:Jensen inequality}) is a direct derivation
from Bregman definition (\ref{eq:BREGMAN}), with $\bBa=\bBx$ and
$\bBb=\EXPECTATION[\bBx]$, as follows: $\BD(\bBx||\EXPECTATION[\bBx])=\Bfunc(\bBx)-\Bfunc(\EXPECTATION[\bBx])-\left\langle \bBx-\EXPECTATION[\bBx],\nabla\Bfunc(\EXPECTATION[\bBx])\right\rangle $
and, hence, $\EXPECTATION[\BD(\bBx||\EXPECTATION[\bBx])]=\EXPECTATION[\Bfunc(\bBx)]-\EXPECTATION[\Bfunc(\EXPECTATION[\bBx])]-\left\langle \underset{=0}{\underbrace{\EXPECTATION[\bBx]-\EXPECTATION[\bBx]}},\nabla\Bfunc(\EXPECTATION[\bBx])\right\rangle $. 
\end{IEEEproof}
\begin{rem}
\label{rem:E=00005Bx=00005D}Although we have $\VARIANCE[\bBx]\neq\text{\ensuremath{\VARIANCE}}_{\Bfunc}[\bBx]$
in general, the mean $\EXPECTATION[\bBx]$ is the same minimum point
for any expected Bregman divergence, as shown in (\ref{eq:Bregman_mean}).
This notable property of the mean has been exploited extensively for
Bregman k-means algorithms in literature \cite{Bregman:cluster:Voronoi:10,Bregman:cluster:affine:05}. 
\end{rem}
A list of Bregman divergences, corresponding to different functional
forms of $\Bfunc(\bBx)$, can be found feasibly in literature, e.g.
in \cite{Shun-ichiAmari:BOOK:16,Bregman:cluster:centroid:09}. Let
us recall two most popular forms below.

\subsubsection{Euclidean distance}

A special case of Bregman divergence is squared Euclidean distance
\cite{Bregman:cluster:affine:05}: 
\begin{equation}
\BD_{\Bfunc_{E}}(\bBa||\bBb)=||\bBa-\bBb||^{2},\ \text{with}\ \Bfunc_{E}(\bBx)\TRIANGLEQ||\bBx||^{2},\label{eq:Euclidean}
\end{equation}
where $||\cdot||$ denotes $\mathcal{L}_{2}$-norm for elements of
a vector or matrix. In this case, the Bregman pythagorean theorem
(\ref{eq:PYTHAGORE}) becomes the traditional Pythagorean theorem
and the Bregman variance (\ref{eq:Jensen inequality}) becomes the
traditional variance theorem, i.e. $\text{\ensuremath{\VARIANCE}}_{\Bfunc_{E}}[\bBx]=\VARIANCE[\bBx]=\EXPECTATION[||\bBx||^{2}]-||\EXPECTATION[\bBx]||^{2}$.

\subsubsection{Kullback-Leibler (KL) divergence\label{subsec:Kullback-Leibler}}

Another popular case of Bregman divergence is the KL divergence \cite{Bregman:cluster:affine:05}:
\[
\KL(\bBa||\bBb)\TRIANGLEQ\BD_{\KL}(\bBa||\bBb)=\sum_{\idim=1}^{\ndim}\Ba_{\idim}\log\frac{\Ba_{\idim}}{\Bb_{\idim}}-\sum_{\idim=1}^{\ndim}\Ba_{\idim}+\sum_{\idim=1}^{\ndim}\Bb_{\idim},
\]
with $\Bfunc_{\KL}(\bBx)\TRIANGLEQ\sum_{\idim=1}^{\ndim}\Bx_{\idim}\log\Bx_{\idim}$,
$\forall\Bx_{\idim}\in\REAL^{+}$. More generally, it can be shown
that \cite{Bregman:cluster:centroid:09}: 
\begin{equation}
\KLff\TRIANGLEQ\BD_{\KL}(\ftilde||f)=\EXPECTATION_{\ftilde(\para)}\log\frac{\ftilde(\para)}{f(\para)},\label{eq:KL}
\end{equation}
where $\Bfunc_{\KL}(f(\para))\TRIANGLEQ\Entropy(\para)=\EXPECTATION_{f(\para)}\log f(\para)$
is the continuous entropy and $\BD_{\KL}(\ftilde||f)$ is the Bregman
divergence between two density distributions $\ftilde(\para)$ and
$f(\para)$, as presented below.

\subsection{Bregman divergence for functional space}

\begin{figure}
\centering{}\includegraphics[width=0.33\linewidth]{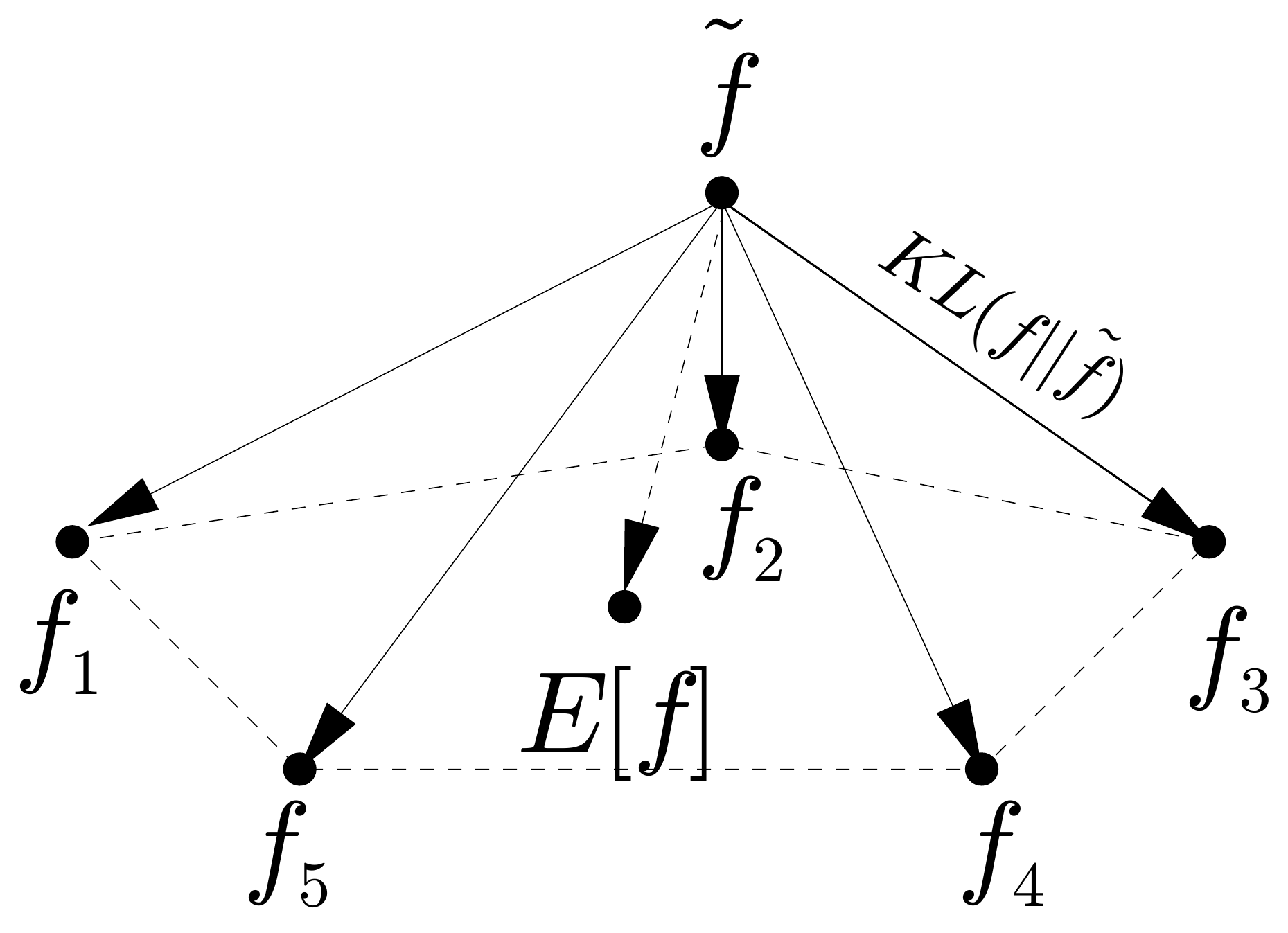}\includegraphics[width=0.67\linewidth]{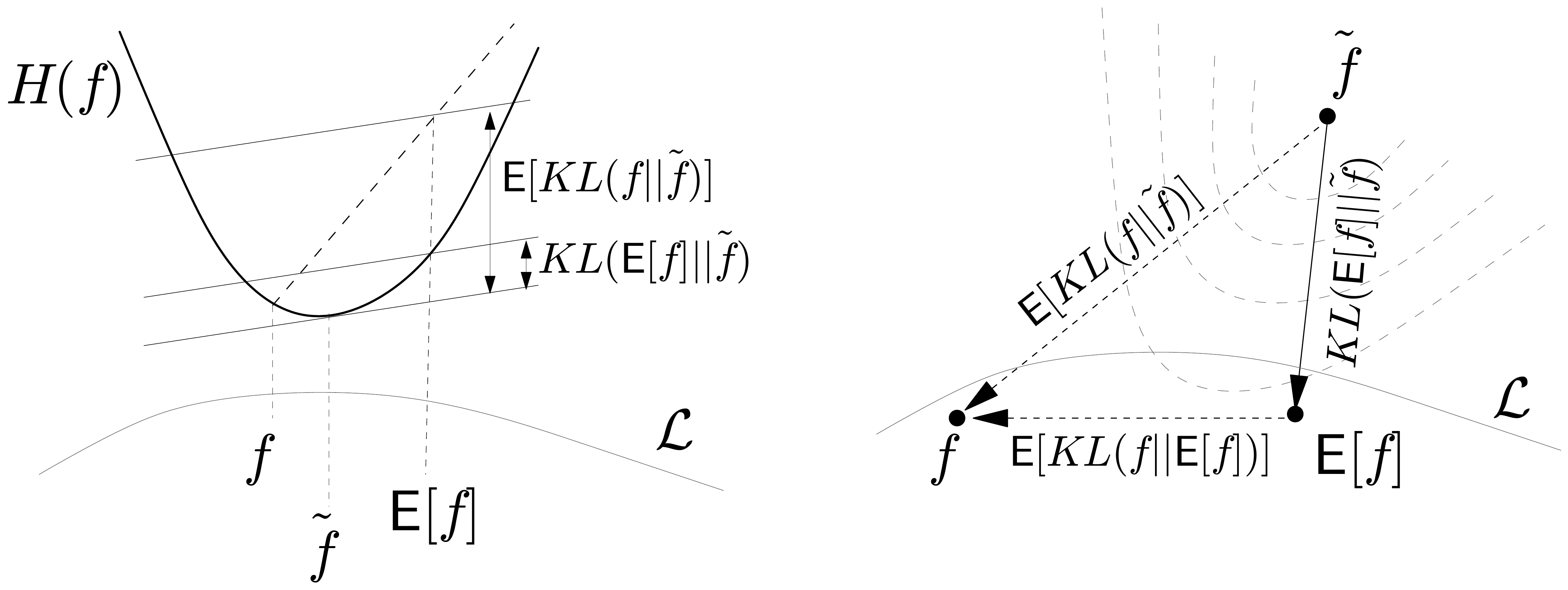}\caption{\label{fig:mixtureKLD}Application of Bregman variance theorem (\ref{eq:Bregman_mixture})
to KL divergence in distribution space $\protect\pdf\in\mathcal{L}$,
with the same convention in Fig. \ref{fig:Jensen}. As an example,
the mixture $\protect\pdf_{0}(\protect\para)\protect\TRIANGLEQ\protect\EXPECTATION[f]=\sum_{\protect\itime=1}^{5}\protect\Bp_{\protect\itime}\protect\pdf_{\protect\itime}(\protect\para)$
in (\ref{eq:mean_mixture}) must lie inside the polytope $\mathcal{L}=\{\protect\pdf_{1}(\protect\para),\ldots,\protect\pdf_{5}(\protect\para)\}$.
In middle sub-figure, $H(\protect\pdf)$ denotes the continuous entropy
over p.d.f. $\protect\pdf$. The mixture $\protect\ftilde=\protect\pdf_{0}=\protect\EXPECTATION[f]$
is then the minimum functional point of $\protect\EXPECTATION[\protect\KL(\protect\pdf||\protect\ftilde)]$,
which is also an upper bound of $\protect\KL(\protect\EXPECTATION[\protect\pdf]||\protect\ftilde)$
over $\protect\ftilde\in\mathcal{L}$, as shown in (\ref{eq:Bregman_mixture}-\ref{eq:argmin_mixture}).}
\end{figure}

In the calculus of variations, the Bregman divergence for vector space
is a special case of the Bregman divergence for functional space,
defined as follows:
\begin{defn}
(Bregman divergence for functional space) \cite{Bregman:BayesFunction:08}\\
Let $\Bfunc:\Lp(\theta)\rightarrow\REAL$ be a strictly convex and
twice Fréchet-differentiable functional over $\Lp$-normed space.
The Bregman divergence $\BD:\Lp(\theta)\times\Lp(\theta)\rightarrow\REAL^{+}$
between two functions $f,g\in\Lp(\theta)$ is defined as follows:
\begin{align}
\BD(f||g) & \TRIANGLEQ\Bfunc(f)-\Bfunc(g)-\delta\Bfunc(f-g;g),\label{eq:BREGMAN_func}
\end{align}
where $\delta\Bfunc(\cdot;g)$ is Fréchet derivative of $\Bfunc$
at $g$.
\end{defn}
Apart from gradient form, all well-known properties of Bregman divergence
in Proposition \ref{prop:Bregman-properties} are also valid for functional
space \cite{Bregman:BayesFunction:08,Bregman:FunctionalProof:08}.
Hence, we can feasibly derive the Bregman variance theorem for probabilistic
functional space, as follows:
\begin{prop}
(Bregman variance theorem for functions)\label{prop:Bregman-variance-function}\\
Let functional point $f(\para)$ be a r.v. drawn from the functional
space $\Lp(\theta)$ with functional mean $\EXPECTATION[f]\TRIANGLEQ\EXPECTATION[f(\para)]$
and functional variance $\VARIANCE[f]\text{\ensuremath{\TRIANGLEQ}}\EXPECTATION[||f(\para)-\EXPECTATION[f]||^{2}]$.
Then we have:
\[
\text{\ensuremath{\VARIANCE}}_{\Bfunc}[f]\text{\ensuremath{\TRIANGLEQ}}\EXPECTATION\left[\BD(f||\EXPECTATION(f))\right]=\EXPECTATION[\phi(f)]-\phi(\EXPECTATION[f])\geq0.
\]
Equivalently, we have:
\[
\text{\ensuremath{\VARIANCE}}_{\Bfunc}[f]\TRIANGLEQ\EXPECTATION\left[\BD(f||\EXPECTATION(f))\right]=\EXPECTATION[D(f||\ftilde)]-\BD(\EXPECTATION[f]||\ftilde)\geq0,
\]
for any functional point $\ftilde\TRIANGLEQ\ftilde(\para)\in\Lp(\theta)$
and: 
\begin{equation}
\pdf_{0}\TRIANGLEQ\EXPECTATION[f]=\argmin_{\ftilde}\EXPECTATION[D(f||\ftilde)].\label{eq:fmean}
\end{equation}
\end{prop}
\begin{IEEEproof}
Because the Fréchet derivative in (\ref{eq:BREGMAN_func}) is a linear
operator like gradient in (\ref{eq:BREGMAN}), we can derive the above
results in the same manner of the proof of Theorem \ref{thm:Bregman-variance-Jensen}.
\end{IEEEproof}
\begin{rem}
From Proposition~\ref{prop:Bregman-variance-function}, we have $\VARIANCE[f]=\text{\ensuremath{\VARIANCE}}_{\Bfunc_{E}}[f]$
for Euclidean case $\phi_{E}(f)=||f(\para)-\EXPECTATION[f]||^{2}$,
but $\VARIANCE[f]\neq\text{\ensuremath{\VARIANCE}}_{\Bfunc}[f]$ in
general. The functional mean $\pdf_{0}\TRIANGLEQ\EXPECTATION[f]$
is also the same minimum function for any expected Bregman divergence,
similarly to Remark \ref{rem:E=00005Bx=00005D}. 
\end{rem}
For later use, let us apply Proposition \ref{prop:Bregman-variance-function}
and show here the Bregman variance for a probabilistic mixture:
\begin{cor}
(Bregman variance theorem for mixture) \\
Let functional point $f(\theta)$ be a r.v. drawn from a functional
set $\fbold\TRIANGLEQ\{f_{`1}(\para),\ldots,f_{`\ntime}(\para)\}$
of $\ntime$ distributions over $\theta$, with probabilities $\Bp_{\itime}\in\UNIT\TRIANGLEQ[0,1]$,
$\sum_{\itime=1}^{\ntime}\Bp_{\itime}=1$. The functional mean (\ref{eq:fmean})
is then regarded as a mixture, as follows:
\begin{equation}
\pdf_{0}(\para)\TRIANGLEQ\EXPECTATION[f]=\sum_{\itime=1}^{\ndim}\Bp_{\itime}\pdf_{\itime}(\para),\label{eq:mean_mixture}
\end{equation}
with variance $\VARIANCE[f]\text{=}\sum_{\itime=1}^{\ndim}\Bp_{\itime}||f(\para)-\bar{f}(\para)||^{2}$.
The Bregman variance is then:

\begin{equation}
\text{\ensuremath{\VARIANCE}}_{\Bfunc}[f]=\sum_{\itime=1}^{\ndim}\Bp_{\itime}D(f_{\itime}||\pdf_{0})=\sum_{\itime=1}^{\ndim}\Bp_{\itime}D(f_{\itime}||\ftilde)-D(\pdf_{0}||\ftilde)\geq0,\label{eq:Bregman_mixture}
\end{equation}
for any distribution $\ftilde\TRIANGLEQ\ftilde(\para)$ and, hence,
from (\ref{eq:fmean}-\ref{eq:mean_mixture}), we have: 

\begin{equation}
\pdf_{0}(\para)=\EXPECTATION[f]=\sum_{\itime=1}^{\ndim}\Bp_{\itime}\pdf_{\itime}(\para)=\argmin_{\ftilde}\sum_{\itime=1}^{\ndim}\Bp_{\itime}D(\pdf_{\itime}||\ftilde).\label{eq:argmin_mixture}
\end{equation}
\end{cor}
\begin{IEEEproof}
This case is a consequence of Proposition \ref{prop:Bregman-variance-function}.
\end{IEEEproof}
The case of KL divergence, which is a special case of Bregman variance
with $\Bfunc=\Bfunc_{KL}$ in (\ref{eq:Bregman_mixture}), is illustrated
in Fig. \ref{fig:mixtureKLD}.
\begin{rem}
\label{rem:KLvariance}The computation of KL variance via (\ref{eq:Bregman_mixture})
for a mixture is often more feasible than the computation of Euclidean
variance in practice. Indeed, the KL form corresponds to geometric
mean \cite{Bregman:cluster:centroid:09}, which can yield linearly
computational complexity over exponential coordinates (particularly
for exponential family \cite{Shun-ichiAmari:BOOK:16,Bregman:cluster:centroid:09}),
while the Euclidean form corresponds to arithmetic mean, which would
yield exponentially computational complexity for exponential family
distributions over Euclidean coordinates in general, as shown in section
\ref{subsec:Conditionally-exponential-family}.
\end{rem}

\section{Copula theory \label{sec:Copula-theory}}

The copula concept was firstly defined in \cite{copula:Sklar:59},
although it was also defined under different names such as ``uniform
representation'' or ``dependence function'' \cite{copula:BOOK:Carlo:15}.
The copula has been studied intensively in many decades in statistics,
particularly for finances \cite{copula:cite:book:04,copula:Book:figs:14}.
Yet the application of copula in information theory is still limited
at the moment. In this section, we will review the basic concept of
copula and its direct connection to mutual information of a system.
The KL divergence for copula, which is the nutshell of CVB approximation
in next section, will be provided at the end of this section.

\subsection{Sklar's theorem}

Because the Sklar's paper \cite{copula:Sklar:59} is the beginning
of copula's history, let us recall the Sklar's theorem first. 

\begin{figure}
\begin{centering}
\includegraphics[width=0.5\linewidth]{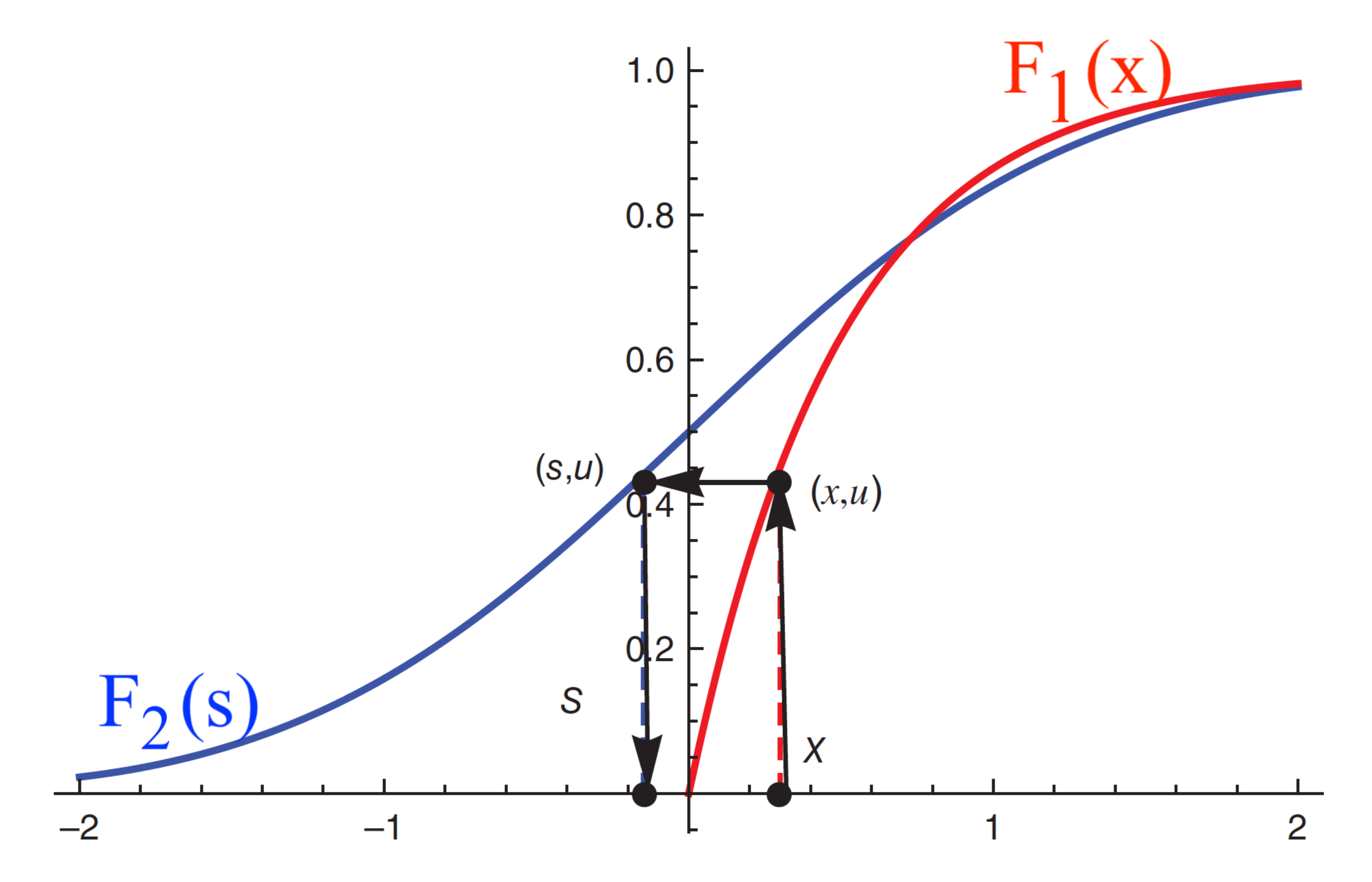}\includegraphics[width=0.5\linewidth]{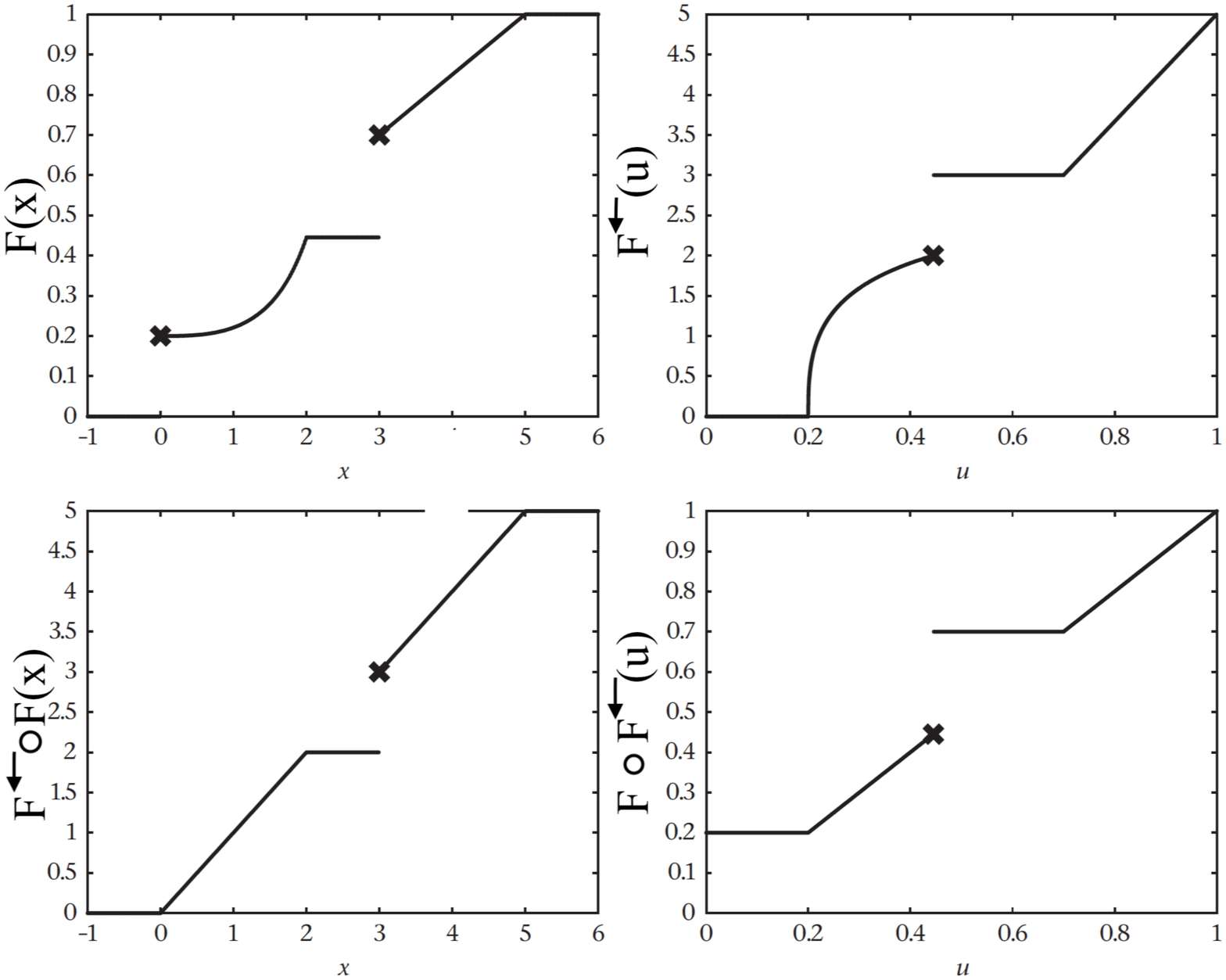}
\par\end{centering}
\centering{}\caption{\label{fig:transformCDF}Illustration of variable transformation from
$x$ to $s$ in the case of continuous c.d.f. \cite{copula:cite:book:17}
(left), together with pseudo-inverse $\protect\pseudo(\protect\Cu)$
of a non-continuous c.d.f. $\protect\cmf(x)$ and their concatenations
$\protect\pseudo\circ\protect\cmf(x)$, $\protect\cmf\circ\protect\pseudo(\protect\Cu)$
\cite{copula:Book:figs:14} (right). We can see that the uniqueness
of copula requires the continuous property of c.d.f., since non-continuous
c.d.f. does not preserve the inverse transformation.}
\end{figure}

\begin{defn}
(Pseudo-inverse function)\\
Let $F:\REAL\rightarrow\UNIT$ be a cumulative distributional function
(c.d.f.) of a r.v. $\theta\in\REAL$. Since $F(\theta)$ is not strictly
increasing in general, as illustrated in Fig.~\ref{fig:transformCDF},
a pseudo-inverse function (also called quantile function) $\pseudo:\UNIT\rightarrow\REAL$
is defined as follows: 
\[
\pseudo(\Cu)\TRIANGLEQ\inf\{\theta\in\REAL:\cmf(\theta)\geq\Cu\},\ \Cu\in\UNIT.
\]
\end{defn}
Note that, the quasi-inverse $\pseudo$ coincides with the inverse
function $\inverse$ if $F(\theta)$ is continuous and strictly increasing,
as illustrated in Fig. \ref{fig:transformCDF}.
\begin{thm}
(Sklar's theorem) \cite{copula:Sklar:59,copula:Sklar:96}\\
For any r.v. $\bpara=\setv{\para}{\ndim}^{\transpose}\in\REAL^{\ndim}$
with joint c.d.f. $F(\bpara)$ and marginal c.d.f. $F_{\idim}(\para_{\idim})$,
$\forall\seti{\idim}{\ndim}$, there always exists an equivalent joint
c.d.f., namely copula $C$, whose all marginal c.d.f. $C_{\idim}(F_{\idim}(\para_{\idim}))$
are uniform over $\UNIT$ as follows: 
\begin{equation}
F(\bpara)=C(F_{1}(\para_{1}),\ldots,F_{\ndim}(\para_{\ndim}))\label{eq:COPULA}
\end{equation}
In general, the copula form $C$ of a joint c.d.f.\textup{ $F$} is
not unique, but its value on the range $\bCu\in\Ran\{F_{1}\}\times\ldots\times\Ran\{F_{\ndim}\}\subseteq\UNIT^{\ndim}$
is always unique, as follows: 
\begin{equation}
C(\bCu)=F(\pseudo_{1}(\Cu_{1}),\ldots,\pseudo_{\ndim}(\Cu_{\ndim}))\label{eq:COPULA_pseudo}
\end{equation}
with $\bCu\TRIANGLEQ\setv{\Cu}{\ndim}^{\transpose}$and $\Cu_{\idim}\TRIANGLEQ F_{\idim}(\para_{\idim}):\REAL\rightarrow\UNIT$,
$\forall\idim$. If all marginals $F_{1},\ldots F_{\ndim}$ are continuous,
the copula $C$ in (\ref{eq:COPULA}) is uniquely determined by quantile
transformation (\ref{eq:COPULA_pseudo}), in which $\pseudo$ coincides
with the inverse function $\inverse$. 
\end{thm}

\subsubsection{Bound of copula}

\begin{figure}
\centering{}\includegraphics[width=0.4\linewidth]{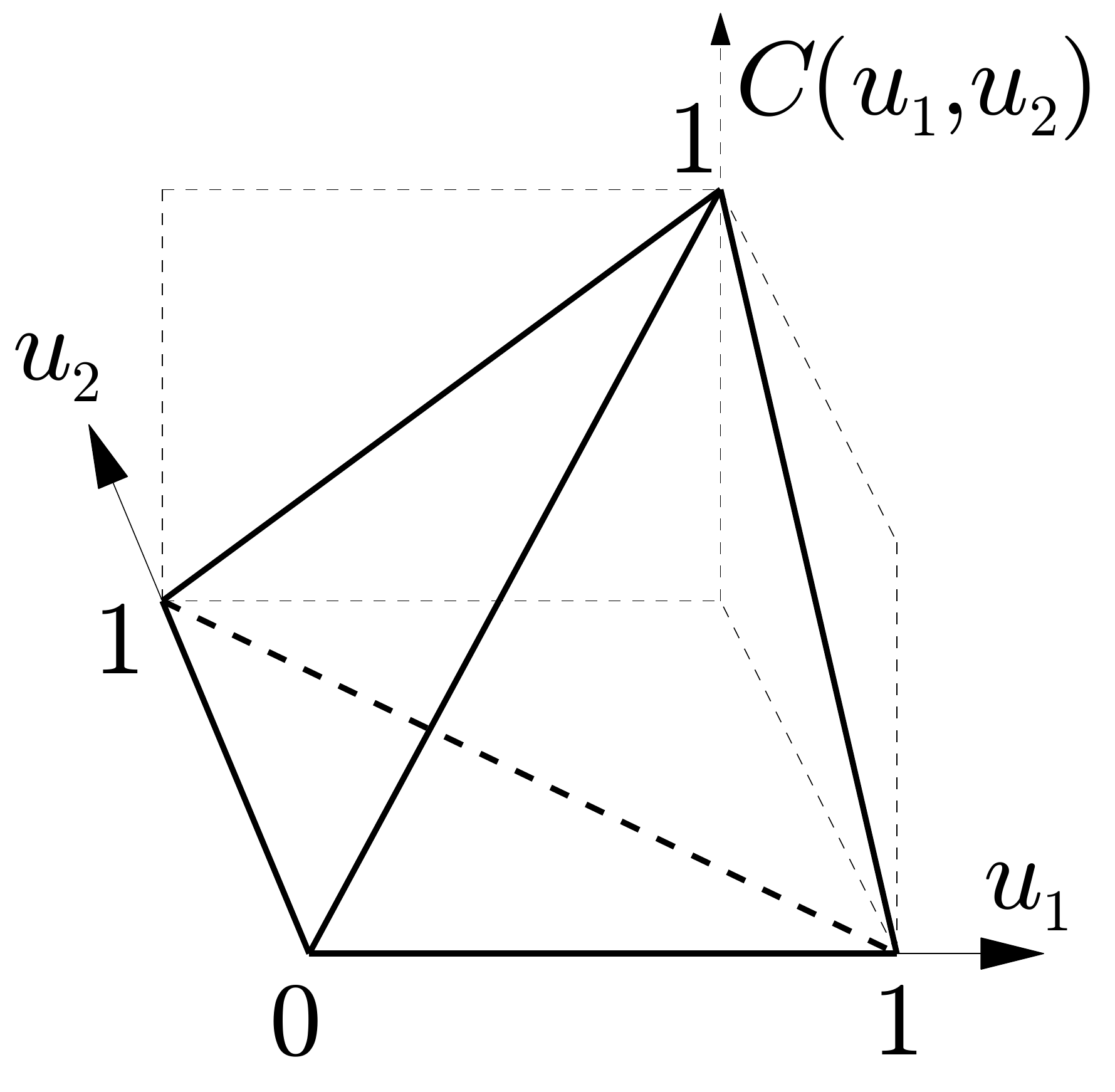}\includegraphics[width=0.6\linewidth]{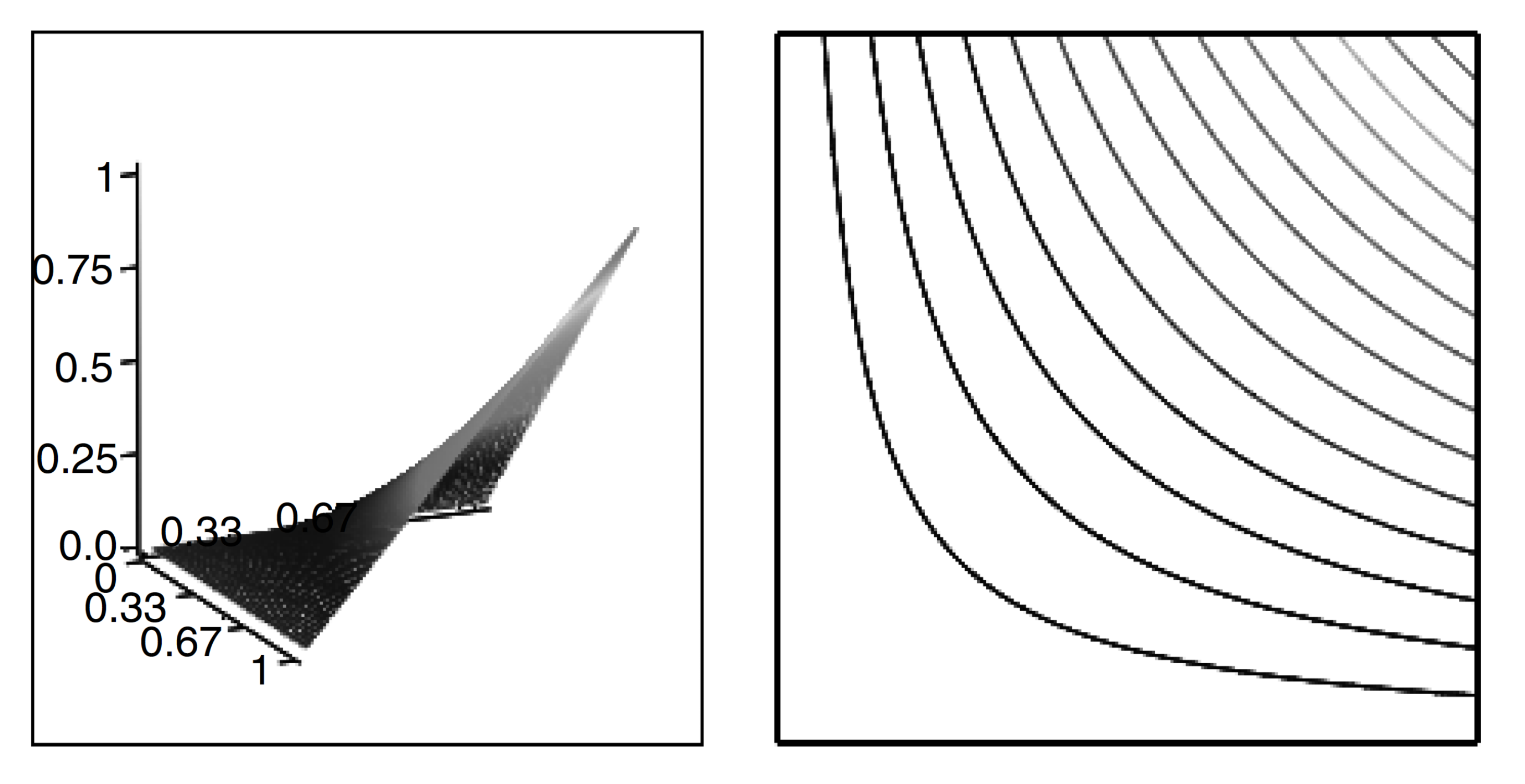}\caption{\label{fig:Frechet_bound}All bivariate copulas must lie inside the
pyramid of Fréchet-Hoeffding bound. Both marginal c.d.f. $C(\protect\Cu_{1},1)$
and $C(1,\protect\Cu_{2})$ must be uniform over $[0,1]$ by definition
and, hence, plotted in the left sub-figure as straight lines. The
two sub-figures on the right illustrates the contours of independent
copula $C(\protect\Cu_{1},\protect\Cu_{2})=\protect\Cu_{1}\protect\Cu_{2}$
\cite{copula:cite:book:04}.}
\end{figure}

For rough visualization of copula, let us recall the Fréchet-Hoeffding
bound of copula \cite{copula:BOOK:Carlo:15,copula:Book:figs:14}:

\[
\max\{0,1-\ndim+\sum_{\idim=1}^{\ndim}\Cu_{\idim})\}\leq C(\Cu_{1},\ldots,\Cu_{\ndim})\leq\min\{\Cu_{1},\ldots,\Cu_{\ndim}\}
\]
where $\Cu_{\idim}\TRIANGLEQ F_{\idim}(\para_{\idim}):\REAL\rightarrow\UNIT$.
This bound is illustrated in Fig.~\ref{fig:Frechet_bound} for the
case of two dimensions. 

\subsubsection{Discrete copula}

The pseudo-inverse form (\ref{eq:COPULA_pseudo}) is often called
sub-copula in literature \cite{copula:BOOK:Carlo:15}, since its values
are only defined on a possibly subset of $\UNIT^{\ndim}$. This mostly
happens in the case of discrete distributions, where marginal $F(\bpara)$
is not continuous, as illustrated in Fig.~\ref{fig:transformCDF}.
Hence, there are possibly more than one continuous copula (\ref{eq:COPULA})
satisfying the discrete sub-copula form (\ref{eq:COPULA}) at specific
values $\bCu\in\Ran\{F_{1}\}\times\ldots\times\Ran\{F_{\ndim}\}$. 

As illustrated in Fig. \ref{fig:transformCDF}, the Sklar's theorem
only guarantees the uniqueness of copula form $C$ for a strictly
increasing continuous $F$ (c.f. \cite{copula:BOOK:Carlo:15} for
examples of non-unique copulas $C$ associated with a discrete $F$).
Nonetheless, the quantile function in (\ref{eq:COPULA_pseudo}) is
still useful to compute copula values in the discrete range of $F$.
For example, in \cite{copula:BOOK:Carlo:15,copula:discrete:bimatrix:06},
the copula form of any discrete bivariate distribution was shown to
be equivalent to a bi-stochastic non-negative matrix, whose sum of
any row or column is equal to one. 

\subsubsection{Continuous copula}

For simplicity, let us focus on copula form of continuous c.d.f. $F$,
although the results in this paper can be extended to discrete case
via pseudo-inverse function in (\ref{eq:COPULA_pseudo}). For continuous
case, the quantile transformation (\ref{eq:COPULA_pseudo}) yields
the density form of copula $C$, as follows:
\begin{cor}
(Copula density function) \cite{copula:BOOK:Carlo:15}\\
If all marginals  $F_{1},\ldots F_{\ndim}$ are absolutely continuous
with respect to Lebesgue measure on $\REAL^{\ndim}$, the density
$c(\bCu)\TRIANGLEQ\frac{\partial C(\bCu)}{\partial\Cu_{1}\ldots\partial\Cu_{\ndim}}$
of copula $C$ in (\ref{eq:COPULA}) is given by: 
\begin{equation}
f(\bpara)=c(\bCu(\bpara))\prod_{\idim=1}^{\ndim}f_{\idim}(\para_{\idim})\label{eq:COPULA_pdf}
\end{equation}
where $f$ is density of c.d.f. $F$ and $\bCu\TRIANGLEQ\bCu(\bpara)\in\UNIT^{\ndim}$,
with $\Cu_{\idim}\TRIANGLEQ F_{\idim}(\para_{\idim})$, $\forall\seti{\idim}{\ndim}$.
\end{cor}
\begin{IEEEproof}
By chain rule, we have $f(\bpara)=\frac{\partial F(\bpara)}{\partial\para_{1}\ldots\partial\para_{\ndim}}=\frac{\partial C(\bCu)}{\partial\Cu_{1}\ldots\partial\Cu_{\ndim}}\prod_{\idim=1}^{\ndim}\frac{\partial\Cu_{\idim}}{\partial\para_{\idim}}=c(\bCu)\prod_{\idim=1}^{\ndim}f_{\idim}(\para_{\idim})$.
\end{IEEEproof}
The density (\ref{eq:COPULA_pdf}) shows that a joint p.d.f. can be
factorized into two parts: the dependent part represented by copula
and the independent part represented by product of its marginals.
Hence, the copula fully extracts all dependent relationships between
r.v. $\para_{\idim}$, $\seti{\idim}{\ndim}$, from joint p.d.f. $f(\bpara)$.
\begin{rem}
\label{rem:indepedentCopula}Note that, since copula $C$ is essentially
a c.d.f. by definition (\ref{eq:COPULA}), the copula $C(\bCu)$ of
independent c.d.f. $F(\bpara)=\prod_{\idim=1}^{\ndim}F_{\idim}(\para_{\idim})$
is also factorable, i.e. $C(\bCu)=\prod_{\idim=1}^{\ndim}\Cu_{\idim}$,
and, hence, $c(\bCu)=1$ by (\ref{eq:COPULA_pdf}), as illustrated
in Fig.~\ref{fig:Frechet_bound}.
\end{rem}

\subsection{Copula's invariant transformations \label{subsec:Copula's-invariant}}

Let us focus on continuous copula and its useful transformation's
properties in this subsection. These properties are also satisfied
with discrete copulas via pseudo-inverse function~(\ref{eq:COPULA_pseudo}). 

\subsubsection{Copula's rescaling transformation}

By copula's density definition (\ref{eq:COPULA_pdf}), we can see
that a copula $c(\bCu(\bpara))$ is merely a rescaled coordinate form
of original joint p.d.f. $f(\bpara)$, as follows:
\begin{cor}
(Copula's rescaling property)
\end{cor}
\begin{equation}
1=\int_{\bCu(\bpara)\in\UNIT^{\ndim}}c(\bCu)d\bCu=\int_{\bpara\in\REAL^{\ndim}}f(\bpara)d\bpara\label{eq:re-scaling}
\end{equation}

\begin{IEEEproof}
By definition in (\ref{eq:COPULA_pdf}), we have $\Cu_{\idim}\TRIANGLEQ F_{\idim}(\para_{\idim})$
and $d\bpara\TRIANGLEQ\prod_{\idim=1}^{\ndim}d\para_{\idim}$, which
yields: $d\bCu=\prod_{\idim=1}^{\ndim}d\Cu_{\idim}=\prod_{\idim=1}^{\ndim}f_{\idim}(\para_{\idim})d\bpara=\frac{f(\bpara)}{c(\bCu(\bpara))}d\bpara.$
Q.E.D.
\end{IEEEproof}
The rescaling property (\ref{eq:re-scaling}) will be useful later
when we wish to change the integrated variables from $\bpara$ to
$\bCu$ in copula's manipulation.

\subsubsection{Copula's monotone transformation}

Under generally monotonic transformation, which is not necessarily
strictly increasing, the copula is linearly invariant (c.f. \cite{copula:BOOK:Carlo:15}
for details). In this paper, let us recall here the useful rank-invariant
property of copula under increasing transformation, as follows:
\begin{thm}
(Copula's rank-invariant property) \cite{copula:Book:figs:14,copula:BOOK:Carlo:15}
\\
Let $\tbpara\TRIANGLEQ\setv{\tpara}{\ndim}^{\transpose}\in\REAL^{\ndim}$,
in which $\tpara_{\idim}\TRIANGLEQ\varphi_{\idim}(\para_{\idim})$
is a strictly increasing function of r.v. $\para_{\idim}\in\REAL$,
$\forall\seti{\idim}{\ndim}$. Then the density copulas $\ctilde$
and $\copula$ of $\tbpara$ and $\bpara$, respectively, have the
same form, i.e. $\ctilde(\bCu)=c(\bCu)$, $\forall\bCu\in\UNIT^{\ndim}$. 
\end{thm}
Intuitively, the copula's rank-invariant property is merely a consequence
of natural rank-invariant property of marginal c.d.f. under increasing
transformation, as implied by definition of copula (\ref{eq:COPULA})
and illustrated in Fig. \ref{fig:transformCDF}.

\subsubsection{Copula's marginal transformation}

For later use, let us emphasize a very special case of rank-invariant
property, namely marginal transformation. By definition (\ref{eq:COPULA_pdf}),
we can see that copula separates the dependence part of joint p.d.f.
from its marginals. Hence, we can freely replace any marginal $\cmf_{\idim}$
with new marginal $\cmftilde_{\idim}$, $\forall\seti{\idim}{\ndim}$,
without changing the form of copula, as shown below: 
\begin{cor}
(Copula's marginal-invariant property) \label{cor:Copula's-marginal-invariant}\\
Let $\tbpara(\bpara)\TRIANGLEQ\setx{\para_{1}}{\tpara_{\idim}(\para_{\idim})}{\para_{\ndim}}^{\transpose}\in\REAL^{\ndim}$,
in which r.v. $\para_{\idim}$ in $\bpara$ is replaced by a continuously
transformed r.v. $\tpara_{\idim}(\para_{\idim})\TRIANGLEQ\tpseudo_{\idim}(\cmf_{\idim}(\para_{\idim}))$
, for any $\seti{\idim}{\ndim}$. Then the density copulas $\ctilde_{\idim}$
and $\copula$ of $\tbpara(\bpara)$ and $\bpara$, respectively,
have the same form, i.e. $\ctilde_{\idim}(\bCu)=c(\bCu)$, $\forall\bCu\in\UNIT^{\ndim}$. 
\end{cor}
\begin{IEEEproof}
This corollary is a direct consequence of the copula's rank-invariant
property, since the continuous c.d.f. functions $\tpseudo_{\idim}$
and $\cmf_{\idim}$ are both strictly increasing function for continuous
variables.
\end{IEEEproof}
The marginal-invariant property shows that when we replace the marginal
distribution $f_{\idim}(\para_{\idim})$ of joint p.d.f. $f(\bpara)$
in (\ref{eq:COPULA_pdf}) by another marginal distribution $\ftilde_{\idim}(\para_{\idim})$,
the resulted joint distribution $\ftilde(\bpara)$ does not change
its original copula form, i.e.:
\begin{equation}
\left.\begin{array}{cc}
f(\bpara) & =f(\para_{\backslash\idim}|\para_{\idim})f_{\idim}(\para_{\idim})\\
\ftilde(\bpara) & =f(\para_{\backslash\idim}|\para_{\idim})\ftilde_{\idim}(\para_{\idim})
\end{array}\right\} \Rightarrow\ctilde(\bCu)=c(\bCu),\forall\bCu\in\UNIT^{\ndim}\label{eq:marginal_invariant}
\end{equation}
Indeed, by Corollary \ref{cor:Copula's-marginal-invariant}, we have
$f(\bpara)=\ftilde(\tbpara(\bpara))$, i.e. the distribution $\ftilde(\bpara)$
is merely a marginally rescaling form of $f(\bpara)$ and, hence,
does not change the form of copula. 

\subsection{Copula's divergence}

Because the copula is essentially a distribution itself, the KL divergence
(\ref{eq:KL}) can be applied directly to any two copulas. Let us
show the relationship between joint p.d.f. and its copula via KL divergence
in this subsection.

\subsubsection{Mutual information}

Because all dependencies in a joint p.d.f. $f$ in (\ref{eq:COPULA_pdf})
is captured by its copula, it is natural that the mutual information
of joint p.d.f. $f$ can also be computed via its copula form $c$
in (\ref{eq:COPULA_pdf}), as shown below.
\begin{prop}
(Mutual information)\label{cor:(Mutual-information)}\\
For continuous copula $c$ in (\ref{eq:COPULA_pdf}), the mutual information
$\Mutual(\bpara)$ of joint p.d.f. $f(\bpara)$ is equal to continuous
entropy $\Entropy$ of copula density $c(\bCu(\bpara))$, as follows:
\begin{equation}
\Mutual(\bpara)=\Entropy(c(\bCu)).\label{eq:MUTUAL}
\end{equation}
\end{prop}
\begin{IEEEproof}
The proof is straight-forward from definition of KL divergence (\ref{eq:KL})
and copula density (\ref{eq:COPULA_pdf}), as follows: $\Mutual(\bpara)\TRIANGLEQ\text{\ensuremath{\KL}}(f(\bpara)||\prod_{\idim=1}^{\ndim}f_{\idim}(\para_{\idim}))=\EXPECTATION_{f(\bpara)}\log\frac{f(\bpara)}{\prod_{\idim=1}^{\ndim}f_{\idim}(\para_{\idim})}=\EXPECTATION_{f(\bpara)}\log c(\bCu(\bpara))=\EXPECTATION_{c(\bCu)}\log c(\bCu)=\text{H}(c(\bCu))$,
in which $\bpara$ was transformed to $\bCu$ via rescaling property
(\ref{eq:re-scaling}). For a special case of bivariate copula density,
another proof was given in \cite{copula:Entropy:11}.
\end{IEEEproof}

\subsubsection{KL divergence (KLD)}

In literature, the below copula-based KL divergence for a joint p.d.f.
was already given for a special case of conditional structure \cite{copula:KLD:proof:16}.
For later use, let us recall their proof here in a slightly more generally
form, via pseudo-inverse (\ref{eq:COPULA_pseudo}) and rescaling property
(\ref{eq:re-scaling}).
\begin{prop}
(Copula's divergence) \cite{copula:KLD:proof:16} \\
The KLD of two joint p.d.f. $f$, $\ftilde$ in (\ref{eq:COPULA_pdf})
is the sum of KLD of their copulas $c$, $\ctilde$ and KLDs of their
marginals $f_{\idim}$, $\ftilde_{\idim}$, as follows: 

\begin{equation}
\text{\ensuremath{\KL}}_{\ftilde||f}=\text{\ensuremath{\KL}}(\ctilde(\bCu)||c(\cmf(\tpseudo(\bCu))))+\sum_{\idim=1}^{\ndim}\text{\ensuremath{\KL}}_{\ftilde_{\idim}||f_{\idim}}\label{eq:KLD_copula}
\end{equation}
in which the copula $\ctilde$ of $\ftilde$ was rescaled back to
marginal coordinates of $f$, i.e. $\ctilde(\cmftilde(\pseudo(\bCu)))\TRIANGLEQ\ctilde(\cmftilde_{1}(\pseudo_{1}(\Cu_{1})),\ldots,\cmftilde_{\ndim}(\pseudo_{\ndim}(\Cu_{\ndim}))).$
\end{prop}
\begin{IEEEproof}
By definition of KLD (\ref{eq:KL}) and copula density (\ref{eq:COPULA_pdf}),
we have: $\text{\ensuremath{\KL}}(f(\bpara)||\ftilde(\bpara))=\EXPECTATION_{f(\bpara)}\log\frac{f(\bpara)}{\ftilde(\bpara)}=\EXPECTATION_{f(\bpara)}\log\frac{c(\bCu(\bpara))}{\ctilde(\tilde{\bCu}(\bpara))}+\sum_{\idim=1}^{\ndim}\EXPECTATION_{f(\bpara)}\log\frac{f_{\idim}(\para_{\idim})}{\ftilde_{\idim}(\para_{\idim})}$,
of which the second term in r.h.s. is actually KLDs of marginal, i.e.
$\EXPECTATION_{f(\bpara)}\log\frac{f_{\idim}(\para_{\idim})}{\ftilde_{\idim}(\para_{\idim})}=\EXPECTATION_{f_{\idim}(\para_{\idim})}\log\frac{f_{\idim}(\para_{\idim})}{\ftilde_{\idim}(\para_{\idim})}=\text{\ensuremath{\KL}}(f_{\idim}(\para_{\idim}))||\ftilde_{\idim}(\para_{\idim}))$
and the first term in r.h.s. is actually KLD of copulas, via rescaling
property (\ref{eq:re-scaling}), as follows: $\EXPECTATION_{f(\bpara)}\log\frac{c(\bCu(\bpara))}{\ctilde(\tilde{\bCu}(\bpara))}=\EXPECTATION_{f(\bpara)}\log\frac{c(F_{1}(\para_{1}),\ldots,F_{\ndim}(\para_{\ndim}))}{\ctilde(\cmftilde_{1}(\para_{1}),\ldots,\cmftilde_{\ndim}(\para_{\ndim}))}=\EXPECTATION_{c(\bCu)}\log\frac{c(\bCu)}{\ctilde(\cmftilde(\pseudo(\bCu)))}=\text{\ensuremath{\KL}}(c(\bCu)||\ctilde(\cmftilde(\pseudo(\bCu))))$.
\end{IEEEproof}
Note that, by copula's marginal- and rank-invariant properties in
section \ref{subsec:Copula's-invariant}, we can see that the marginal
rescaling form $\ctilde(\cmftilde(\pseudo(\bCu)))$ of $\ctilde$
in (\ref{eq:KLD_copula}) does not change the original form of copula
$\ctilde$.
\begin{rem}
If all $\cmftilde_{\idim}$ are exact marginals of $\cmf(\bpara)$,
i.e. $\cmftilde_{\idim}=\cmf_{\idim}$ in (\ref{eq:KLD_copula}),
$\forall\seti{\idim}{\ndim}$, we have $\text{\ensuremath{\KL}}(f(\bpara)||\ftilde(\bpara))=\text{\ensuremath{\KL}}(c(\bCu)||\ctilde(\bCu))$.
Furthermore, if $\ctilde(\bCu)$ is also an independent copula, as
noted in Remark \ref{rem:indepedentCopula}, the KL divergence in
(\ref{eq:KLD_copula}) will be equal to mutual information $\Mutual(\bpara)$
in (\ref{eq:MUTUAL}).
\end{rem}

\section{Copula Variational Bayes approximation \label{sec:Copula-Variational-Bayes}}

As shown in (\ref{eq:KLD_copula}), the KL divergence between any
two distributions can always be factorized as the sum of KL divergence
of their copulas and KL divergences of their marginals. Exploiting
this property, we will design a novel iterative copula VB (CVB) algorithm
in this section, such that the CVB distribution is closest to the
true distribution in terms of KL divergence, under constraint of initially
approximated copula's form. The mean-field approximations, which are
special cases of CVB, will also be revisited later in this section. 

\subsection{Motivation of marginal approximation}

Let us now consider a joint p.d.f. $f(\bpara)$, of which the true
marginals $f_{\idim}(\para_{\idim})=\int_{\para_{\backslash\idim}}f(\bpara)d\para_{\backslash\idim}$,
$\seti{\idim}{\ndim},$ are either unknown or too complicated to compute.
A natural approximation of $f_{\idim}(\para_{\idim})$ is then to
seek a closed form distribution $\ftilde_{\idim}(\para_{\idim})$
such that their KL divergences $\sum_{\idim=1}^{\ndim}\text{\ensuremath{\KL}}_{f_{\idim}||\ftilde_{\idim}}$
in (\ref{eq:KLD_copula}) is minimized. This direct approach is, however,
not feasible if the integration for true marginal $f_{\idim}(\para_{\idim})$
is very hard to compute at the beginning. 

A popular approach in literature is to find an approximation $\ftilde(\bpara)$
of the joint distribution $f(\bpara)$ such that their KL divergence
$\KLff\TRIANGLEQ\KL(\ftilde(\bpara)||\pdf(\bpara))$ can be minimized.
This indirect approach is more feasible since it circumvents the explicit
form of $f_{\idim}(\para_{\idim})$. Also, since $\KLff$ is the upper
bound of $\sum_{\idim=1}^{\ndim}\text{\ensuremath{\KL}}_{\ftilde_{\idim}||f_{\idim}}$,
as shown in (\ref{eq:KLD_copula}), it would yield good approximated
marginals $\ftilde_{\idim}(\para_{\idim})$ if $\KLff$ could be set
low enough. This is the objective of CVB algorithm in this section.
\begin{rem}
Another approximation approach is to find $\ftilde(\bpara)$ such
that the copula's KL divergence $\text{\ensuremath{\KL}}(\ctilde(\bCu)||c(\cmf(\tpseudo(\bCu))))$
in (\ref{eq:KLD_copula}) is as close as possible to $\text{\ensuremath{\KL}}(c(\bCu)||\ctilde(\bCu))$,
which is equivalent to the exact case $\ftilde_{\idim}=f_{\idim}$,
$\forall\seti{\idim}{\ndim}.$ This copula's analysis approach is
promising, since the original copula form can be extracted from mutual
dependence part of the original $f(\bpara)$, without the need of
marginal's normalization, as shown in \cite{copula:KLD:proof:16}
for a simple case of a Gaussian copula function. However, this approach
would generally involve copula's explicit analysis, which is not a
focus of this paper and will be left for future work. 
\end{rem}

\subsection{Copula Variational approximation}

\begin{figure}
\centering{}\includegraphics[width=1\linewidth]{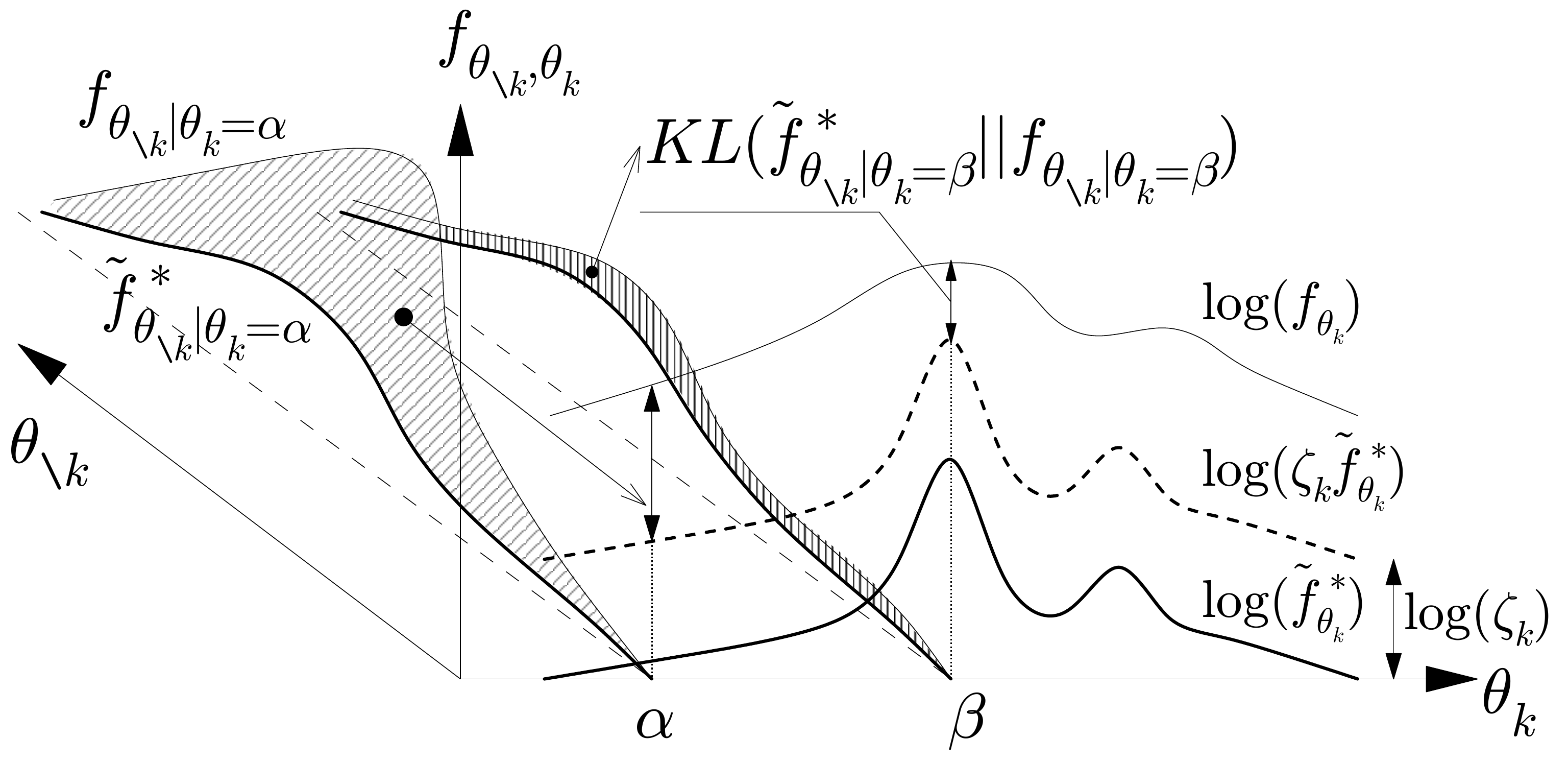}\caption{\label{fig:fCVA}Illustration of Conditionally Variational approximation
(CVA), as defined in (\ref{eq:CV_KLDform}). The lower KL divergence,
the better approximation. Given initially a conditional form $\protect\ftilde_{\protect\para_{\backslash\protect\idim}|\protect\para_{\protect\idim}}^{*}$
for $\protect\ftilde_{\protect\bpara}=\protect\ftilde_{\protect\para_{\backslash\protect\idim}|\protect\para_{\protect\idim}}^{*}\protect\ftilde_{\protect\para_{\protect\idim}}$,
the optimally approximated marginal $\protect\ftilde_{\protect\para_{\protect\idim}}^{*}$
minimizing $\protect\KL(\protect\ftilde_{\protect\bpara}||\protect\pdf_{\protect\bpara})$
is proportional to the true marginal $\protect\pdf_{\protect\para_{\protect\idim}}$
in $\protect\pdf_{\protect\bpara}=\protect\pdf_{\protect\para_{\backslash\protect\idim}|\protect\para_{\protect\idim}}\protect\pdf_{\protect\para_{\protect\idim}}$
by a fraction of normalized conditional divergence $\text{\ensuremath{\protect\VBzeta}}_{\protect\idim}\exp\protect\KL(\protect\ftilde_{\protect\para_{\backslash\protect\idim}|\protect\para_{\protect\idim}}^{*}||\protect\pdf_{\protect\para_{\backslash\protect\idim}|\protect\para_{\protect\idim}})$,
where $\text{\ensuremath{\protect\VBzeta}}_{\protect\idim}$ is the
normalizing constant. In traditional VB approximation (\ref{eq:(VB)}),
we simply set $\protect\ftilde_{\protect\para_{\backslash\protect\idim}|\protect\para_{\protect\idim}}^{*}=\protect\ftilde_{\protect\para_{\backslash\protect\idim}}^{*}$,
which is independent of $\protect\para_{\protect\idim}$.}
\end{figure}

Since the CVB algorithm is actually an iterative procedure of many
Conditionally Variational approximation (CVA) steps, let us define
the CVA step first, which is also illustrated in Fig.~\ref{fig:fCVA}.

\subsubsection{Conditionally Variational approximation (CVA)}

For a good approximation $\ftilde_{\idim}$ of $f_{\idim}$, let us
initially pick a closed form p.d.f. $\ftilde(\bpara)=\ftilde^{*}(\para_{\backslash\idim}|\para_{\idim})\ftilde_{\idim}(\para_{\idim})$,
in which the conditional distribution $\ftilde_{\backslash\idim|\idim}^{*}\TRIANGLEQ\ftilde^{*}(\para_{\backslash\idim}|\para_{\idim})$
is fixed and given. The optimal approximation $\ftilde_{\idim}^{*}\TRIANGLEQ\ftilde_{\idim}^{*}(\para_{\idim})$
is then found by the following Theorem, which is also the foundational
idea of this paper: 
\begin{thm}
(Conditionally Variational approximation)\label{thm:(Conditionally-Variational)}\\
Let $\ftilde=\ftilde_{\backslash\idim|\idim}^{*}\ftilde_{\idim}$
be a family of distributions with fixed-form conditional $\ftilde_{\backslash\idim|\idim}^{*}$.
Then $\ftilde$ is convex over marginals $\ftilde_{\idim}$, which
yields:

\begin{equation}
\KL_{\ftilde||\pdf}=\KL_{\ftilde||\ftilde^{*}}+\KL_{\ftilde^{*}||\pdf}\geq\KL_{\ftilde^{*}||\pdf}=\log\frac{1}{\VBzeta_{\idim}}\label{eq:CV_pythagore}
\end{equation}
owing to Bregman pythagorean property (\ref{eq:PYTHAGORE}) for functional
space (\ref{eq:KL}-\ref{eq:BREGMAN_func}). The distribution $\ftilde=\ftilde^{*}$
minimizing $\KL_{\ftilde||\pdf}$ and the value $\VBzeta_{\idim}$
in (\ref{eq:CV_pythagore}) are given as follows:
\begin{align}
\ftilde_{\idim}^{*}(\para_{\idim}) & =\frac{\pdf_{\idim}(\para_{\idim})}{\text{\ensuremath{\VBzeta}}_{\idim}\exp(\KL_{\ftilde_{\backslash\idim|\idim}^{*}||\pdf_{\backslash\idim|\idim}})}\label{eq:CV_KLDform}\\
 & =\frac{1}{\text{\ensuremath{\VBzeta}}_{\idim}}\exp\EXPECTATION_{\ftilde^{*}(\para_{\backslash\idim}|\para_{\idim})}\log\frac{f(\bpara)}{\ftilde^{*}(\para_{\backslash\idim}|\para_{\idim})}\nonumber 
\end{align}
in which $\VBzeta_{\idim}$ is the normalizing constant of $\ftilde_{\idim}^{*}$
in (\ref{eq:CV_KLDform}) and $\KL_{\ftilde_{\backslash\idim|\idim}^{*}||\pdf_{\backslash\idim|\idim}}\TRIANGLEQ\KL(\ftilde^{*}(\para_{\backslash\idim}|\para_{\idim})||\pdf(\para_{\backslash\idim}|\para_{\idim}))$. 

Note that, if the marginal $\ftilde_{\idim}=\ftilde_{\idim}$ is initially
fixed instead, $\ftilde$ is then convex over $\ftilde_{\backslash\idim|\idim}$
and, hence, the conditional $\ftilde_{\backslash\idim|\idim}^{*}$
minimizing $\KL_{\ftilde||\pdf}$ in (\ref{eq:CV_pythagore}) is the
true conditional distribution $\pdf_{\backslash\idim|\idim}$, i.e.
$\ftilde_{\backslash\idim|\idim}^{*}=\pdf_{\backslash\idim|\idim}$. 
\end{thm}
\begin{IEEEproof}
Firstly, we note that, for any mixture $\ftilde_{\idim}(\para_{\idim})=p_{1}\ftilde_{1}(\para_{\idim})+p_{2}\ftilde_{2}(\para_{\idim})$,
we always have $\ftilde(\bpara)=p_{1}\ftilde_{1}(\bpara)+p_{2}\ftilde_{2}(\bpara)$.
Hence, $\ftilde$ is convex over $\ftilde_{\idim}$ with fixed $\ftilde_{\backslash\idim|\idim}$
and satisfies the Bregman pythagorean equality (\ref{eq:CV_pythagore}),
since KL divergence is a special case of Bregman divergence (\ref{eq:KL}).
We can also verify the pythagorean equality (\ref{eq:CV_pythagore})
directly, similarly to the proof of copula's KL divergence (\ref{eq:KLD_copula}),
as follows: 
\begin{align}
\text{\ensuremath{\KL}}_{\ftilde||f} & =\EXPECTATION_{\ftilde_{\idim}}\text{\ensuremath{\KL}}_{\ftilde_{\backslash\idim|\idim}^{*}||\pdf_{\backslash\idim|\idim}}+\text{\ensuremath{\KL}}_{\ftilde_{\idim}||f_{\idim}}\nonumber \\
 & =\EXPECTATION_{\ftilde_{\idim}}\log\frac{\ftilde_{\idim}}{\frac{1}{\text{\ensuremath{\VBzeta}}_{\idim}}\frac{f_{\idim}}{\exp(\text{\ensuremath{\KL}}_{\ftilde_{\backslash\idim|\idim}^{*}||\pdf_{\backslash\idim|\idim}})}}+\EXPECTATION_{\ftilde_{\idim}}\log\frac{1}{\text{\ensuremath{\VBzeta}}_{\idim}}\nonumber \\
 & =\underset{\KL_{\ftilde||\ftilde^{*}}}{\underbrace{\text{\ensuremath{\KL}}_{\ftilde_{\idim}||\ftilde_{\idim}^{*}}}}+\underset{\KL_{\ftilde^{*}||\pdf}}{\underbrace{\log\frac{1}{\text{\ensuremath{\VBzeta_{\idim}}}}}}\label{eq:CV_proof}
\end{align}
in which the form $\ftilde_{\idim}^{*}$ is defined in (\ref{eq:CV_KLDform})
and $\VBzeta_{\idim}$ is independent of $\para_{\idim}$. Also, we
have $\KL_{\ftilde||\ftilde^{*}}=\text{\ensuremath{\KL}}_{\ftilde_{\idim}||\ftilde_{\idim}^{*}}$
in the first term of r.h.s. of (\ref{eq:CV_proof}) since $\ftilde$
and $\ftilde^{*}$ only differ in marginals $\ftilde_{\idim}$, $\ftilde_{\idim}^{*}$.
For the second term, by definition (\ref{eq:CV_KLDform}), we have
$\text{\ensuremath{\KL}}_{\ftilde_{\backslash\idim|\idim}^{*}||\pdf_{\backslash\idim|\idim}}=\log\frac{1}{\text{\ensuremath{\VBzeta}}_{\idim}}\frac{\pdf_{\idim}(\para_{\idim})}{\ftilde_{\idim}^{*}(\para_{\idim})}$,
which yields: $\EXPECTATION_{\ftilde_{\idim}^{*}}\text{\ensuremath{\KL}}_{\ftilde_{\backslash\idim|\idim}^{*}||\pdf_{\backslash\idim|\idim}}=\log\frac{1}{\text{\ensuremath{\VBzeta}}_{\idim}}-\text{\ensuremath{\KL}}_{\ftilde_{\idim}^{*}||\pdf_{\idim}}\Leftrightarrow\KL_{\ftilde^{*}||\pdf}=\log\frac{1}{\text{\ensuremath{\VBzeta}}_{\idim}}$
in (\ref{eq:CV_pythagore}) and (\ref{eq:CV_proof}). Lastly, the
second equality in (\ref{eq:CV_KLDform}) is given as follows: $\pdf_{\idim}(\para_{\idim})/\exp\KL_{\ftilde_{\backslash\idim|\idim}^{*}||\pdf_{\backslash\idim|\idim}}=\pdf_{\idim}(\para_{\idim})\exp\EXPECTATION_{\ftilde_{\backslash\idim|\idim}^{*}}\log\frac{\pdf_{\backslash\idim|\idim}}{\ftilde_{\backslash\idim|\idim}^{*}}=\exp\EXPECTATION_{\ftilde_{\backslash\idim|\idim}^{*}}\log\frac{f(\bpara)}{\ftilde_{\backslash\idim|\idim}^{*}}$. 

If $\ftilde_{\idim}=\ftilde_{\idim}^{*}$ is fixed instead, $\ftilde$
is then convex over a mixture of $\ftilde_{\backslash\idim|\idim}$
as shown above. Then, $\text{\ensuremath{\KL}}_{\ftilde||f}$ in (\ref{eq:CV_proof})
is minimum at $\ftilde_{\backslash\idim|\idim}^{*}=\pdf_{\backslash\idim|\idim}$,
since the term $\text{\ensuremath{\KL}}_{\ftilde_{\idim}||f_{\idim}}=\text{\ensuremath{\KL}}_{\ftilde_{\idim}^{*}||f_{\idim}}$
in (\ref{eq:CV_proof}) is now fixed and the term $\EXPECTATION_{\ftilde_{\idim}^{*}}\text{\ensuremath{\KL}}_{\ftilde_{\backslash\idim|\idim}^{*}||\pdf_{\backslash\idim|\idim}}$
is minimum at zero with $\ftilde_{\backslash\idim|\idim}^{*}=\pdf_{\backslash\idim|\idim}$. 
\end{IEEEproof}
In Theorem \ref{thm:(Conditionally-Variational)}, the conditional
$\ftilde_{\backslash\idim|\idim}^{*}$ is fixed beforehand and $\ftilde_{\idim}^{*}$
is found in a free-form variational space, hence the name Conditionally
Variational approximation (CVA). The case of fixed marginal $\ftilde_{\idim}^{*}$
is not interesting, since $\KL_{\ftilde||\pdf}$ in this case is only
minimized at the true conditional $\pdf_{\backslash\idim|\idim}$,
which is often unknown initially. 
\begin{rem}
The CVA form above is a generalized form of the so-called Conditional
Variational Bayesian inference \cite{name:CVB:01} or Conditional
mean-field \cite{name:CVB:06} in literature, which are merely applications
of mean-field approximations to a conditionally independent structure,
i.e. $\ftilde(\bpara|\hyperpara)=\prod_{\idim=1}^{\ndim}\ftilde_{\idim}(\para_{\idim}|\hyperpara)$,
given a latent variable $\hyperpara$ in this case.
\end{rem}

\subsubsection{Copula Variational algorithm\label{subsec:iterative_CVA}}

\begin{figure}
\centering{}\includegraphics[width=0.9\linewidth]{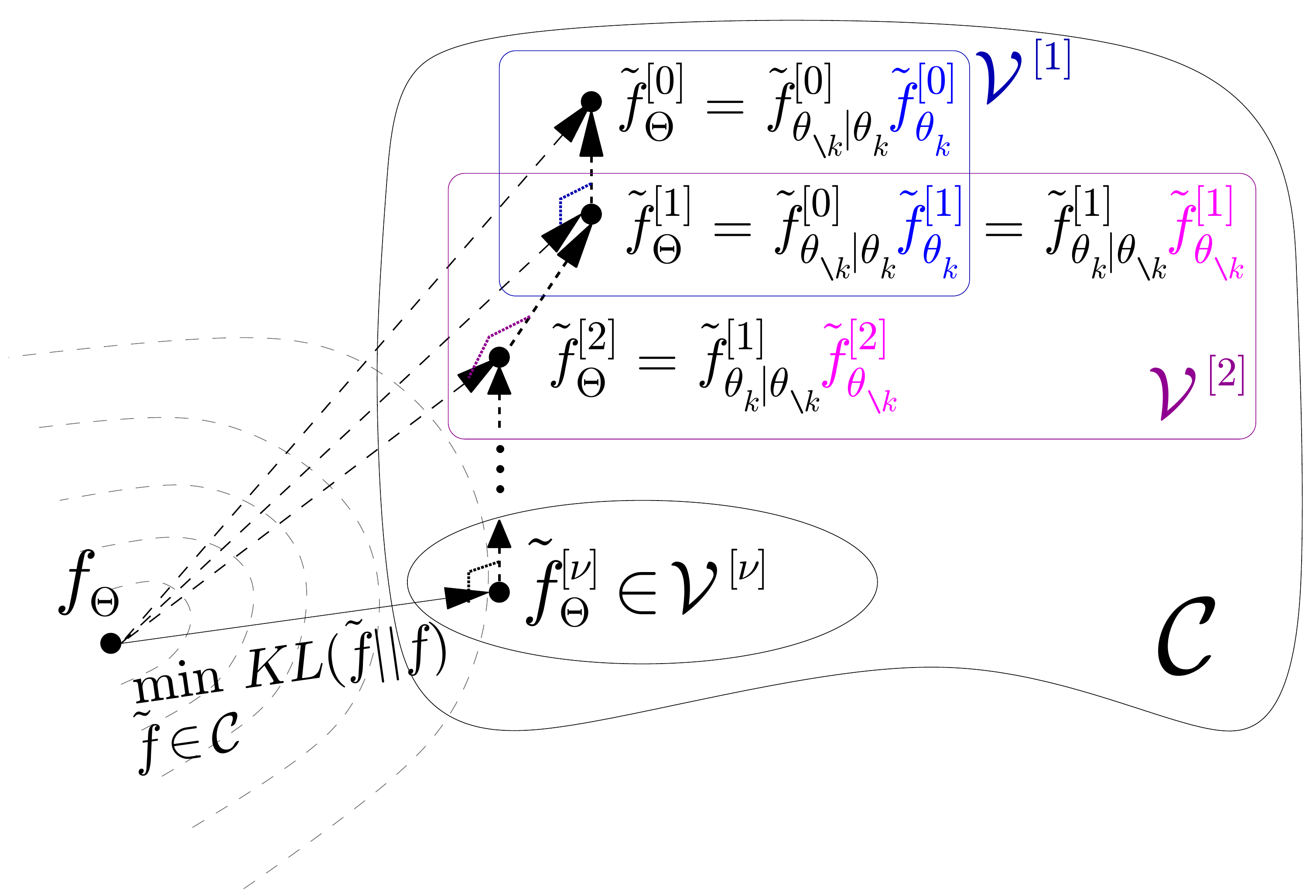}\caption{\label{fig:iterative_CVA}Venn diagram for iterative Copula Variational
approximation (CVA), given in (\ref{eq:CVB_kform}). The dashed contours
represent the convexity of $\protect\KL(\protect\ftilde_{\protect\bpara}||\protect\pdf_{\protect\bpara})$
over distributional points $\protect\ftilde_{\protect\bpara}$. The
set $\mathcal{C}$, possibly nonconvex, denotes a class of distributions
with the same copula form. Given initial form $\protect\ftilde_{\protect\para_{\backslash\protect\idim}|\protect\para_{\protect\idim}}^{[0]}$,
the joint distributions $\protect\ftilde_{\protect\bpara}^{[0]}$
and $\protect\ftilde_{\protect\bpara}^{[1]}$ belong to the same convex
set $\mathcal{V}^{[1]}\subseteq\mathcal{C}$, by Theorem \ref{thm:(Conditionally-Variational)}
and Corollary \ref{cor:(Copula-Variational-algorithm)}. The CVA $\protect\ftilde_{\protect\bpara}^{[1]}$
is the Bregman projection of the true distribution $\protect\pdf_{\protect\bpara}$
onto $\mathcal{V}^{[1]}$, with $\protect\ftilde_{\protect\para_{\protect\idim}}^{[1]}=\protect\argmin_{\protect\ftilde_{\protect\para_{\protect\idim}}\in\mathcal{V}^{[1]}}\protect\KL(\protect\ftilde_{\protect\bpara}||\protect\pdf_{\protect\bpara})$,
as shown in (\ref{eq:CV_pythagore}) and illustrated in Fig. \ref{fig:Pythagore}.
By interchanging the role of $\protect\para_{\backslash\protect\idim}$
and $\protect\para_{\protect\idim}$, the $\protect\KL(\protect\ftilde_{\protect\bpara}^{[\protect\iVB]}||\protect\pdf_{\protect\bpara})$
never increases over iterations $\protect\iVB$ and, hence, converges
to a local minimum inside copula set $\mathcal{C}$. In traditional
VB algorithm, we set $\protect\ftilde_{\protect\para_{\backslash\protect\idim}|\protect\para_{\protect\idim}}^{[\protect\iVB]}=\protect\ftilde_{\protect\para_{\backslash\protect\idim}}^{[\protect\iVB]}$,
which belongs to the independent copula class at all iterations $\protect\iVB$.}
\end{figure}

In CVA form above, we can only find one approximated marginal $\ftilde_{\idim}^{*}(\para_{\idim})$,
given conditional form $\ftilde_{\backslash\idim|\idim}^{*}(\para_{\backslash\idim}|\para_{\idim})$.
In the iterative form below, we will iteratively multiply $\ftilde_{\idim}^{*}(\para_{\idim})$
back to $\ftilde_{\backslash\idim|\idim}^{*}(\para_{\backslash\idim}|\para_{\idim})$
in order to find the reverse conditional $\ftilde_{\idim|\backslash\idim}^{*}(\para_{\idim}|\para_{\backslash\idim})$
for the next $\ftilde_{\backslash\idim}^{*}(\para_{\backslash\idim})$
via (\ref{eq:CV_KLDform}). At convergence, we can find a set of approximations
$\ftilde_{\idim}$, $\forall\seti{\idim}{\ndim}$, such that the $\KL_{\ftilde||\pdf}$
is locally minimized, as follows:
\begin{cor}
(Copula Variational approximation) \label{cor:(Copula-Variational-algorithm)}\\
Let $\ftilde=\ftilde_{\backslash\idim|\idim}^{[0]}\ftilde_{\idim}$
be the initial approximation with initial form $\ftilde_{\backslash\idim|\idim}^{[0]}$.
At iteration $\seti{\iVB}{\nVB}$, the approximation $\ftilde^{[\iVB]}=\ftilde_{\backslash\idim|\idim}^{[\iVB-1]}\ftilde_{\idim}^{[\iVB]}=\ftilde_{\idim|\backslash\idim}^{[\iVB]}\ftilde_{\backslash\idim}^{[\iVB]}$
is given by (\ref{eq:CV_KLDform}), as follows: 
\begin{equation}
\ftilde_{\idim}^{[\iVB]}(\para_{\idim})=\frac{\pdf_{\idim}(\para_{\idim})}{\VBzeta_{\idim}^{[\iVB]}\exp(\KL_{\ftilde_{\backslash\idim|\idim}^{[\iVB-1]}||\pdf_{\backslash\idim|\idim}})}\label{eq:CVB_kform}
\end{equation}
in which the reverse conditional is $\ftilde_{\idim|\backslash\idim}^{[\iVB]}=\frac{\ftilde_{\backslash\idim|\idim}^{[\iVB-1]}\ftilde_{\idim}^{[\iVB]}}{\ftilde_{\backslash\idim}^{[\iVB]}}$
and $\ftilde_{\backslash\idim}^{[\iVB]}\TRIANGLEQ\int_{\para_{\idim}}\ftilde_{\backslash\idim|\idim}^{[\iVB-1]}\ftilde_{\idim}^{[\iVB]}$,
$\forall\seti{\idim}{\ndim}$. Then, the value $\KL_{\ftilde^{[\iVB]}||\pdf}=\log\frac{1}{\VBzeta_{\idim}^{[\iVB]}}$
in (\ref{eq:CV_pythagore}), where $\VBzeta_{\idim}^{[\iVB]}$ is
the normalizing constant of marginal $\ftilde_{\idim}^{[\iVB]}$,
monotonically decreases to a local minimum at convergence $\iVB=\nVB$,
as illustrated in Fig. \ref{fig:iterative_CVA}. 

Note that, by copula's marginal-invariant property (\ref{eq:marginal_invariant}),
the copula's form of the iterative joint distribution $\ftilde^{[\iVB]}(\bpara)$
is invariant with any updated marginals $\ftilde_{\idim}^{[\iVB]}(\para_{\idim})$,
$\forall\idim$, hence the name Copula Variational approximation. 
\end{cor}
\begin{IEEEproof}
Since the calculation of reverse form $\ftilde_{\idim|\backslash\idim}^{[\iVB]}$
does not change $\ftilde^{[\iVB]}(\bpara)$, the value $\KL_{\ftilde^{[\iVB]}||\pdf}$
only decreases with marginal update $\ftilde_{\idim}^{[\iVB]}$ via
(\ref{eq:CV_pythagore}-\ref{eq:CV_KLDform}) and, hence, converges
monotonically. 
\end{IEEEproof}
If the initial form $\ftilde_{\backslash\idim|\idim}^{[0]}$ belongs
to the independent space, i.e. $\ftilde_{\backslash\idim|\idim}^{[0]}=\ftilde_{\backslash\idim}^{[0]}$,
the copula of the joint $\ftilde_{\bpara}^{[0]}$ will have independent
form, as noted in Remark \ref{rem:indepedentCopula}, and cannot leave
this independence space via dual iterations of (\ref{eq:CVB_kform}).
Hence, for a binary partition $\bpara=\{\para_{\backslash\idim},\para_{\idim}\}$,
an initially independent copula will lead to a mean-field approximation. 

Nonetheless, this is not true in general for ternary partition $\bpara=\{\para_{\idim},\para_{\ipick},\para_{\istate}\}$
or for a generic network of parameters, since the iterative CVA (\ref{eq:CVB_kform})
can be implemented with different partitions of a network at any iteration,
without changing the joint network's copula or increasing the joint
$\KL$ divergence $\KL_{\ftilde^{[\iVB]}||\pdf}$. 

For example, in ternary partition, even if we initially set $\ftilde_{\idim|\backslash\idim}^{[0]}=\ftilde_{\idim}^{[0]}$
independent of $\para_{\backslash\idim}=\{\para_{\ipick},\para_{\istate}\}$
and yield the updated $\ftilde_{\backslash\idim}^{[1]}=\ftilde^{[1]}(\para_{\ipick},\para_{\istate})$
for $\ftilde^{[1]}=\ftilde_{\idim}^{[0]}\ftilde_{\backslash\idim}^{[1]}=\ftilde_{\istate|\ipick}^{[1]}\ftilde_{\backslash\istate}^{[1]}$
via (\ref{eq:CVB_kform}), the reverse form $\ftilde_{\istate|\ipick}^{[1]}$
yields $\ftilde_{\backslash\istate}^{[2]}\TRIANGLEQ\ftilde^{[2]}(\para_{\idim},\para_{\ipick})$
via (\ref{eq:CVB_kform}) again and, hence $\ftilde_{\idim|\backslash\idim}^{[2]}=\ftilde^{[2]}(\para_{\idim}|\para_{\ipick})$
dependent on $\para_{\ipick}$ again, which does not yield a mean-field
approximation in subsequent iterations of (\ref{eq:CVB_kform}). This
ternary partition scheme will be implemented in (\ref{eq:cluster_CVB})
and clarified further in Remark \ref{rem:init-CVB}.

\subsubsection{Conditionally exponential family (CEF) approximation\label{subsec:Conditionally-exponential-family}}

The computation in above approximations will be linearly tractable,
if the true joint $f(\bpara)$ and the approximated conditional $\ftilde_{\backslash\idim|\idim}$
can be linearly factorized with respect to $\log$-operator in (\ref{eq:CV_KLDform})
and (\ref{eq:CVB_kform}). The distributions satisfying this property
belong to a special class of distributions, namely CEF, defined as
follows:
\begin{defn}
(Conditionally Exponential Family)\\
A joint distribution $\pdf(\bpara)$ is a member of CEF if it has
the following form: 
\begin{equation}
\pdf(\bpara)\propto\exp\left\langle \boldsymbol{g}_{\idim}(\para_{\idim}),\boldsymbol{g}_{\backslash\idim}(\para_{\backslash\idim})\right\rangle \label{eq:CEF}
\end{equation}
where $\boldsymbol{g}_{\idim}$, $\boldsymbol{g}_{\backslash\idim}$
are vectors dependent on $\para_{\idim}$, $\para_{\backslash\idim}$
element-wise, respectively. Note that, the form (\ref{eq:CEF}) is
similar to the well-known Exponential Family in literature \cite{AQ::Book:06,Bayes:BOOK:Bernado},
hence the name CEF. 
\end{defn}
From (\ref{eq:CEF}), the marginal of a joint CEF distribution is:
\begin{equation}
\pdf(\para_{\idim})\propto\int_{\para_{\backslash\idim}}\exp\left\langle \boldsymbol{g}_{\idim}(\para_{\idim}),\boldsymbol{g}_{\backslash\idim}(\para_{\backslash\idim})\right\rangle d\para_{\backslash\idim}\label{eq:CEF_marginal}
\end{equation}
which may not be tractable, since the CEF form is not factorable in
general. In contrast, the CVA (\ref{eq:CV_KLDform}) for CEF distributions
(\ref{eq:CEF}) is more tractable, as follows:
\begin{cor}
(CEF approximation) \\
Let $\ftilde=\ftilde_{\backslash\idim|\idim}^{*}\ftilde_{\idim}$
be a distribution with $\ftilde_{\backslash\idim|\idim}^{*}=\exp\left\langle \boldsymbol{h}_{\idim}(\para_{\idim}),\boldsymbol{h}_{\backslash\idim}(\para_{\backslash\idim})\right\rangle $
given by CEF form in (\ref{eq:CEF}). If the true distribution $\pdf(\bpara)$
also takes the CEF form (\ref{eq:CEF}), the approximation $\ftilde^{*}$
minimizing $\KL_{\ftilde||\pdf}$ in (\ref{eq:CV_pythagore}), as
given by (\ref{eq:CV_KLDform}), also belongs to CEF, as follows:
\begin{equation}
\ftilde_{\idim}^{*}(\para_{\idim})\propto\exp\left\langle \boldsymbol{\eta}_{\idim}(\para_{\idim}),\boldsymbol{\eta}_{\backslash\idim}^{*}(\para_{\idim})\right\rangle \label{eq:CEF_approx}
\end{equation}
where $\boldsymbol{\eta}_{\backslash\idim}^{*}(\para_{\idim})\TRIANGLEQ\EXPECTATION_{\ftilde_{\backslash\idim|\idim}^{*}}\boldsymbol{\eta}_{\backslash\idim}(\para_{\backslash\idim})$,
with $\boldsymbol{\eta}_{\idim}\TRIANGLEQ\boldsymbol{g}_{\idim}-\boldsymbol{h}_{\idim}$
and $\boldsymbol{\eta}_{\backslash\idim}\TRIANGLEQ\boldsymbol{g}_{\backslash\idim}-\boldsymbol{h}_{\backslash\idim}$.
\end{cor}
\begin{IEEEproof}
The form (\ref{eq:CEF_approx}) is a direct consequence of (\ref{eq:CV_KLDform}),
since both $\ftilde_{\backslash\idim|\idim}^{*}$ and $\pdf(\bpara)$
in (\ref{eq:CV_KLDform}) now have CEF form (\ref{eq:CEF}).
\end{IEEEproof}
From (\ref{eq:CEF_marginal}-\ref{eq:CEF_approx}), we can see that
the integral in (\ref{eq:CEF_marginal}) has moved inside the non-linear
$\exp$ operator in (\ref{eq:CEF_approx}) and, hence, become linear
and numerically tractable. Then, substituting (\ref{eq:CEF_approx})
into iterative CVA (\ref{eq:CVB_kform}), we can see that the iterative
CVA for CEF is also tractable, since we only have to update the parameters
of CEF iteratively in (\ref{eq:CEF_approx}) until convergence. 
\begin{rem}
In the nutshell, the key advantage of KL divergence is to approximate
the originally intractable arithmetic mean (\ref{eq:CEF_marginal})
by the tractable geometric mean in exponential domain (\ref{eq:CEF_approx}),
as noted in Remark \ref{rem:KLvariance}.
\end{rem}

\subsubsection{Backward KLD and minimum-risk (MR) approximation}

In above approximations, we have used the forward $\text{\ensuremath{\KL}}_{\ftilde||f}$
(\ref{eq:CV_pythagore}) as the approximation criterion, since the
Bregman pythagorean property (\ref{eq:PYTHAGORE}) is only valid for
forward $\text{\ensuremath{\KL}}_{\ftilde||f}$. Moreover, the approximation
via backward $\text{\ensuremath{\KL}}_{f||\ftilde}$ is not interesting
since the minimum is only achieved with the true distributions, as
shown below:
\begin{cor}
(Conditionally minimum-risk approximation) \label{cor:(Conditionally-minimum-risk)}\\
The approximation $\ftilde^{*}=\ftilde_{\backslash\idim|\idim}\ftilde_{\idim}$
minimizing backward $\text{\ensuremath{\KL}}_{f||\ftilde}$ is either
$\ftilde^{*}=\ftilde_{\backslash\idim|\idim}^{\text{MR}}\pdf_{\idim}$
or $\ftilde^{*}=\pdf_{\backslash\idim|\idim}\ftilde_{\idim}^{\MR}$
for fixed $\ftilde_{\backslash\idim|\idim}^{\text{MR}}$ or fixed
$\ftilde_{\idim}^{\text{MR}}$, respectively, where $\pdf_{\idim}$
and $\pdf_{\backslash\idim|\idim}$ are the true marginal and conditional
distributions.
\end{cor}
\begin{IEEEproof}
Similar to proof of Theorem \ref{thm:(Conditionally-Variational)},
the backward form is $\text{\ensuremath{\KL}}_{f||\ftilde}=\EXPECTATION_{\pdf_{\idim}}\text{\ensuremath{\KL}}_{\pdf_{\backslash\idim|\idim}||\ftilde_{\backslash\idim|\idim}}+\text{\ensuremath{\KL}}_{\pdf_{\idim}||\ftilde_{\idim}}$.
Hence, $\text{\ensuremath{\KL}}_{f||\ftilde^{*}}$ is minimum at $\ftilde_{\idim}^{\text{MR}}=\pdf_{\idim}$
for fixed $\text{\ensuremath{\KL}}_{\pdf_{\backslash\idim|\idim}||\ftilde_{\backslash\idim|\idim}^{\MR}}$
and minimum at $\ftilde_{\backslash\idim|\idim}^{\text{MR}}=\pdf_{\backslash\idim|\idim}$
for fixed $\text{\ensuremath{\KL}}_{\pdf_{\idim}||\ftilde_{\idim}^{\MR}}$.
\end{IEEEproof}
\begin{rem}
The Corollary \ref{cor:(Conditionally-minimum-risk)} is the generalized
form of the minimum-risk approximation in \cite{AQ::Book:06}, which
minimizes backward KL divergence in the context of VB approximation
in mean-field theory. The name ``minimum-risk'' refers to the fact
that the true distribution always yields minimum-risk estimation in
Bayesian theory (c.f. Appendix \ref{subsec:Bayesian-minimum-risk}).
\end{rem}

\subsection{Mean-field approximations \label{subsec:Mean-field}}

If we confine the conditional form $\ftilde=\ftilde_{\backslash\idim|\idim}\ftilde_{\idim}$
in above approximations by independent form, i.e. $\ftilde=\ftilde_{\backslash\idim}\ftilde_{\idim}$,
we will recover the so-called mean-field approximations in literature.
Four cases of them, namely VB, EM, ICM and k-means algorithms, will
be presented below.

\subsubsection{Variational Bayes (VB) algorithm\label{subsec:Variational-Bayes-(VB)}}

From CVA (\ref{eq:CV_KLDform}), the VB algorithm is given as follows:
\begin{cor}
(VB approximation)\\
The independent distribution $\ftilde^{*}=\ftilde_{\backslash\idim}^{*}\ftilde_{\idim}^{*}$
minimizing $\KL_{\ftilde||\pdf}$ in (\ref{eq:CV_pythagore}) is given
by (\ref{eq:CV_KLDform}), as follows: 
\begin{align}
\ftilde_{\idim}^{*}(\para_{\idim}) & \propto\frac{\pdf_{\idim}(\para_{\idim})}{\exp(\KL_{\ftilde_{\backslash\idim}^{*}||\pdf_{\backslash\idim|\idim}})}\propto\exp\EXPECTATION_{\ftilde_{\backslash\idim}^{*}(\para_{\backslash\idim})}\log f(\bpara),\label{eq:(VB)}
\end{align}
$\forall\seti{\idim}{\ndim}$, as illustrated in Fig. \ref{fig:fCVA}.
\end{cor}
\begin{IEEEproof}
Since $\ftilde_{\backslash\idim|\idim}=\ftilde_{\backslash\idim}$
does not depends on $\para_{\idim}$ in this case, substituting $\ftilde_{\backslash\idim|\idim}=\ftilde_{\backslash\idim}$
into (\ref{eq:CV_KLDform}) yields (\ref{eq:(VB)}).
\end{IEEEproof}
Since there is no conditional form $\ftilde_{\backslash\idim|\idim}$
to be updated, the iterative VB algorithm simply updates (\ref{eq:(VB)})
iteratively for all marginals $\ftilde_{\idim}$ and $\ftilde_{\backslash\idim}$,
similar to (\ref{eq:CVB_kform}), until convergence. Hence, VB algorithm
is a special case of Copula Variational algorithm in Corollary~\ref{cor:(Copula-Variational-algorithm)},
in which the approximated copula is of independent form, as noted
in Remark~\ref{rem:indepedentCopula}. 

\subsubsection{Expectation-Maximization (EM) algorithm}

\begin{figure}
\centering{}\includegraphics[width=1\linewidth]{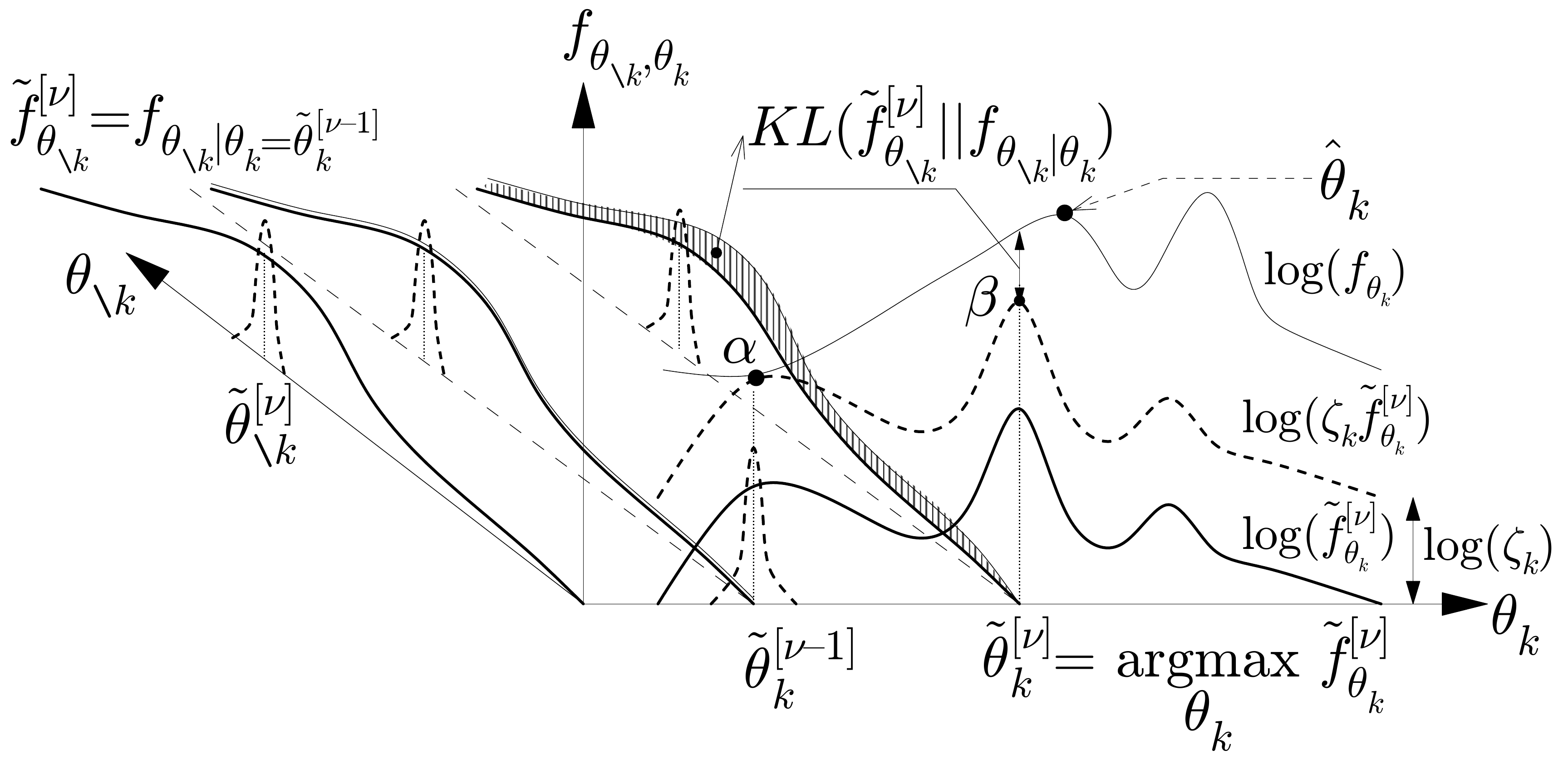}\caption{\label{fig:fEM}Illustration of Expectation-Maximization (EM) algorithm~(\ref{eq:EM})
as a special case of VB approximation. The lower KL divergence, the
better approximation. Given restricted form $\protect\ftilde_{\protect\para_{\backslash\protect\idim}}^{[\protect\iVB]}=\protect\pdf(\protect\para_{\backslash\protect\idim}|\protect\tpara_{\protect\idim}^{[\protect\iVB-1]})$
at iteration $\protect\iVB$, the approximated $\protect\ftilde_{\protect\para_{\protect\idim}}^{[\protect\iVB]}$
minimizing $\protect\KL(\protect\ftilde_{\protect\para_{\backslash\protect\idim}}^{[\protect\iVB]}\protect\ftilde_{\protect\para_{\protect\idim}}^{[\protect\iVB]}||\protect\pdf_{\protect\bpara})$
is proportional to the true marginal $\protect\pdf_{\protect\para_{\protect\idim}}$
by a fraction of conditional divergence, similar to Fig.~\ref{fig:fCVA}.
Note that, $\protect\tpara_{\protect\idim}^{[\protect\iVB]}$ might
fail to converge to a local mode $\widehat{\protect\para}_{\protect\idim}$
of the true marginal $\protect\pdf_{\protect\para_{\protect\idim}}$,
if the peak $\beta$ is lower than point $\alpha$. For ICM algorithm
(\ref{eq:ICM}), we further restrict $\protect\ftilde_{\protect\para_{\backslash\protect\idim}}^{[\protect\iVB]}$
to a Dirac delta distribution concentrating around its mode $\protect\tpara_{\backslash\protect\idim}^{[\protect\iVB]}$
and, hence, $\tilde{\protect\bpara}^{[\protect\iVB]}=\{\protect\tpara_{\backslash\protect\idim}^{[\protect\iVB]},\protect\tpara_{\protect\idim}^{[\protect\iVB]}\}$
always converges to a joint local mode $\widehat{\protect\bpara}$
of the true distribution $\protect\pdf_{\protect\bpara}$. }
\end{figure}

If we restrict the independent form $\ftilde=\ftilde_{\backslash\idim}\ftilde_{\idim}$
in VB algorithm with Dirac delta form $\ftilde_{\EM}\TRIANGLEQ\ftilde_{\backslash\idim}\dtilde_{\idim}$,
where $\dtilde_{\idim}\TRIANGLEQ\delta(\para_{\idim}-\tpara_{\idim})$,
we will recover the EM algorithm, as follows:
\begin{cor}
(EM algorithm) \\
At iteration $\seti{\iVB}{\nVB}$, the EM approximation of $\pdf(\bpara)$
is $\ftilde_{\EM}^{[\iVB]}\TRIANGLEQ\ftilde_{\backslash\idim}^{[\iVB]}\dtilde_{\idim}^{[\iVB]}$,
in which $\ftilde_{\backslash\idim}^{[\iVB]}=f(\para_{\backslash\idim}|\tpara_{\idim}^{[\iVB]})$
and $\dtilde_{\idim}^{[\iVB]}\TRIANGLEQ\delta(\para_{\idim}-\tpara_{\idim}^{[\iVB]})$,
as given by (\ref{eq:(VB)}): 

\begin{align}
\tpara_{\idim}^{[\iVB]} & \TRIANGLEQ\argmax_{\para_{\idim}}\ftilde_{\idim}^{[\iVB]}(\para_{\idim})=\argmax_{\para_{\idim}}\EXPECTATION_{f(\para_{\backslash\idim}|\tpara_{\idim}^{[\iVB-1]})}\log\pdf(\bpara)\label{eq:EM}\\
 & =\argmax_{\para_{\idim}}\frac{\pdf_{\idim}(\para_{\idim})}{\exp(\KL_{f(\para_{\backslash\idim}|\tpara_{\idim}^{[\iVB-1]})||f(\para_{\backslash\idim}|\para_{\idim})})}.\nonumber 
\end{align}
If $\tpara_{\idim}^{[\iVB]}$ converges to a true local maximum $\hpara_{\idim}$
of the original marginal $\pdf_{\idim}(\para_{\idim})$, as illustrated
in Fig. \ref{fig:fEM}, then $\KL_{\ftilde_{\EM}^{[\iVB]}||\pdf}=-\log\pdf_{\idim}(\tpara_{\idim}^{[\iVB]})$
converges to a local minimum.
\end{cor}
\begin{IEEEproof}
Substituting the Dirac delta function $\ftilde_{\idim}^{*}(\para_{\idim})=\delta(\para_{\idim}-\tpara_{\idim}^{[\iVB-1]})$
to VB approximation (\ref{eq:(VB)}), we have $\ftilde_{\backslash\idim}^{[\iVB-1]}(\para_{\backslash\idim})=f(\para_{\backslash\idim}|\tpara_{\idim}^{[\iVB-1]})$,
which yields (\ref{eq:EM}) owing to (\ref{eq:(VB)}). 

Since the $\KL$ value in (\ref{eq:EM}) is never negative, we have
$g(\para_{\idim})\TRIANGLEQ\pdf_{\idim}(\para_{\idim})/\exp(\KL_{f(\para_{\backslash\idim}|\tpara_{\idim}^{[\iVB-1]})||f(\para_{\backslash\idim}|\para_{\idim})})\leq\pdf_{\idim}(\para_{\idim})$
and the equality $g(\tpara_{\idim}^{[\iVB]})=\pdf_{\idim}(\tpara_{\idim}^{[\iVB]})$
happens at $\para_{\idim}=\tpara_{\idim}^{[\iVB]}$, which means:
$\pdf_{\idim}(\tpara_{\idim}^{[\iVB-1]})=g(\tpara_{\idim}^{[\iVB-1]})\leq g(\tpara_{\idim}^{[\iVB]})=\max g(\para_{\idim})\leq\max\pdf_{\idim}(\para_{\idim})$.
Then, as illustrated in Fig. \ref{fig:fEM}, if $\ftilde_{\idim}^{[\iVB]}(\tpara_{\idim}^{[\iVB]})$
strictly increases over $\iVB$, $\tpara_{\idim}^{[\iVB]}$ will converge
to a local mode $\widehat{\para}_{\idim}$ of $\pdf_{\idim}(\para_{\idim})$,
owing to majorization-maximization (MM) principle \cite{EM:majorization:04,EM:majorization:17}.
Otherwise, $\tpara_{\idim}^{[\iVB]}$ might fail to converge to $\widehat{\para}_{\idim}$. 

Lastly, from (\ref{eq:CV_proof}), we have: $\KL_{\ftilde_{\EM}^{[\iVB]}||\pdf}=\EXPECTATION_{\dtilde_{\idim}^{[\iVB]}}\KL_{f(\para_{\backslash\idim}|\tpara_{\idim}^{[\iVB]})||f(\para_{\backslash\idim}|\para_{\idim})}+\text{\ensuremath{\KL}}_{\dtilde_{\idim}^{[\iVB]}||f_{\idim}}=\text{\ensuremath{\KL}}_{\dtilde_{\idim}^{[\iVB]}||f_{\idim}}=-\log\pdf_{\idim}(\tpara_{\idim}^{[\iVB]})$
by sifting property of Dirac delta function. 
\end{IEEEproof}
From (\ref{eq:(VB)}-\ref{eq:EM}), we can see that EM algorithm is
a special case of VB algorithm. Both of them minimizes the KL divergence
within the independent distribution space, namely mean-field space. 

Since EM algorithm is a fixed-form approximation, it has low computational
complexity. Nonetheless, as illustrated in Fig. \ref{fig:fEM}, the
point estimate $\tpara_{\idim}^{[\iVB]}$ in EM algorithm (\ref{eq:EM})
might fail to converge to a local mode $\hpara_{\idim}$ of true marginal
$\pdf_{\idim}(\para_{\idim})$ in practice. In contrast, VB approximation
is a free-form distribution and capable of approximating higher-order
moments of true marginal $\pdf_{\idim}(\para_{\idim})$. 
\begin{rem}
\label{rem:EM}Note that, EM algorithm is also a special case of Copula
Variational algorithm (\ref{eq:CVB_kform}) in conditional space.
Indeed, if the marginal $\ftilde_{\idim}$ of $\ftilde=\ftilde_{\backslash\idim|\idim}\ftilde_{\idim}$
in (\ref{eq:CVB_kform}) is restricted to Dirac delta form, i.e. $\ftilde_{\idim}=\dtilde_{\idim}$,
the joint $\ftilde$ will become a degenerated independent distribution,
i.e. $\ftilde=\ftilde(\para_{\backslash\idim}|\para_{\idim}=\tpara_{\idim})\delta(\para_{\idim}-\tpara_{\idim})=\ftilde_{\backslash\idim}\dtilde_{\idim}$,
owing to sifting property of Dirac delta. Hence, EM algorithm is a
very special approximation, since it belongs to both mean-field and
copula-field approximations.
\end{rem}

\subsubsection{Iterative conditional mode (ICM) algorithm}

If we further restrict the independent form $\ftilde=\ftilde_{\backslash\idim}\ftilde_{\idim}$
in VB algorithm fully to Dirac delta form $\ftilde_{\ICM}\TRIANGLEQ\dtilde_{\backslash\idim}\dtilde_{\idim}$,
we will recover the iterative plug-in algorithm, also called Iterative
Conditional Mode (ICM) in literature \cite{ch4:art:ICM:Besag86,ch4:art:ICM:localMAP_2006},
as follows: 
\begin{cor}
(ICM algorithm)\label{cor:(ICM-algorithm)}\\
At iteration $\seti{\iVB}{\nVB}$, the ICM approximation of $\pdf(\bpara)$
is $\ftilde_{\ICM}^{[\iVB]}\TRIANGLEQ\ftilde_{\backslash\idim}^{[\iVB]}\ftilde_{\idim}^{[\iVB]}=\dtilde_{\backslash\idim}^{[\iVB]}\dtilde_{\idim}^{[\iVB]}$,
where $\dtilde_{\backslash\idim}^{[\iVB]}\TRIANGLEQ\delta(\para_{\backslash\idim}-\tpara_{\backslash\idim}^{[\iVB]})$
and $\dtilde_{\idim}^{[\iVB]}\TRIANGLEQ\delta(\para_{\idim}-\tpara_{\idim}^{[\iVB]})$
is given by (\ref{eq:(VB)}), as follows: 

\begin{align}
\tpara_{\idim}^{[\iVB]} & \TRIANGLEQ\argmax_{\para_{\idim}}f(\para_{\idim}|\tpara_{\backslash\idim}^{[\iVB-1]})=\argmax_{\para_{\idim}}f(\para_{\idim},\para_{\backslash\idim}=\tpara_{\backslash\idim}^{[\iVB-1]})\label{eq:ICM}\\
\tpara_{\backslash\idim}^{[\iVB]} & \TRIANGLEQ\argmax_{\para_{\backslash\idim}}f(\para_{\backslash\idim}|\tpara_{\idim}^{[\iVB]})=\argmax_{\para_{\backslash\idim}}f(\para_{\idim}=\tpara_{\idim}^{[\iVB]},\para_{\backslash\idim})\nonumber 
\end{align}
From (\ref{eq:ICM}), we can see that $\tilde{\bpara}^{[\iVB]}=\{\tpara_{\backslash\idim}^{[\iVB]},\tpara_{\idim}^{[\iVB]}\}$
iteratively converges to a local maximum $\hat{\bpara}$ of the true
distribution $\pdf(\bpara)$ and, hence, $\KL_{\ftilde_{\ICM}^{[\iVB]}||\pdf}=-\log\pdf(\tilde{\bpara}^{[\iVB]})$
converges to a local minimum. 
\end{cor}
\begin{IEEEproof}
The proof is a straight-forward derivation from either VB (\ref{eq:(VB)})
or EM (\ref{eq:EM}) algorithms, by sifting property of Dirac delta
forms $\ftilde_{\backslash\idim}^{[\iVB]}=\delta(\para_{\backslash\idim}-\tpara_{\backslash\idim}^{[\iVB]})$
and $\ftilde_{\idim}^{[\iVB]}=\delta(\para_{\idim}-\tpara_{\idim}^{[\iVB]})$.
\end{IEEEproof}
Since we merely plug the value $\{\tpara_{\backslash\idim}^{[\iVB]},\tpara_{\idim}^{[\iVB]}\}$
into the true distribution $\pdf(\bpara)$ iteratively in (\ref{eq:ICM})
until it reaches a local maximum, the performance of this naive hit-or-miss
approach is strongly influenced by the initial points $\{\tpara_{\backslash\idim}^{[0]},\tpara_{\idim}^{[0]}\}$.
Hence it is often used in practice when very low computational complexity
is required or when the true distribution $\pdf(\bpara)$ does not
have tractable CEF form (\ref{eq:CEF}). 
\begin{rem}
Similar to the Remark \ref{rem:EM}, we can see that ICM is a degenerated
form of VB, EM and Copula Variational approximations, owing to its
very simple form (\ref{eq:ICM}).
\end{rem}

\subsubsection{K-means algorithm}

In section \ref{subsec:ICM-and-k-means}, we will show that the popular
k-means algorithm is equivalent to ICM algorithm being applied to
a mixture of independent Gaussian distributions. Hence, k-means is
also a member of mean-field approximations.

\subsection{Copula Variational Bayes (CVB) approximation}

In a model with unknown multi-parameters $\bpara=\{\para_{\idim},\para_{\backslash\idim}\}$,
the minimum-risk estimation of $\para_{\idim}$ can be evaluated from
the marginal posterior $f(\para_{\idim}|\xbold)=\int_{\para_{\backslash\idim}}f(\bpara|\xbold)d\para_{\backslash\idim}$
(c.f. Appendix \ref{subsec:Bayesian-minimum-risk}), in which the
posterior distribution $f(\bpara|\xbold)$ is then given via Bayes'
rule: $f(\bpara|\xbold)\propto f(\xbold,\bpara)=f(\xbold|\bpara)f(\bpara)$.
In practice, however, the computational complexity of the normalizing
constant of $f(\bpara|\xbold)$ involves all possible values of $\bpara$
and typically grows exponentially with number of data's dimension,
which is termed the curse of dimensionality \cite{Bayes:CurseOfDimension:97}.
Then, without normalizing constant of $f(\bpara|\xbold)$, the computation
of moments of $f(\para_{\idim}|\xbold)$ is also intractable. 

In this subsection, we will apply both copula-field and mean-field
approximations to the joint posterior distribution $\pdf(\bpara|\xbold)\propto f(\xbold,\bpara)$
and, then, return all marginal approximations $\ftilde(\para_{\idim}|\xbold)$
directly from the joint model $f(\xbold,\bpara)$, without computing
the normalizing constant of $\pdf(\bpara|\xbold)$, as explained below.
\begin{cor}
(Copula Variational Bayes algorithm)\\
At iteration $\seti{\iVB}{\nVB}$, the CVB approximation $\ftilde^{[\iVB]}(\bpara|\xbold)=\ftilde^{[\iVB-1]}(\para_{\backslash\idim}|\para_{\idim},\xbold)\ftilde_{\idim}^{[\iVB]}(\para_{\idim}|\xbold)$
for the joint posterior $\pdf(\bpara|\xbold)$ is given by (\ref{eq:CV_KLDform})
and (\ref{eq:CVB_kform}), as follows:

\begin{align}
\ftilde_{\idim}^{[\iVB]}(\para_{\idim}|\xbold) & =\frac{1}{\VBzeta_{\idim}^{[\iVB]}(\xbold)}\frac{\pdf(\para_{\idim}|\xbold)}{\KL(\ftilde^{[\iVB-1]}(\para_{\backslash\idim}|\para_{\idim},\xbold)||\pdf(\para_{\backslash\idim}|\para_{\idim},\xbold))}\label{eq:CVB_posterior}\\
 & =\frac{1}{\VBzeta_{\idim}^{[\iVB]}(\xbold)}\exp\EXPECTATION_{\ftilde^{[\iVB-1]}(\para_{\backslash\idim}|\para_{\idim},\xbold)}\log\frac{\pdf(\xbold,\bpara)}{\ftilde^{[\iVB-1]}(\para_{\backslash\idim}|\para_{\idim},\xbold)}\nonumber 
\end{align}
in which $\ftilde^{[\iVB]}(\para_{\idim}|\para_{\backslash\idim},\xbold)=\ftilde^{[\iVB]}(\bpara|\xbold)/\ftilde_{\backslash\idim}^{[\iVB]}(\para_{\backslash\idim}|\xbold)$,
$\forall\seti{\idim}{\ndim}$. For stopping rule, the evidence lower
bound (ELBO) for CVB is defined similarly to (\ref{eq:CV_pythagore}),
as follows: $\KL_{\ftilde^{[\iVB]}||\pdf}=-\text{ELBO}^{[\iVB]}+\log\pdf(\xbold)\geq0$,
i.e. we have:

\begin{equation}
\log\pdf(\xbold)\geq\text{ELBO}^{[\iVB]}\TRIANGLEQ-\KL_{\ftilde^{[\iVB]}(\bpara|\xbold)||\pdf(\xbold,\bpara)}=\log\VBzeta_{\idim}^{[\iVB]}(\xbold)\label{eq:ELBO}
\end{equation}
Since the evidence $\pdf(\xbold)$ is a constant, $\text{\ensuremath{\KL}}_{\ftilde^{[\iVB]}||f}\TRIANGLEQ\text{\ensuremath{\KL}}_{\ftilde^{[\iVB]}(\bpara|\xbold)||\pdf(\bpara|\xbold)}$
monotonically decreases to a local minimum, while the marginal normalizing
constant $\VBzeta_{\idim}^{[\iVB]}(\xbold)$ in (\ref{eq:CVB_posterior})
and $\text{ELBO}^{[\iVB]}$ in (\ref{eq:ELBO}) monotonically increase
to a local maximum at convergence $\iVB=\nVB$. 

Note that, the copula's form of the iterative CVB $\ftilde^{[\iVB]}(\bpara|\xbold)$
is invariant with any updated marginal $\ftilde_{\idim}^{[\iVB]}(\para_{\idim}|\xbold)$,
$\forall\seti{\idim}{\ndim}$, as shown in (\ref{eq:marginal_invariant}),
hence the name Copula Variational Bayes approximation. 
\end{cor}
\begin{IEEEproof}
Firstly, we have $\KL_{\ftilde^{[\iVB]}(\bpara|\xbold)||\pdf(\xbold,\bpara)}=\text{\ensuremath{\KL}}_{\ftilde^{[\iVB]}||f}-\log\pdf(\xbold)$
by definition of KL divergence (\ref{eq:KL}), hence the definition
of $\text{ELBO}^{[\iVB]}$ in (\ref{eq:ELBO}). Then, similar to (\ref{eq:CV_proof}),
the value $\text{\ensuremath{\KL}}_{\ftilde||f}\TRIANGLEQ\text{\ensuremath{\KL}}_{\ftilde(\bpara|\xbold)||\pdf(\bpara|\xbold)}$
for arbitrary $\ftilde$ in this case is:

\begin{equation}
\text{\ensuremath{\KL}}_{\ftilde||f}=\underset{-\text{ELBO}}{\underbrace{\text{\ensuremath{\KL}}_{\ftilde_{\idim}||\ftilde_{\idim}^{[\iVB]}}+\log\frac{1}{\text{\ensuremath{\VBzeta_{\idim}^{[\iVB]}}(\ensuremath{\xbold})}}}}+\log\pdf(\xbold),\label{eq:KLD_ELBO}
\end{equation}
in which $\ftilde_{\idim}^{[\iVB]}$ is defined in \ref{eq:CVB_posterior},
the form $\pdf(\bpara)$ in (\ref{eq:CV_proof}) is now replaced by
$\pdf(\bpara|\xbold)=\pdf(\xbold,\bpara)/\pdf(\xbold)$, hence the
term $\pdf(\xbold,\bpara)$ in (\ref{eq:CVB_posterior}) and the constant
evidence $\log\pdf(\xbold)$ in (\ref{eq:KLD_ELBO}). Since $\text{\ensuremath{\KL}}_{\ftilde_{\idim}||\ftilde_{\idim}^{[\iVB]}}=0$
for the case $\ftilde_{\idim}=\ftilde_{\idim}^{[\iVB]}$, the value
$\text{ELBO}$ in (\ref{eq:KLD_ELBO}) is equal to $\log\VBzeta_{\idim}^{[\iVB]}(\xbold)$,
which yields (\ref{eq:ELBO}). The rest of proof is similar to the
proof of Corollary \ref{cor:(Copula-Variational-algorithm)}. 
\end{IEEEproof}
Note that, CVB algorithm (\ref{eq:CVB_posterior}) is essentially
the same as the Copula Variational algorithm in (\ref{eq:CVB_kform}).
The key difference is that the former is applied to a joint posterior
$\pdf(\bpara|\xbold)$, while the latter is applied to a joint distribution
$\pdf(\bpara)$. Hence, in CVB, the joint model $\pdf(\xbold,\bpara)$
and ELBO  (\ref{eq:ELBO}) are preferred, since the evidence $\log\pdf(\xbold)$
is often hard to compute in practice. Nevertheless, for notational
simplicity, let us call both of them CVB hereafter. By this way, the
name CVA (\ref{eq:CV_KLDform}) also implies that it is the first
step of CVB algorithm.
\begin{rem}
Although the iterative CVB form (\ref{eq:CVB_posterior}) is novel,
the definition of ELBO via KL divergence in (\ref{eq:ELBO}) was recently
proposed in \cite{CVB:Blei:gradient:15}. Nevertheless, the value
$\log\VBzeta_{\idim}^{[\iVB]}(\xbold)$ of ELBO in (\ref{eq:ELBO})
was not given therein. Also, the so-called copula variational inference
in \cite{CVB:Blei:gradient:15} was to locally minimize ELBO (\ref{eq:ELBO})
via a sampling-based stochastic-gradient decent for copula's parameters,
rather than via a deterministic expectation operator in (\ref{eq:CVB_posterior}).
No explicit CVB's marginal form at convergence was given in \cite{CVB:Blei:gradient:15}. 
\end{rem}

\subsubsection{Conditionally Exponential Family (CEF) for posterior distribution}

Similar to (\ref{eq:CEF_approx}), the computation of CVB algorithm
(\ref{eq:CVB_posterior}) will be linearly tractable if the true posterior
$\pdf(\bpara|\xbold)$ belongs to CEF (\ref{eq:CEF}), as follows:
\begin{equation}
\pdf(\bpara|\xbold)\propto\pdf(\xbold,\bpara)=\frac{1}{\Normalizing}\exp\left\langle \boldsymbol{g}_{\idim}(\para_{\idim},\xbold),\boldsymbol{g}_{\backslash\idim}(\para_{\backslash\idim},\xbold)\right\rangle .\label{eq:CEF_posterior}
\end{equation}
Since $\Normalizing$ is merely a normalizing constant in (\ref{eq:CEF_posterior}),
we can also replace $\pdf(\xbold,\bpara)$ in CVB algorithm (\ref{eq:CVB_posterior})
by its unnormalized form $q(\xbold,\bpara)\TRIANGLEQ\exp\left\langle \boldsymbol{g}_{\idim},\boldsymbol{g}_{\backslash\idim}\right\rangle $
in (\ref{eq:CEF_posterior}). Since the parameters $\para_{\idim}$
and $\para_{\backslash\idim}$ in (\ref{eq:CEF_posterior}) are separable,
the CVB form (\ref{eq:CEF_approx}) is tractable and conjugate to
the original distribution (\ref{eq:CEF_posterior}). For this reason,
the CEF form (\ref{eq:CEF_posterior}) was also called the conditionally
conjugate model for exponential family \cite{VB:ELBO:Blei:17}, the
conjugate-exponential class \cite{VB:proof:Lagrange:VEM:03} or the
separable-in-parameter family \cite{AQ::Book:06} in mean-field context.

\subsubsection{Mean-field approximations for posterior distribution}

Similar to CVB (\ref{eq:CVB_posterior}), the mean-field algorithms
in section \ref{subsec:Mean-field} can be applied to the posterior
$\pdf(\bpara|\xbold)$, except that the original joint distribution
$\pdf(\bpara)$ in those mean-field algorithms is now replaced by
the joint model $\pdf(\xbold,\bpara)$. By this way, the EM and ICM
algorithms are also able to return a local maximum-a-posteriori (MAP)
estimate of the true marginal $\pdf(\para_{\idim}|\xbold)$ and the
true joint $\pdf(\bpara|\xbold)$, respectively, either directly from
joint model $\pdf(\xbold,\bpara)$ or indirectly from its unnormalized
form $q(\xbold,\bpara)$.

In literature, there are three main approaches for proof of VB approximation
(\ref{eq:(VB)}) when applied to the joint model $\pdf(\xbold,\bpara)$
in (\ref{eq:CVB_posterior}), as briefly summarized below. All VB's
proofs were, however, confined within independent space $\ftilde=\ftilde_{\backslash\idim}\ftilde_{\idim}$
and, hence, did not yield the CVB form (\ref{eq:CVB_posterior}):
\begin{itemize}
\item The first approach (e.g. in \cite{AQ::Book:06,VB:proof:KLD:Bishop:05,VB:proof:KLD:08})
is to expand $\text{\ensuremath{\KL}}_{\ftilde||f}$ directly, i.e.
similar to CVA's proof (\ref{eq:CV_proof}). 
\item The second approach (e.g. in \cite{VB:proof:MichealJordan:08,VB:proof:Jensen:15})
is to start with Jensen's inequality for the so-called energy \cite{VB:FreeEnergy:MacKay:03,VB:FreeEnergy:Yedidia:05}:
$\log Z(\xbold)=\log\EXPECTATION_{\ftilde(\bpara)}\frac{q(\xbold,\bpara)}{\ftilde(\bpara)}\geq\EXPECTATION_{\ftilde(\bpara)}\log\frac{q(\xbold,\bpara)}{\ftilde(\bpara)}$,
which is equivalent to the ELBO's inequality in (\ref{eq:ELBO}),
since the term $Z(\xbold)\TRIANGLEQ\int_{\bpara}q(\xbold,\bpara)d\bpara$
is proportional to $\pdf(\xbold)$, i.e. $\pdf(\xbold)=\int_{\bpara}\pdf(\xbold,\bpara)d\bpara=\frac{1}{\Normalizing}\int_{\bpara}q(\xbold,\bpara)d\bpara=\frac{Z(\xbold)}{\Normalizing}$,
owing to (\ref{eq:CEF_posterior}). Note that, the Jensen's inequality
is merely a consequence of Bregman variance theorem, of which KL divergence
is a special case, as shown in Theorem \ref{thm:Bregman-variance-Jensen}.
\item The third approach (e.g. in \cite{VB:proof:Lagrange:VEM:03,VB:proof:Lagrange:book:15})
is to derive the functional derivative of $\text{\ensuremath{\KL}}_{\ftilde||f}$
via Lagrange multiplier in calculus of variations (hence the name
``variational'' in VB). In this paper, however, the Bregman pythagorean
projection for functional space (\ref{eq:PYTHAGORE}, \ref{eq:BREGMAN_func})
was applied instead and it gave a simpler proof for CVA (\ref{eq:CV_pythagore})
and VB (\ref{eq:(VB)}), since the gradient form of Bregman divergence
in (\ref{eq:3POINTs}) is more concise than traditional functional
derivative. 
\end{itemize}
In practice, since the evidence $\pdf(\xbold)$ is hard to compute,
the ELBO term in (\ref{eq:ELBO}) was originally defined as a feasible
stopping rule for iterative VB algorithm \cite{VB:ELBO:Blei:17}.
The ELBO for CVB in (\ref{eq:ELBO}), computed via conditional form
$\ftilde_{\backslash\idim|\idim}^{[\iVB-1]}$ in (\ref{eq:CVB_posterior}),
can also be used as a stopping rule for CVB algorithm.

\begin{figure*}
\begin{centering}
\includegraphics[width=0.8\linewidth]{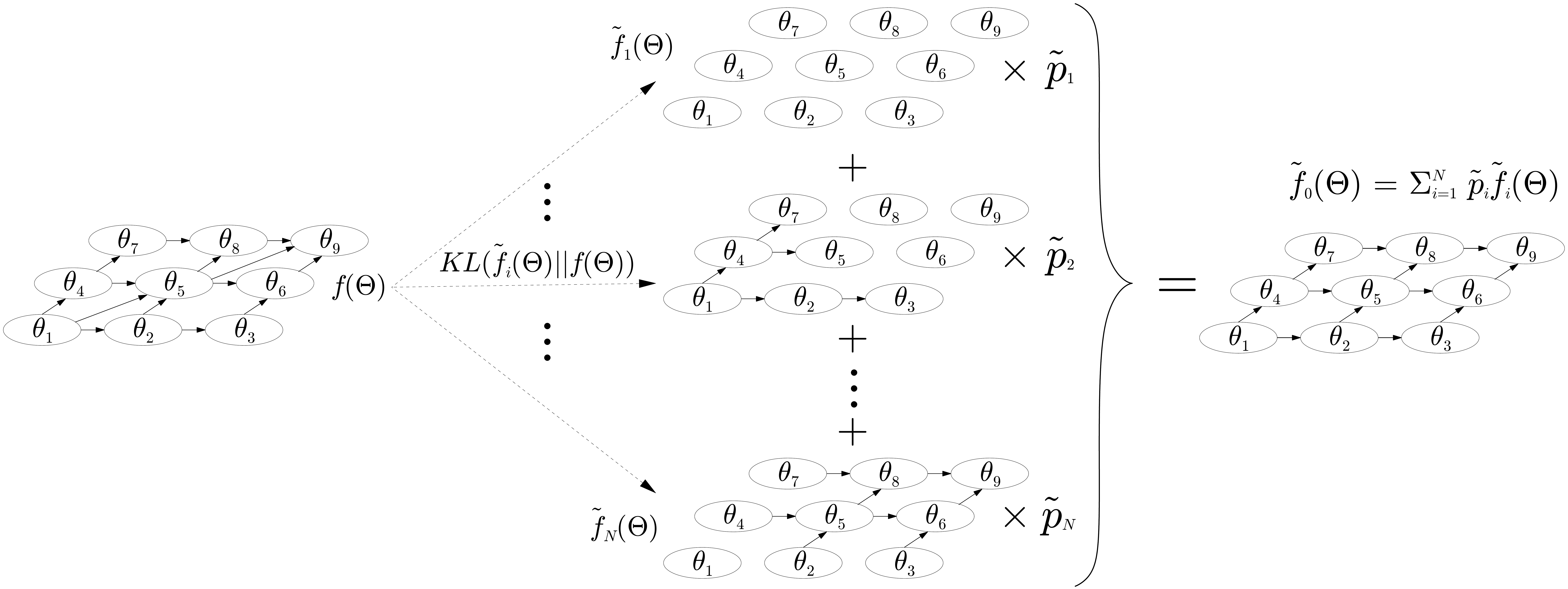}
\par\end{centering}
\caption{\label{fig:networkCVB}Augmented CVB approximation $\protect\ftilde_{0}(\protect\bpara)$
for a complicated joint distribution $\protect\pdf(\protect\bpara)$,
illustrated via directed acyclic graphs (DAG). Each $\protect\ftilde_{\protect\itime}(\protect\bpara)$
is a converged CVB approximation of $\protect\pdf(\protect\bpara)$
with simpler structure. The weight vector $\protect\tbweight\protect\TRIANGLEQ\protect\setv{\protect\tweight}{\protect\ntime}^{\protect\transpose}$,
with $\sum_{\protect\itime=1}^{\protect\ntime}\protect\tweight_{\protect\itime}=1$,
is then calculated via (\ref{eq:optimal_weight}) and yields the optimal
mixture $\protect\ftilde_{0}(\protect\bpara)\protect\TRIANGLEQ\sum_{\protect\itime=1}^{\protect\ntime}\protect\tweight_{\protect\itime}\protect\ftilde_{\protect\itime}(\protect\bpara)$
minimizing the upper bounds (\ref{eq:KL_p}-\ref{eq:KL_i}) of $\protect\KL_{\protect\ftilde_{0}||\protect\pdf}$.
Since $\protect\KL_{\protect\ftilde_{0}||\protect\pdf}$ is convex
over $\protect\ftilde_{0}$, the mixture $\protect\ftilde_{0}$ would
be close to the original $\protect\pdf$, if we can design a set of
$\protect\ftilde_{\protect\itime}$ such that $\protect\pdf$ stays
inside a polytope bounded by vertices $\protect\ftilde_{\protect\itime}$,
as illustrated in Fig. \ref{fig:mixtureKLD}. Hence, a good choice
of $\protect\ftilde_{\protect\itime}$ might be a set of overlapped
sectors of the original network $\protect\pdf$, such that its mixture
would have a similar structure of $\protect\pdf$, as illustrated
in above DAGs.}
\end{figure*}

\section{Hierarchical CVB for Bayesian network \label{sec:Bayesian-network}}

In this section, let us apply the CVB approximation to a joint posterior
$\pdf(\bpara|\xbold)$ of a generic Bayesian network. Since the network
structure of $\pdf(\bpara|\xbold)$ is often complicated in practice,
an intuitive approach is to approximate $\pdf(\bpara|\xbold)$ with
a simpler CEF structure $\ftilde(\bpara|\xbold)$, such that the $\KL_{\ftilde||\pdf}$
can be locally minimized via iterative CVB algorithm. 

Nevertheless, since CVB approximation $\ftilde^{[\iVB]}(\bpara|\xbold)$
in (\ref{eq:CVB_posterior}) cannot change its copula form at any
iteration $\iVB$, a natural approach is to design initially a set
of simple network structures $\ftilde_{\itime}^{[0]}$, $\seti{\itime}{\ntime}$,
and then combine them into a more complex structure with lowest $\KL_{\ftilde^{[\nVB]}||\pdf}$,
or equivalently, highest ELBO (\ref{eq:ELBO}) at convergence $\iVB=\nVB$.
An augmented hierarchy method for merging potential CVB's structures,
as illustrated in Fig. \ref{fig:networkCVB}, will be studied below. 

For simplicity, let us consider the case of joint distribution $\pdf(\bpara)$
first, before applying the augmented approach to joint posterior $\pdf(\bpara|\xbold)$.

\subsection{Augmented CVB for mixture model}

Let us firstly consider a mixture model, which is the simplest structure
of a hierarchical network. The traditional mixture $\pdf(\bpara|\bweight)=\sum_{\itime=1}^{\ntime}\weight_{\itime}\pdf_{\itime}(\bpara)=\sum_{\bLabel}\pdf(\bpara,\bLabel|\bweight)$
and its approximation $\ftilde(\bpara|\tbweight)=\sum_{\itime=1}^{\ntime}\tweight_{\itime}\ftilde_{\itime}(\bpara)=\sum_{\bLabel}\ftilde(\bpara,\bLabel|\tbweight)$
can be written in augmented form via a boolean label vector $\bLabel\TRIANGLEQ\setv{\Label}{\ntime}^{\transpose}\in\UNIT^{\ntime}$,
as follows: 
\begin{align}
\pdf(\bpara,\bLabel|\bweight) & =\pdf(\bpara|\bLabel)\pdf(\bLabel|\bweight)=\prod_{\itime=1}^{\ntime}\pdf_{\itime}^{\Label{}_{\itime}}(\bpara)\weight_{\itime}^{\Label{}_{\itime}},\label{eq:augmented_mixture}\\
\ftilde(\bpara,\bLabel|\tbweight) & =\ftilde(\bpara|\bLabel)\ftilde(\bLabel|\tbweight)=\prod_{\itime=1}^{\ntime}\ftilde_{\itime}^{\Label{}_{\itime}}(\bpara)\tweight_{\itime}^{\Label{}_{\itime}},\nonumber 
\end{align}
where $\bLabel\in\setd{\element}{\ntime}$ and $\element_{\itime}\TRIANGLEQ[0,\ldots,1,\ldots0]^{\transpose}$
is a $\ntime\times1$ element vector with all zero elements except
the unit value at $\itime$-th position, $\forall\seti{\itime}{\ntime}$.
Each $\ftilde_{\itime}$ is then assumed to be the converged CVB approximation
of each original component $\pdf_{\itime}$. 

Ideally, our aim is to pick the weight vector $\tbweight\TRIANGLEQ\setv{\tweight}{\ntime}^{\transpose}$
such that $\KL(\ftilde(\bpara|\tbweight)||\pdf(\bpara|\bweight))$
is minimized. Nevertheless, it is not feasible to directly factorize
the mixture form $\pdf(\bpara|\bweight)$ and $\ftilde(\bpara|\tbweight)$
via non-linear form of KL divergence. Instead, let us minimize the
KL divergence of their augmented forms in (\ref{eq:augmented_mixture}),
as follows:
\begin{equation}
\tbweight^{*}\TRIANGLEQ\argmin_{\tbweight}\KL(\ftilde(\bpara,\bLabel|\tbweight)||\pdf(\bpara,\bLabel|\bweight)),\label{eq:augmented_weight}
\end{equation}
which is also an upper bound of $\KL(\ftilde(\bpara|\tbweight)||\pdf(\bpara|\bweight))$,
as shown in (\ref{eq:KLD_copula}). The solution for (\ref{eq:augmented_weight})
can be found via CVA (\ref{eq:CV_KLDform}), as follows:
\begin{cor}
(CVA for mixture model)\\
Applying CVA (\ref{eq:CV_KLDform}) to (\ref{eq:augmented_weight}),
we can compute the optimal weight $\tbweight^{*}\TRIANGLEQ\setv{\tweight^{*}}{\ntime}^{\transpose}$
minimizing (\ref{eq:augmented_weight}), as follows:
\begin{equation}
\tweight_{\itime}^{*}\propto\frac{\weight_{\itime}}{\exp(\text{\ensuremath{\KL}}_{\ftilde_{\itime}||f_{\itime}})},\ \forall\seti{\itime}{\ntime}.\label{eq:optimal_weight}
\end{equation}
From (\ref{eq:CV_proof}), the minimum value of (\ref{eq:augmented_weight})
is then: 
\begin{equation}
\KL_{\tbweight^{*}}\TRIANGLEQ\sum_{\itime=1}^{\ntime}\tweight_{\itime}^{*}\text{\ensuremath{\KL}}_{\ftilde_{\itime}||f_{\itime}}+\sum_{\itime=1}^{\ntime}\tweight_{\itime}^{*}\log\frac{\weight_{\itime}}{\tweight_{\itime}^{*}}\label{eq:KL_p}
\end{equation}
\end{cor}
\begin{IEEEproof}
From CVA (\ref{eq:CV_KLDform}), the marginal $\ftilde(\bLabel|\tbweight)$
minimizing (\ref{eq:augmented_weight}) is $\ftilde(\bLabel|\tbweight)\propto\pdf(\bLabel|\bweight)/\exp(\text{\ensuremath{\KL}}(\ftilde(\bpara|\bLabel)||\pdf(\bpara|\bLabel))$,
which yields (\ref{eq:optimal_weight}), since $\text{\ensuremath{\KL}}(\ftilde(\bpara|\bLabel)||\pdf(\bpara|\bLabel))=\sum_{\itime=1}^{\ntime}\Label_{\itime}\text{\ensuremath{\KL}}(\ftilde_{\itime}(\bpara)||\pdf_{\itime}(\bpara))$.
\end{IEEEproof}

\subsection{Augmented CVB for Bayesian network\label{subsec:Augmented-CVB}}

Let us now apply the above approach to a generic network $\pdf(\bpara)$.
In (\ref{eq:augmented_mixture}), let us set $\pdf_{\itime}(\bpara)=\pdf(\bpara)$,
$\forall\itime$, together with uniform weight $\bweight=\bar{\bweight}\TRIANGLEQ\setv{\bar{\weight}}{\ntime}^{\transpose}=[\frac{1}{\ntime},\ldots,\frac{1}{\ntime}]^{\transpose}$.
Each $\ftilde_{\itime}$ in (\ref{eq:optimal_weight}) is now a CVB
approximation, with possibly simpler structures, of the same original
network $\pdf(\bpara)$, as illustrated in Fig. \ref{fig:networkCVB}. 

Owing to Bregman's property~4 in Proposition \ref{prop:Bregman-properties},
$\text{\ensuremath{\KL}}_{\ftilde||f}$ is convex over $\ftilde$.
Hence, there exists a linear mixture $\ftilde_{0}(\bpara|\tbweight)=\sum_{\itime=1}^{\ntime}\tweight_{\itime}\ftilde_{\itime}(\bpara)$,
such that: 
\begin{equation}
\KL_{\ftilde_{0}||\pdf}\leq\KL_{\element_{\itime^{*}}}\TRIANGLEQ\min_{\seti{\itime}{\ntime}}\KL_{\ftilde_{\itime}||\pdf}\label{eq:KL_i}
\end{equation}
in which the equality is reached if we set $\tbweight=\element_{\itime^{*}}$,
with $\itime^{*}\TRIANGLEQ\arg\min_{\itime}\KL_{\ftilde_{\itime}||\pdf}$.

Since minimizing $\KL_{\ftilde_{0}||\pdf}$ directly is not feasible,
as explained above, we can firstly minimize $\text{\ensuremath{\KL}}_{\ftilde_{\itime}||f}$
in (\ref{eq:KL_i}) via iterative CVB algorithm for each approximated
structure $\ftilde_{\itime}$. We then compute the optimal weights
$\tbweight^{*}$ in (\ref{eq:augmented_weight}, \ref{eq:optimal_weight})
for the minimum upper bound $\KL_{\tbweight^{*}}$ of $\KL_{\ftilde_{0}||\pdf}$.
Note that $\KL_{\tbweight^{*}}$in (\ref{eq:KL_p}) and $\KL_{\element_{\itime^{*}}}$
in (\ref{eq:KL_i}) are two different upper bounds of $\KL_{\ftilde_{0}||\pdf}$
and may not yield the global minimum solution for $\KL_{\ftilde_{0}||\pdf}$
in general. The choice $\tbweight=\element_{\itime^{*}}$ might yield
lower $\KL_{\ftilde_{0}||\pdf}$ than $\tbweight=\tbweight^{*}$,
even when we have $\KL_{\element_{\itime^{*}}}>\KL_{\tbweight^{*}}$. 

Although we can only find the minimum upper bound solution for the
mixture $\ftilde_{0}$ in this paper, the key advantage of the mixture
form is that the moments of $\ftilde_{0}$ are simply a mixture of
moments of $\ftilde_{\itime}$, i.e.: 
\begin{equation}
\widehat{\bpara}_{0}=\EXPECTATION_{\ftilde_{0}}(\bpara)=\sum_{\itime=1}^{\ntime}\tweight_{\itime}\EXPECTATION_{\ftilde_{\itime}}(\bpara)=\sum_{\itime=1}^{\ntime}\tweight_{\itime}\widehat{\bpara}_{\itime}.\label{eq:mixture_moments}
\end{equation}
By this way, the true moments $\widehat{\bpara}$ of complicated network
$\pdf(\bpara)$ can be approximated by a mixture of moments $\widehat{\bpara}_{\itime}$
of simpler CVB's network structure $\ftilde_{\itime}(\bpara)$. 

Another advantage of mixture form is that the optimal weight vector
$\tilde{\bweight}$ can be evaluated tractably, without the need of
normalizing constant of $\pdf(\bpara|\xbold)$ in Bayesian context.
Indeed, for a posterior Bayesian network $\pdf(\bpara|\xbold)$, we
can simply replace the value $\text{\ensuremath{\KL}}_{\ftilde_{\itime}||f}$
in (\ref{eq:optimal_weight}-\ref{eq:KL_i}) by ELBO's value in (\ref{eq:ELBO}),
since the evidence $f(\xbold)$ is a constant. 

\subsection{Hierarchical CVB approximation }

In principle, if we keep augmenting the above CVB's augmented mixture,
it is possible to establish an $\istate$-order hierarchical CVB approximation
$\ftilde^{\{\istate\}}(\bpara)$ for a complicated network $\pdf(\bpara)$,
$\forall\istate\in\{0,1,\ldots,\nstate\}$. For example, each zero-order
mixture $\ftilde_{\itime}^{\{0\}}(\bpara|\tbweight_{\itime}^{*})=\sum_{\istate=1}^{\nstate}\tweight_{\itime,\istate}^{*}\ftilde_{\itime,\istate}(\bpara)=\sum_{\bLabel_{\itime}}\ftilde(\bpara,\bLabel_{\itime}|\tbweight_{\itime}^{*})$,
$\forall\seti{\itime}{\ntime}$, can be considered as a component
of the first-order mixture $\ftilde_{0}^{\{1\}}(\bpara|\tbqweight,\tbWeight^{*})=\sum_{\itime=1}^{\ntime}\tqweight_{\itime}\ftilde_{\itime}^{\{0\}}(\bpara|\tbweight_{\itime}^{*})$,
where $\tbWeight^{*}\TRIANGLEQ\setv{\tbweight^{*}}{\ntime}$ and $\tbqweight\TRIANGLEQ\setv{\tqweight}{\ntime}^{\transpose}$. 

If $\ftilde_{\itime,\istate}(\bpara)$ are all tractable CVB's approximations
with simpler and possibly overlapped sectors of the network $\pdf(\bpara)$,
the optimal vectors $\tbweight_{\itime}^{*}$ can be evaluated feasibly
via $\KL_{\ftilde_{\itime,\istate}||\pdf}$ in (\ref{eq:optimal_weight}).
Nonetheless, the computation of the optimal vector $\tbqweight^{*}$
via $\KL_{\ftilde_{\itime}^{\{0\}}||\pdf}$ in (\ref{eq:optimal_weight})
might be intractable in practice, because $\KL_{\ftilde_{\itime}^{\{0\}}||\pdf}$
is a KL divergence of a mixture of distributions and, hence, it is
difficult to evaluate $\KL_{\ftilde_{\itime}^{\{0\}}||\pdf}$ directly
in closed form. 

An intuitive solution for this issue might be to apply CVB again to
the augmented form $\KL(\ftilde(\bpara,\bLabel_{\itime}|\tbweight_{\itime})||\pdf(\bpara,\bLabel_{\itime}|\bar{\bweight})),$
similar to (\ref{eq:augmented_weight}). By this way, we could avoid
the mixture form $\ftilde_{\itime}^{\{0\}}(\bpara|\tbweight_{\itime})=\sum_{\bLabel_{\itime}}\ftilde(\bpara,\bLabel_{\itime}|\tbweight_{\itime}^{*})$
and directly derive a CVB's closed form for $\ftilde_{\itime}^{\{0\}}(\bpara|\tbweight_{\itime})$.
This hierarchical CVB approach is, however, outside the scope of this
paper and will be left for future work.
\begin{rem}
In literature, the idea of augmented hierarchy was mentioned briefly
in \cite{CVB:Blei:hierarchical:16,CVB:Blei:hierarchical:17}, in which
the potential approximations $\ftilde_{\itime}$ are confined to a
set of mean-field approximations and the prior $\ftilde(\bLabel|\tbweight)$
is extended from a mixture to a latent Markovian model. Nevertheless,
the ELBO minimization in \cite{CVB:Blei:hierarchical:16,CVB:Blei:hierarchical:17}
was implemented via stochastic-gradient decent methods and did not
yield an explicit form for the mixture's weights in (\ref{eq:optimal_weight}). 
\end{rem}

\section{Case study \label{sec:Case-study}}

In this section, let us illustrate the superior performance of CVB
to mean-field approximations for two canonical scenarios in practice:
the bivariate Gaussian distribution and Gaussian mixture clustering.
These two cases belong to CEF class (\ref{eq:CEF}) and, hence, their
CVB approximation is tractable, as shown below. 

\begin{figure*}
\centering{}\includegraphics[width=0.33\linewidth]{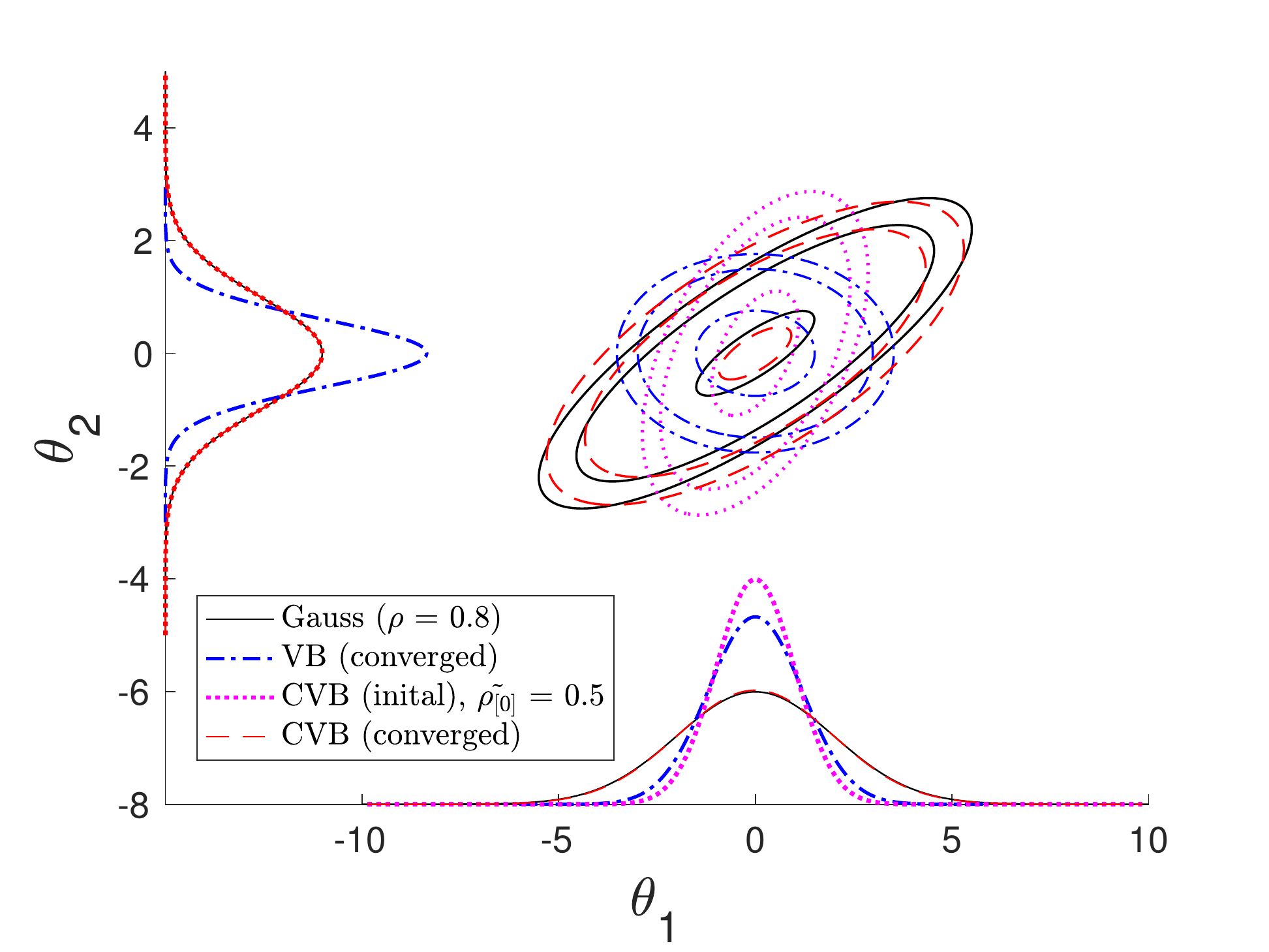}\includegraphics[width=0.33\linewidth]{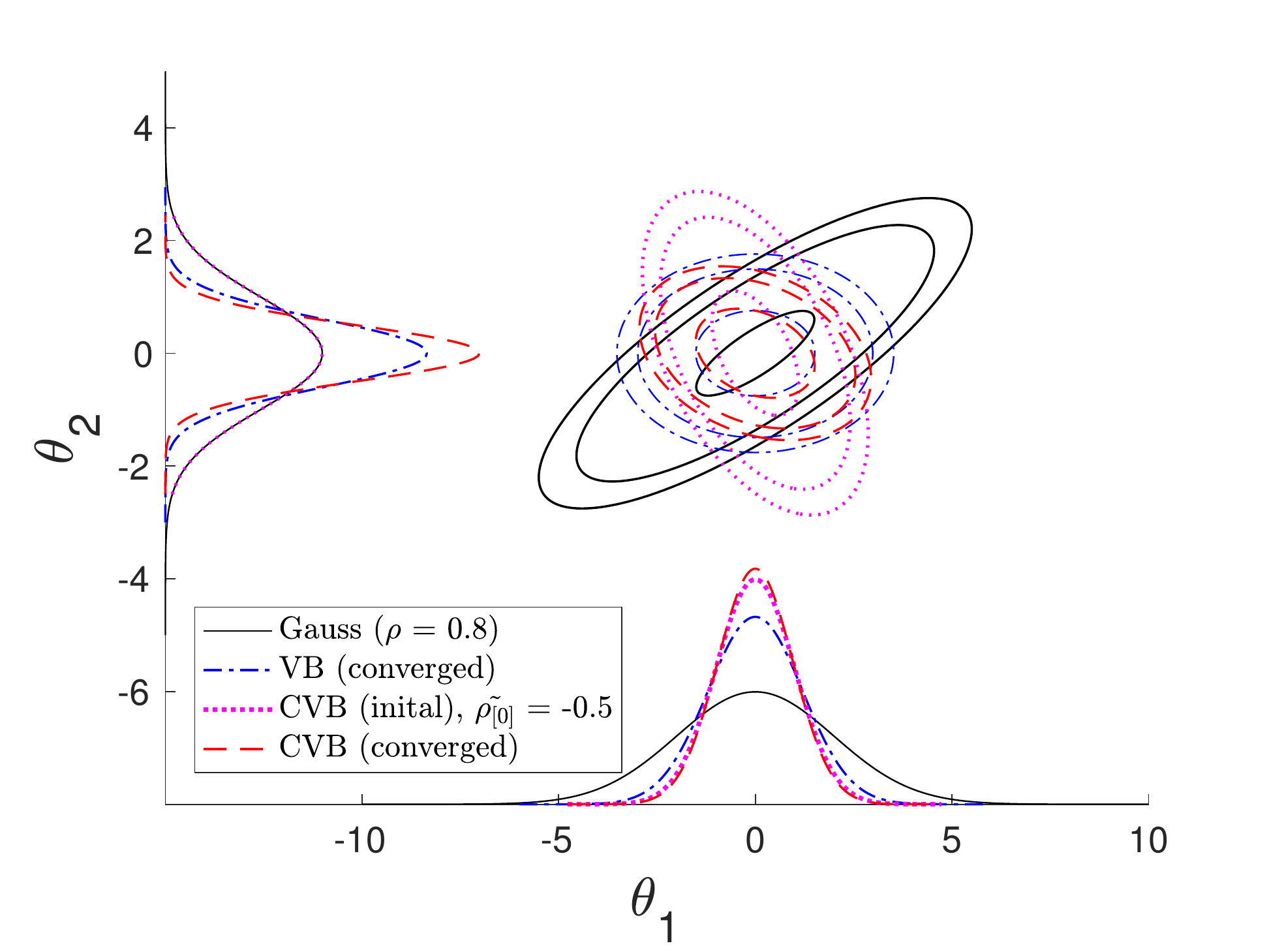}\includegraphics[width=0.33\linewidth]{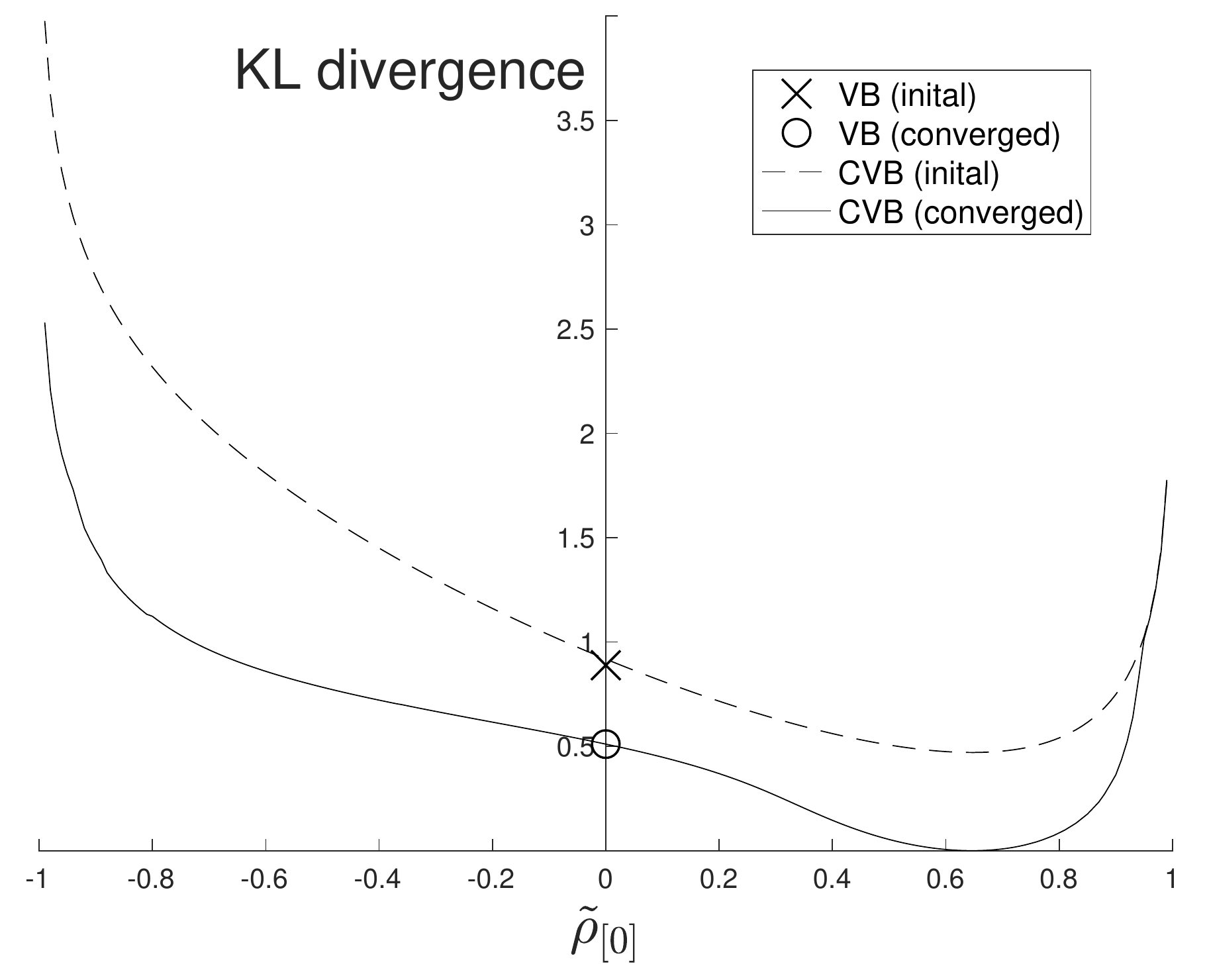}\caption{\label{fig:Gauss}CVB and VB approximations $\protect\ftilde_{\protect\bpara}$
for a zero-mean bivariate Gaussian distribution $\protect\pdf_{\protect\bpara}$,
with true variances $\protect\std_{1}^{2}=4$, $\protect\std_{2}^{2}=1$
and correlation coefficient $\protect\correff=0.8$. The initial guess
values for CVB and VB are $\protect\tstd_{1}^{[0]}=\protect\tstd_{2}^{[0]}=1$,
together with various $\protect\tcorreff_{[0]}\in(-1,1)$ for CVB.
The cases $\protect\tcorreff_{[0]}=0.5$ and $\protect\tcorreff_{[0]}=-0.5$
are shown on the left and middle panel, respectively. The marginal
distributions, which are also Gaussian, are plotted on two axes in
these two panels. The lower KL divergence $\protect\KL(\protect\ftilde_{\protect\bpara}||\protect\pdf_{\protect\bpara})$
on the right panel, the better approximation, as illustrated in Fig.
\ref{fig:fCVA}, \ref{fig:fEM}. The CVB will be exact, i.e. $\protect\KL(\protect\ftilde_{\protect\bpara}||\protect\pdf_{\protect\bpara})\approx0$
at convergence, if the initial guess values $\protect\tcorreff_{[0]}$
are in range $\protect\tcorreff_{[0]}\in[0.6,0.7]$, which is close
to the true value $\protect\correff=0.8$. If $\protect\tcorreff_{[0]}=0$,
the CVB is equivalent to VB approximation in independent class. The
number $\protect\nVB$ of iterations until convergence for VB and
CVB are, respectively, 8 and $11.1\pm5.2$, averaged over all cases
of $\protect\tcorreff_{[0]}\in(-1,1)$ for CVB. Only one marginal
is updated per iteration.}
\end{figure*}

\subsection{Bivariate Gaussian distribution}

In this subsection, let us approximate a bivariate Gaussian distribution
$\pdf(\bpara)=\calN_{\bpara}(0,\covmatrix)$ with zero mean and covariance
matrix $\covmatrix\TRIANGLEQ\left[\begin{array}{cc}
\std_{1}^{2} & \correff\std_{1}\std_{2}\\
\correff\std_{1}\std_{2} & \std_{2}^{2}
\end{array}\right]$. The purpose is then to illustrate the performance of CVB and VB
approximations for $\pdf(\bpara)$ with different values of correlation
coefficient $\correff\in[-1,1]$. 

For simple notation, let us denote the marginal and conditional distributions
of $\pdf(\bpara)$ by $\pdf_{1}=\calN_{\para_{1}}(0,\std_{1})$ and
$\pdf_{2|1}=\calN_{\para_{2}}(\rcorr_{2|1}\para_{1},\std_{2|1})$,
respectively, in which $\rcorr_{2|1}\TRIANGLEQ\correff\frac{\std_{2}}{\std_{1}}$
and $\std_{2|1}\TRIANGLEQ\std_{2}\sqrt{1-\correff^{2}}$.

\subsubsection{CVB approximation}

Since Gaussian distribution belongs to CEF class (\ref{eq:CEF}),
the CVB form $\ftilde_{\CVB}^{[1]}=\ftilde_{2|1}^{[0]}\ftilde_{1}^{[1]}=\ftilde_{1|2}^{[1]}\ftilde_{2}^{[1]}$
in (\ref{eq:CVB_kform}) is also Gaussian, as shown in (\ref{eq:CEF_approx}).
Then, given initial values $\trcorr_{2|1}^{[0]}\TRIANGLEQ\tcorreff_{[0]}\frac{\tstd_{2}^{[0]}}{\tstd_{1}^{[0]}}$
and $\tstd_{2|1}^{[0]}\TRIANGLEQ\tstd_{2}^{[0]}\sqrt{1-\tcorreff_{[0]}^{2}}$,
we have $\ftilde_{2|1}^{[0]}=\calN_{\para_{2}}(\trcorr_{2|1}^{[0]}\para_{1},\tstd_{2|1}^{[0]})$.
At iteration $\iVB=1$, the CVA form (\ref{eq:CV_KLDform}) yields:

\begin{align*}
\ftilde_{1}^{[1]} & =\frac{1}{\fzeta_{1}^{[1]}}\frac{\pdf_{1}}{\exp(\KL_{\ftilde_{2|1}^{[0]}||\pdf_{2|1}})}\\
 & =\frac{1}{\fzeta_{1}^{[1]}}\frac{\frac{1}{\std_{1}\sqrt{2\pi}}\exp-\frac{\para_{1}^{2}}{2\std_{1}^{2}}}{\frac{\std_{2|1}}{\tstd_{2|1}^{[0]}}\exp\frac{1}{2}\left[\frac{\left(\trcorr_{2|1}^{[0]}-\rcorr_{2|1}\right)^{2}\para_{1}^{2}+(\tstd_{2|1}^{[0]})^{2}}{\std_{2|1}^{2}}-1\right]}\\
 & =\calN_{\para_{1}}(0,\tstd_{1}^{[1]}),
\end{align*}
in which $\KL_{\ftilde_{2|1}^{[0]}||\pdf_{2|1}}$ is KL divergence
between Gaussian distributions and:

\begin{align}
\tstd_{1}^{[1]} & =\frac{1}{\sqrt{\frac{1}{\std_{1}^{2}}+\frac{\left(\trcorr_{2|1}^{[0]}-\rcorr_{2|1}\right)^{2}}{\std_{2}^{2}(1-\correff^{2})}}},\ \fzeta_{1}^{[1]}=\frac{\tstd_{1}^{[1]}}{\std_{1}}\frac{\tstd_{2|1}^{[0]}}{\std_{2|1}}\exp\frac{\std_{2|1}^{2}-(\tstd_{2|1}^{[0]})^{2}}{2\std_{2|1}^{2}}.\label{eq:biGauss_sigma1}
\end{align}
Then, in order to derive the reverse form $\ftilde_{1|2}^{[1]}\ftilde_{2}^{[1]}=\ftilde_{2|1}^{[0]}\ftilde_{1}^{[1]}$,
let us firstly note that $\trcorr_{2|1}^{[0]}=\trcorr_{2|1}^{[1]}$
and $\tstd_{2|1}^{[0]}=\tstd_{2|1}^{[1]}$ , since the conditional
form $\ftilde_{2|1}$ of two distributions $\ftilde_{2|1}^{[0]}\ftilde_{1}^{[0]}$
and $\ftilde_{2|1}^{[0]}\ftilde_{1}^{[1]}=\ftilde_{1|2}^{[1]}\ftilde_{2}^{[1]}$
are still the same. Then, the updated parameters are: 
\[
\begin{cases}
\trcorr_{2|1}^{[0]} & =\trcorr_{2|1}^{[1]}\\
\tstd_{2|1}^{[0]} & =\tstd_{2|1}^{[1]}
\end{cases}\Leftrightarrow\begin{cases}
\tcorreff_{[0]}\frac{\tstd_{2}^{[0]}}{\tstd_{1}^{[0]}} & =\tcorreff_{[1]}\frac{\tstd_{2}^{[1]}}{\tstd_{1}^{[1]}}\\
\tstd_{2}^{[0]}\sqrt{1-\tcorreff_{[0]}^{2})} & =\tstd_{2}^{[1]}\sqrt{1-\tcorreff_{[1]}^{2}}
\end{cases}
\]
which, by solving for $\tcorreff_{[1]}$ and $\tstd_{2}^{[1]}$, yields:

\begin{align*}
\tcorreff_{[1]}^{2} & =\frac{\tcorreff_{[0]}^{2}}{\tcorreff_{[0]}^{2}+\left(\frac{\tstd_{1}^{[0]}}{\tstd_{1}^{[1]}}\right)^{2}(1-\tcorreff_{[0]}^{2})},\\
\tstd_{2}^{[1]} & =\tstd_{2}^{[0]}\sqrt{\tcorreff_{[0]}^{2}\left(\frac{\tstd_{1}^{[1]}}{\tstd_{1}^{[0]}}\right)^{2}+(1-\tcorreff_{[0]}^{2}))}.
\end{align*}
Hence, we have $\trcorr_{1|2}^{[1]}=\tcorreff_{[1]}\frac{\tstd_{1}^{[1]}}{\tstd_{2}^{[1]}}$
and $\tstd_{1|2}^{[1]}=\tstd_{1}^{[1]}\sqrt{1-\tcorreff_{[1]}^{2}}$,
which yield the updated forms $\ftilde_{2}^{[1]}=\calN_{\para_{1}}(0,\tstd_{2}^{[1]})$
and $\ftilde_{1|2}^{[1]}=\calN_{\para_{1}}(\trcorr_{1|2}^{[1]}\para_{2},\tstd_{1|2}^{[1]})$.
Reversing the role of $\para_{1}$ with $\para_{2}$ and repeating
the above steps for iteration $\iVB>1$, we will achieve the CVB approximation
at convergence $\iVB=\nVB$, with $\KL_{\ftilde_{\CVB}^{[\iVB]}||\pdf}=\log\frac{1}{\fzeta_{1}^{[\iVB]}}$.

The CVB approximation will be exact if its conditional mean and variance
are exact, i.e. $\trcorr_{2|1}^{[\nVB]}=\rcorr_{2|1}$ and $\tstd_{2|1}^{[\nVB]}=\std_{2|1}$,
since we have $\KL_{\ftilde_{\CVB}^{[\nVB]}||\pdf}=\log\frac{1}{\fzeta_{1}^{[\nVB]}}=0$
in this case, as shown in (\ref{eq:biGauss_sigma1}).

\subsubsection{VB approximation}

Since VB is a special case of CVB in independence space, we can simply
set $\correff=0$ in above CVB algorithm and the result will be VB
approximation. 

\subsubsection{Simulation's results}

The CVB and VB approximations for the case of $\pdf(\bpara)=\calN_{\bpara}(0,\covmatrix)$
are illustrated in Fig.~\ref{fig:Gauss}. Since $\KL_{\ftilde_{\bpara}^{[\iVB]}||\pdf_{\bpara}}$
monotonically decreases with iteration $\iVB$, the right panel shows
the value of KL divergence at initialization $\iVB=0$ and at convergence
$\iVB=\nVB$, with $0\leq\KL(\ftilde_{\bpara}^{[\nVB-1]}||\pdf_{\bpara})-\KL(\ftilde_{\bpara}^{[\nVB]}||\pdf_{\bpara})\leq0.01$.
We can see that VB is a mean-field approximation and, hence, cannot
accurately approximate a correlated Gaussian distribution. In contrast,
the CVB belongs to a conditional copula class and, hence, can yield
higher accuracy. In this sense, CVB can potentially return a globally
optimal approximation for a correlated distribution, while VB can
only return a locally optimal approximation.

Nevertheless, since the iterative CVB cannot escape its initialized
copula class, its accuracy depends heavily on initialization. A solution
for this issue is to initialize CVB with some information of original
distribution. For example, merely setting the initial sign of $\tcorreff_{[0]}$
equal to the sign of true value $\correff$ would gain tremendously
higher accuracy for CVB at convergence, as shown in the left and middle
panel of Fig.~\ref{fig:Gauss}. 

Another solution for CVB's initialization issue is to generate a lot
of potential structures initially and take the average of the results
at convergence. This CVB's mixture-scheme will be illustrated in the
next subsection.

\subsection{Gaussian mixture clustering\label{subsec:Gauss-cluster}}

\begin{figure}
\begin{centering}
\includegraphics[width=0.4\linewidth]{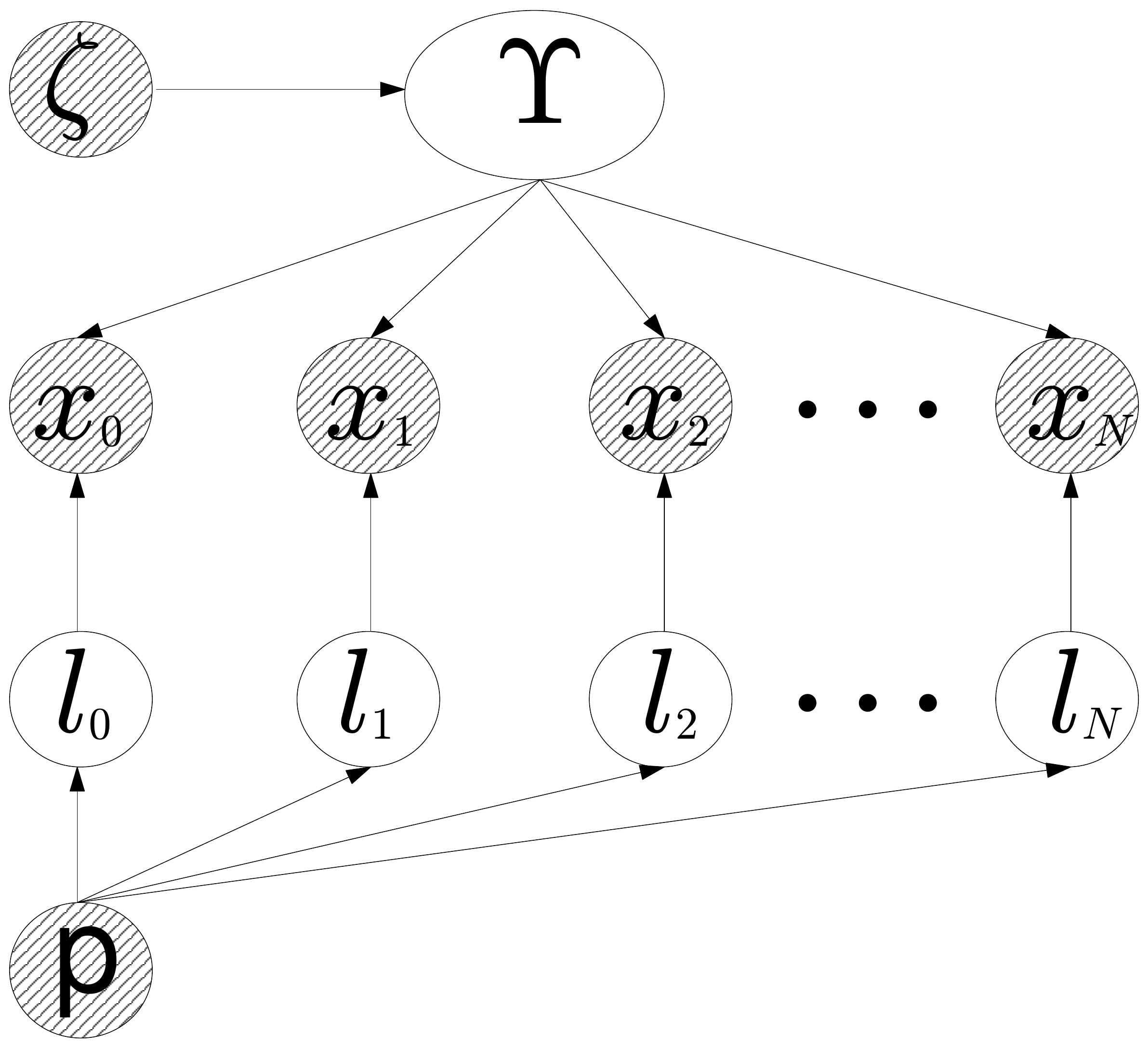} $\ $
\includegraphics[width=0.45\linewidth]{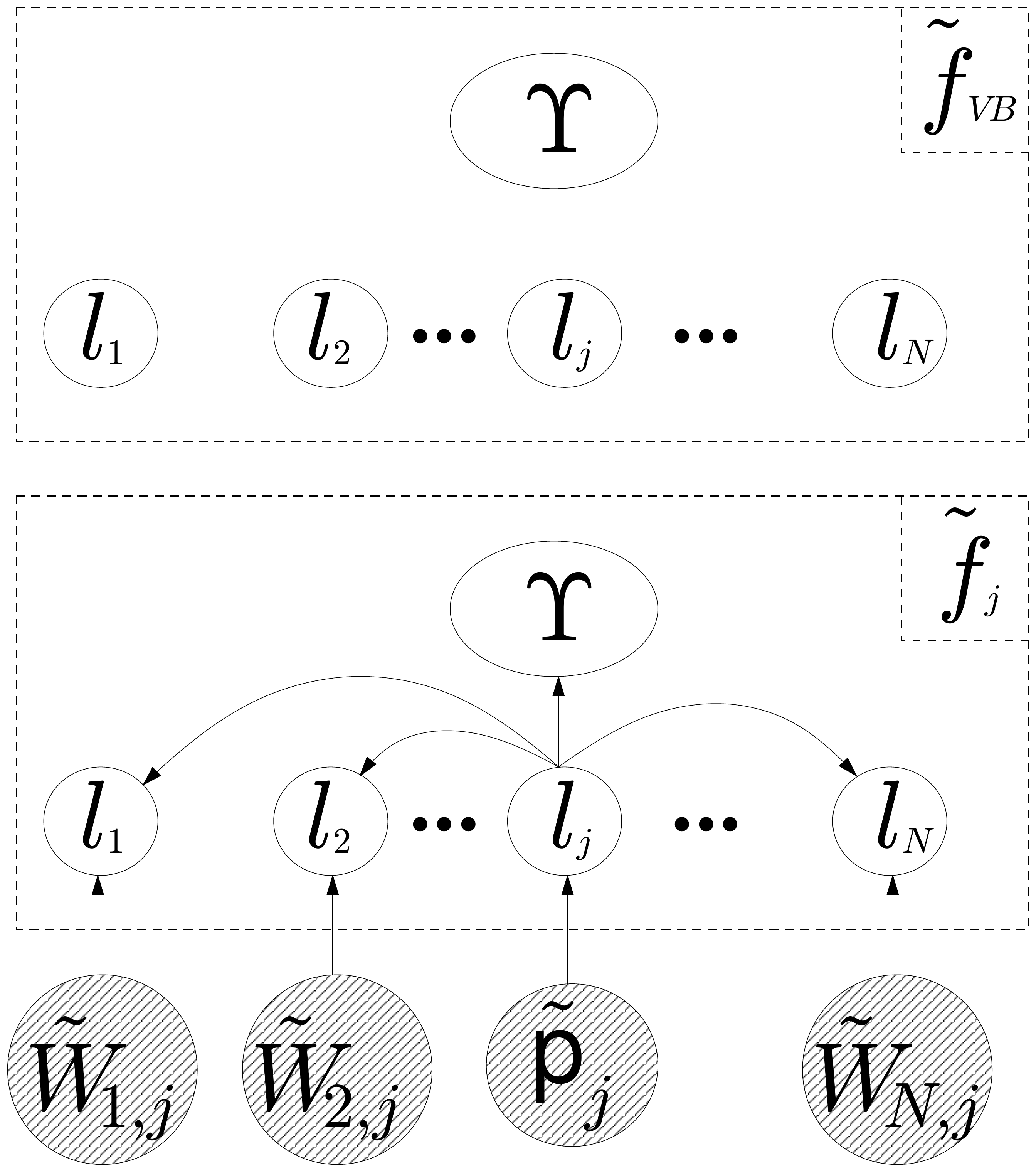}
\par\end{centering}
\caption{\label{fig:cluster}Directed acyclic graphs (DAG) for Gaussian clustering
model with uniform hyper-parameters $\protect\fzeta$, $\protect\bweight$
(left), the VB approximation with independent structure (upper right)
and the CVB approximations (lower right). All variables in shaded
nodes are known, while the others are random variables. Each $\protect\ftilde_{\protect\ipick}$
is a ternary structure centered around $\protect\Label_{\protect\ipick}$,
$\protect\seti{\protect\ipick}{\protect\ntime}$. The augmented CVB
approximation $\protect\ftilde_{0}=\sum_{\protect\ipick=1}^{\protect\ntime}\protect\qweight_{\protect\ipick}^{*}\protect\ftilde_{\protect\ipick}$
is designed in (\ref{eq:CVB_final}) and illustrated in Fig. \ref{fig:networkCVB}.}
\end{figure}
In this subsection, let us illustrate the performance of CVB for a
simple bivariate Gaussian mixture model. For this purpose, let us
consider clusters of bivariate observation data $\Data\TRIANGLEQ\setv{\xbold}{\ntime}\in\REAL^{2\times\ntime}$,
such that $\xbold_{\itime}=[\data_{1,\itime},\data_{2,\itime}]^{\transpose}\in\REAL^{2}$
at each time $\seti{\itime}{\ntime}$ randomly belongs to one of $\ndim$
bivariate independent Gaussian clusters $\calN_{\xbold_{\itime}}(\bmean,\IDEN_{2})$
with equal probability $\bweight\TRIANGLEQ\setv{\weight}{\ndim}^{\transpose}$,
i.e. $\weight_{\idim}=\frac{1}{\ndim}$, $\seti{\forall\idim}{\ndim}$,
at unknown means $\Mean\TRIANGLEQ\setv{\bmean}{\ndim}\in\REAL^{2\times\ndim}$.
$\IDEN_{2}$ denotes the $2\times2$ identity covariance matrix.

Let us also define a temporal matrix $\LABEL\TRIANGLEQ\setv{\bLabel}{\ntime}\in\UNIT^{\ndim\times\ntime}$
of categorical vector labels $\bLabel_{\itime}=\sete{\Label}{\ndim}{\itime}^{\transpose}\in\setd{\element}{\ndim}$,
where $\element_{\idim}=[0,\ldots,1,\ldots0]^{\transpose}\in\UNIT^{\ndim}$
denotes the boolean vector with $\idim$-th non-zero element. By this
way, we set $\bLabel_{\itime}=\element_{\idim}$ if $\xbold_{\itime}$
belongs to $\idim$-th cluster. Then, by probability chain rule, our
model is a Gaussian mixture $\pdf(\Data,\PARA)=\pdf(\Data|\PARA)\pdf(\PARA)$,
in which $\PARA\TRIANGLEQ[\Mean,\LABEL]$ are unknown parameters,
as follows:
\begin{align}
\pdf(\Data,\Mean,\LABEL) & =\pdf(\Data|\Mean,\LABEL)\pdf(\Mean,\LABEL)\nonumber \\
 & =\pdf(\Mean|\LABEL,\Data)\pdf(\LABEL,\Data)\label{eq:cluster_Joint}\\
 & =\pdf(\LABEL|\Mean,\Data)\pdf(\Mean,\Data).\nonumber 
\end{align}

In the first line of (\ref{eq:cluster_Joint}), the distributions
are: 
\begin{align}
\pdf(\Data|\Mean,\LABEL) & =\prod_{\itime=1}^{\ntime}\pdf(\xbold_{\itime}|\Mean,\bLabel_{\itime})=\prod_{\itime=1}^{\ntime}\prod_{\idim=1}^{\ndim}\calN_{\xbold_{\itime}}^{\Label_{\idim,\itime}}(\bmean_{\idim},\IDEN_{2}),\nonumber \\
\pdf(\Mean,\LABEL) & =\pdf(\Mean)\pdf(\LABEL)=\frac{1}{\fzeta\ndim^{\ntime}},\label{eq:cluster_MODEL}
\end{align}
in which the prior $\pdf(\LABEL)=\prod_{\itime=1}^{\ntime}\bLabel_{\itime}^{\transpose}\bweight=\frac{1}{\ndim^{\ntime}}$
is uniform by default and $\pdf(\Mean)$ is the non-informative prior
over $\REAL^{2\times\ndim}$, i.e. $\pdf(\Mean)=\frac{1}{\fzeta}$,
with constant $\fzeta$ being set as high as possible (ideally $\fzeta\rightarrow\infty$). 

The second line of (\ref{eq:cluster_Joint}) can be written as follows: 

\begin{align}
\pdf(\Mean|\LABEL,\Data) & =\prod_{\idim=1}^{\ndim}\calN_{\bmean_{\idim}}\left(\bcmean_{\idim}(\LABEL),\overline{\std}_{\idim}(\LABEL)\IDEN_{2}\right),\label{eq:cluster_posteriors}\\
\pdf(\LABEL,\Data) & =\frac{1}{\fzeta\ndim^{\ntime}}\prod_{\idim=1}^{\ndim}\uweight_{\idim}(\LABEL),\nonumber 
\end{align}
with $\bcmean_{\idim}(\LABEL)$ and $\overline{\std}_{\idim}(\LABEL)$
denoting posterior mean and standard deviation of $\bmean_{\idim}$,
respectively, and $\uweight_{\idim}(\LABEL)$ denoting the updated
form of weight's probability $\weight_{\idim}=\frac{1}{\ndim}$, as
follows:
\begin{align}
\bcmean_{\idim}(\LABEL) & \TRIANGLEQ\frac{\sum_{\itime=1}^{\ntime}\Label_{\idim,\itime}\xbold_{\itime}}{\sum_{\itime=1}^{\ntime}\Label_{\idim,\itime}},\ \overline{\std}_{\idim}(\LABEL)\TRIANGLEQ\frac{1}{\sqrt{\sum_{\itime=1}^{\ntime}\Label_{\idim,\itime}}},\label{eq:cluster_mean}\\
\uweight_{\idim}(\LABEL) & \TRIANGLEQ2\pi\overline{\std}_{\idim}^{2}(\LABEL)\prod_{\itime=1}^{\ntime}\calN_{\xbold_{\itime}}^{\Label_{\idim,\itime}}(\bcmean_{\idim}(\LABEL),\IDEN_{2}),\nonumber 
\end{align}
Note that, the first form (\ref{eq:cluster_MODEL}) is equivalent
to the second form (\ref{eq:cluster_posteriors}-\ref{eq:cluster_mean})
since we have: 
\begin{align}
\frac{\sum_{\itime=1}^{\ntime}\Label_{\idim,\itime}||\xbold_{\itime}-\bmean_{\idim}||^{2}}{\sum_{\itime=1}^{\ntime}\Label_{\idim,\itime}} & =\frac{\sum_{\itime=1}^{\ntime}\Label_{\idim,\itime}||\xbold_{\itime}-\bcmean_{\idim}(\LABEL)||^{2}}{\sum_{\itime=1}^{\ntime}\Label_{\idim,\itime}}\nonumber \\
 & +||\bmean_{\idim}-\bcmean_{\idim}(\LABEL)||^{2},\label{eq:cluster_Bregmann}
\end{align}
owing to Bregman variance theorem in (\ref{eq:Bregman_var}), (\ref{eq:Bregman_mixture}). 

Similarly, the third line in (\ref{eq:cluster_Joint}) can be derived
from (\ref{eq:cluster_MODEL}), as follows: 
\begin{align}
\pdf(\Data,\Mean)=\sum_{\LABEL} & \pdf(\Data,\Mean,\LABEL)=\frac{\prod_{\itime=1}^{\ntime}\sum_{\idim=1}^{\ndim}\calN_{\xbold_{\itime}}(\bmean_{\idim},\IDEN_{2})}{\fzeta\ndim^{\ntime}},\nonumber \\
\pdf(\LABEL|\Mean,\Data)= & \frac{\pdf(\Data,\Mean,\LABEL)}{\pdf(\Mean,\Data)}=\prod_{\itime=1}^{\ntime}\underset{\pdf(\bLabel_{\itime}|\Mean,\xbold_{\itime})}{\underbrace{\frac{\prod_{\idim=1}^{\ndim}\calN_{\xbold_{\itime}}^{\Label_{\idim,\itime}}(\bmean_{\idim},\IDEN_{2})}{\sum_{\idim=1}^{\ndim}\calN_{\xbold_{\itime}}(\bmean_{\idim},\IDEN_{2})}}}.\label{eq:cluster_reverse}
\end{align}
Note that, the model without labels $\pdf(\Data,\Mean)=\pdf(\Data|\Mean)f(\Mean)$
in (\ref{eq:cluster_reverse}) is a mixture of $\ndim^{\ntime}$ Gaussian
components with unknown means $\Mean$, since we have augmented the
model $\pdf(\Data|\Mean)$ with label's form $\pdf(\Data|\Mean,\LABEL)$
above. The posterior form $\pdf(\Mean|\Data)\propto\pdf(\Data,\Mean)=\sum_{\LABEL}\pdf(\Data,\Mean,\LABEL)$
in this case is intractable, since its normalization's complexity
$\mathcal{O}(\ndim^{\ntime})$ grows exponentially with number of
data $\ntime$, hence the curse of dimensionality. 

\subsubsection{ICM and k-means algorithms\label{subsec:ICM-and-k-means}}

From (\ref{eq:cluster_posteriors}-\ref{eq:cluster_mean}), we can
see that the conditional mean $\bcmean_{\idim}(\LABEL)$ is actually
the $\idim$-th clustering sample's mean of $\bmean_{\idim}$, given
all possible boolean values of $\Label_{\idim,\itime}\in\UNIT=\{0,1\}$
over time $\seti{\itime}{\ntime}.$ The probability of categorical
label $\pdf(\LABEL|\Data)\propto\pdf(\LABEL,\Data)$ in (\ref{eq:cluster_posteriors})
is, in turn, calculated as the distance of all observation $\xbold_{\itime}$
to sample's mean $\bcmean_{\idim}$ of each cluster $\seti{\idim}{\ndim}$
via $\uweight_{\idim}(\LABEL)$ in (\ref{eq:cluster_mean}). Nevertheless,
since the weights $\uweight_{\idim}(\LABEL)$ in (\ref{eq:cluster_posteriors}-\ref{eq:cluster_mean})
are not factorable over $\LABEL$, the posterior probability $\pdf(\LABEL|\Data)$
needs to be computed brute-forcedly over all $\ndim^{\ntime}$ possible
values of label matrix $\LABEL$ as a whole and, hence, yields the
curse of dimensionality. 

A popular solution for this case is the k-means algorithm, which is
merely an application of iteratively conditional mode (ICM) algorithm
(\ref{eq:ICM}) to above clustering mixture (\ref{eq:cluster_posteriors}),
(\ref{eq:cluster_reverse}), as follows:
\begin{align}
\hMean^{[\iVB]} & =\argmax_{\Mean}\pdf(\Mean|\hLABEL^{[\iVB-1]},\Data),\label{eq:cluster_ICM}\\
\hLABEL^{[\iVB]} & =\argmax_{\LABEL}\pdf(\LABEL|\hbmean^{[\iVB]},\Data).\nonumber 
\end{align}
where $\hMean^{[\iVB]}\TRIANGLEQ\setu{\hbmean}{\ndim}{[\iVB]}$ and
$\hLABEL^{[\iVB]}\TRIANGLEQ\setu{\hbLabel}{\ntime}{[\iVB]}$. Since
the mode of Gaussian distribution is also its mean value, let us substitute
(\ref{eq:cluster_ICM}) to $\pdf(\Mean|\LABEL,\Data)$ in (\ref{eq:cluster_posteriors}-\ref{eq:cluster_mean})
and $\pdf(\LABEL|\Mean,\Data)$ in (\ref{eq:cluster_reverse}), as
follows: 
\begin{align}
\hbmean_{\idim}^{[\iVB]} & =\bcmean_{\idim}(\hLABEL^{[\iVB-1]})=\frac{\sum_{\itime=1}^{\ntime}\hLabel_{\idim,\itime}^{[\iVB-1]}\xbold_{\itime}}{\sum_{\itime=1}^{\ntime}\hLabel_{\idim,\itime}^{[\iVB-1]}},\nonumber \\
\widehat{\idim}_{\itime}^{[\iVB]} & =\argmax_{\idim}\calN_{\xbold_{\itime}}(\bcmean_{\idim}^{[\iVB]},\IDEN_{2})\label{eq:k-means}\\
 & =\argmin_{\idim}||\xbold_{\itime}-\bcmean_{\idim}^{[\iVB]}||^{2},\nonumber 
\end{align}
in which the form of $\bcmean_{\idim}$ is given in (\ref{eq:cluster_mean}),
$\hbLabel_{\itime}^{[\iVB]}\TRIANGLEQ\setf{\hLabel}{\ndim}{\itime}{[\iVB]}^{\transpose}$
and $\hLabel_{\idim,\itime}^{[\iVB]}=\delta[\idim-\widehat{\idim}_{\itime}^{[\iVB]}]$,
with $\delta[\cdot]$ denoting the Kronecker delta function, $\forall\seti{\itime}{\ntime}$.
By convention, we keep $\hbmean_{\idim}^{[\iVB]}=\hbmean_{\idim}^{[\iVB-1]}$
unchanged if $\sum_{\itime=1}^{\ntime}\hLabel_{\idim,\itime}^{[\iVB-1]}=0$,
since no update for $\idim$-th cluster is found in this case. 

From (\ref{eq:k-means}), we can see that the algorithm starts with
$\ndim$ initial mean values $\bcmean_{\idim}^{[0]}$, $\forall\seti{\idim}{\ndim}$,
then assigns categorical labels to clusters via minimum Euclidean
distance in (\ref{eq:k-means}), which, in turn, yields $\ndim$ new
cluster's means $\bcmean_{\idim}^{[1]}$, $\forall\seti{\idim}{\ndim}$,
and so forth. Hence it is called the k-means algorithm in literature
\cite{kmean:LLoyd:82,kmean:cite:15}. 

At convergence $\iVB=\nVB$, the k-means algorithm returns a locally
joint MAP value $\hPARA^{[\nVB]}=\text{[\ensuremath{\hMean^{[\nVB]}},\ensuremath{\hLABEL^{[\nVB]}}]}$,
which depends on initial guess value $\hPARA^{[0]}$. 

From Corollary \ref{cor:(ICM-algorithm)}, the convergence of $\ELBO$
can be used as a stopping rule, as follows:
\begin{align*}
\ELBO_{\ICM}^{[\iVB]} & =\log\pdf(\Data,\hMean^{[\iVB]},\hLABEL^{[\iVB]})\\
 & =\log\frac{\prod_{\itime=1}^{\ntime}\prod_{\idim=1}^{\ndim}\calN_{\xbold_{\itime}}^{\hLabel_{\idim,\itime}^{[\iVB]}}(\hbmean_{\idim}^{[\iVB]},\IDEN_{2})}{\fzeta\ndim^{\ntime}},
\end{align*}
since $\KL_{\ftilde_{\ICM}^{[\iVB]}||f}=-\ELBO_{\ICM}^{[\iVB]}+\log\pdf(\Data),$
as shown in (\ref{eq:KLD_ELBO}).

\subsubsection{EM algorithms}

Let us now derive two EM approximations for true posterior distribution
$\pdf(\Mean,\LABEL|\Data)$ via (\ref{eq:EM}), as follows:
\begin{align}
\ftilde_{\EM_{1}}(\Mean,\LABEL|\Data) & =\pdf(\Mean|\hLABEL_{\EM_{1}},\Data)\delta[\LABEL-\hLABEL_{\EM_{1}}],\nonumber \\
\ftilde_{\EM_{2}}(\Mean,\LABEL|\Data) & =\pdf(\LABEL|\hMean_{\EM_{2}},\Data)\delta(\Mean-\hMean_{\EM_{2}}).\label{eq:EM1_EM2}
\end{align}
Since our joint model $\pdf(\Mean,\LABEL,\Data)$ in (\ref{eq:cluster_Joint}-\ref{eq:cluster_reverse})
is of CEF form (\ref{eq:CEF}), the EM forms (\ref{eq:EM1_EM2}) can
be feasibly identified via (\ref{eq:CEF_approx}), as follows: 

\begin{align}
\hLABEL_{\EM_{1}}^{[\iVB]} & =\argmax_{\LABEL}\EXPECTATION_{\pdf(\Mean|\hLABEL_{\EM_{1}}^{[\iVB-1]},\Data)}\log\pdf(\Data,\Mean,\LABEL),\label{eq:cluster_EM_mean}\\
\hMean_{\EM_{2}}^{[\iVB]} & =\argmax_{\Mean}\EXPECTATION_{\pdf(\LABEL|\hMean_{\EM_{2}}^{[\iVB-1]},\Data)}\log\pdf(\Data,\Mean,\LABEL),\label{eq:cluster_EM_label}
\end{align}
where $\pdf(\Mean|\hLABEL_{\EM_{1}}^{[\iVB-1]},\Data)=\prod_{\idim=1}^{\ndim}\calN_{\bmean_{\idim}}(\tbmean_{\idim}^{[\iVB]},\tstd_{\idim}^{[\iVB]}\IDEN_{2})$
and $\pdf(\LABEL|\hMean_{\EM_{2}}^{[\iVB-1]},\Data)=\prod_{\itime=1}^{\ntime}\Mul_{\bLabel_{\itime}}(\tbweight_{\itime}^{[\iVB-1]})$.

Replacing $\Mean$ and $\LABEL$ in $\pdf(\Data,\Mean,\LABEL)$ in
(\ref{eq:cluster_EM_mean}) and (\ref{eq:cluster_EM_label}) with
$\tMean^{[\iVB-1]}\TRIANGLEQ\EXPECTATION_{\pdf(\Mean|\hLABEL_{\EM_{1}}^{[\iVB-1]},\Data)}(\Mean)$
and $\tbWeight^{[\iVB-1]}\TRIANGLEQ\EXPECTATION_{\pdf(\LABEL|\hMean_{\EM_{2}}^{[\iVB-1]},\Data)}(\LABEL)$,
we then have, respectively: 
\begin{align}
\tbmean_{\idim}^{[\iVB]} & =\bcmean_{\idim}(\hLABEL_{\EM_{1}}^{[\iVB-1]}),\ \tstd_{\idim}^{[\iVB]}=\overline{\std}_{\idim}(\hLABEL_{\EM_{1}}^{[\iVB-1]}),\nonumber \\
\widehat{\idim}_{\itime}^{[\iVB]} & =\argmax_{\idim}\frac{\calN_{\xbold_{\itime}}(\tbmean_{\idim}^{[\iVB]},\IDEN_{2})}{\exp((\tstd_{\idim}^{[\iVB]})^{2})},\label{eq:kEM1}
\end{align}
and: 

\begin{align}
\tweight_{\idim,\itime}^{[\iVB]} & \propto\calN_{\xbold_{\itime}}(\hbmean_{\idim}^{[\iVB]},\IDEN_{2}),\label{eq:kEM2}\\
\hbmean_{\idim}^{[\iVB]} & =\bcmean_{\idim}(\tbWeight^{[\iVB-1]})=\frac{\sum_{\itime=1}^{\ntime}\tweight_{\idim,\itime}^{[\iVB-1]}\xbold_{\itime}}{\sum_{\itime=1}^{\ntime}\tweight_{\idim,\itime}^{[\iVB-1]}},\nonumber 
\end{align}
where $\hMean_{\EM_{2}}^{[\iVB]}\TRIANGLEQ\setu{\hbmean}{\ndim}{[\iVB]}$,
$\tMean^{[\iVB]}\TRIANGLEQ\setu{\tbmean}{\ndim}{[\iVB]}$, $\hLABEL_{\EM_{1}}^{[\iVB]}\TRIANGLEQ\setu{\hbLabel}{\ntime}{[\iVB]}$
with $\hbLabel_{\itime}^{[\iVB]}\TRIANGLEQ\sete{\hLabel}{\ndim}{\itime}^{\transpose}$
and $\hLabel_{\idim,\itime}^{[\iVB]}=\delta[\idim-\widehat{\idim}_{\itime}^{[\iVB]}]$,
$\tbWeight^{[\iVB]}\TRIANGLEQ\setu{\tbweight}{\ntime}{[\iVB]}$ with
$\sum_{\idim=1}^{\ndim}\tweight_{\idim,\itime}^{[\iVB]}=1$, $\forall\seti{\itime}{\ntime}$. 

The forms $\bcmean_{\idim}$ and $\overline{\std}_{\idim}$ are given
in (\ref{eq:cluster_mean}). By convention, we keep $\tbmean_{\idim}^{[\iVB]}=\tbmean_{\idim}^{[\iVB-1]}$
and $\tstd_{\idim}^{[\iVB]}=\tstd_{\idim}^{[\iVB-1]}$ unchanged if
$\sum_{\itime=1}^{\ntime}\hLabel_{\idim,\itime}^{[\iVB-1]}=0$ in
(\ref{eq:kEM1}). 

Also, since $\pdf(\Mean,\LABEL,\Data)$ is of CEF form (\ref{eq:CEF}),
we can feasibly evaluate $\KL_{\ftilde_{\EM}^{[\iVB]}||\pdf(\Data,\Mean,\LABEL)}$
directly for $\ftilde_{\EM_{1}}^{[\iVB]}$and $\ftilde_{\EM_{2}}^{[\iVB]}$,
as defined in (\ref{eq:EM1_EM2}). The convergence of $\ELBO_{\EM}^{[\iVB]}$,
as given in (\ref{eq:ELBO}), is then computed as follows:

\begin{align*}
\ELBO_{\EM}^{[\iVB]} & =-\KL_{\ftilde_{\EM}^{[\iVB]}||\pdf(\Data,\Mean,\LABEL)}=\log\frac{\zeta_{\EM}^{[\iVB]}}{\fzeta\ndim^{\ntime}},\\
\zeta_{\EM_{1}}^{[\iVB]} & =\frac{\pdf(\Mean=\tMean^{[\iVB]},\LABEL=\hLABEL_{\EM}^{[\iVB]},\Data)}{\exp\left(\sum_{\idim=1}^{\ndim}(\tstd_{\idim}^{[\iVB]})^{2}\sum_{\itime=1}^{\ntime}\hLabel_{\idim,\itime}^{[\iVB]}\right)}\prod_{\idim=1}^{\ndim}((\tstd_{\idim}^{[\iVB]})^{2}2\pi e),\\
\zeta_{\EM_{2}}^{[\iVB]} & =\frac{\pdf(\Mean=\hMean_{\EM}^{[\iVB]},\LABEL=\tbWeight^{[\iVB]},\Data)}{\prod_{\idim=1}^{\ndim}\prod_{\itime=1}^{\ntime}\tweight_{\idim,\itime}^{[\iVB]\tweight_{\idim,\itime}^{[\iVB]}}}.
\end{align*}

\begin{rem}
Comparing (\ref{eq:kEM1}-\ref{eq:kEM2}) with (\ref{eq:k-means}),
we can see that the k-means algorithm only considers the mean (i.e.
the first moment), while EM algorithm takes both mean and variance
(i.e. the first and second moments) into account.
\end{rem}

\subsubsection{VB approximation}

Let us now derive VB approximation $\ftilde_{\VB}(\Mean,\LABEL|\Data)=\ftilde_{\VB}(\Mean|\Data)\ftilde_{\VB}(\LABEL|\Data)$
in (\ref{eq:(VB)}) for true posterior distribution $\pdf(\Mean,\LABEL|\Data)$.
Since $\pdf(\Mean,\LABEL,\Data)$ is of CEF form (\ref{eq:CEF}),
the VB form can be feasibly identified via (\ref{eq:CEF_approx}),
as follows: 
\begin{align}
\ftilde_{\VB}^{[\iVB]}(\Mean|\Data) & \propto\exp\EXPECTATION_{\ftilde_{\VB}^{[\iVB-1]}(\LABEL|\Data)}\log\pdf(\Data,\Mean,\LABEL)\label{eq:cluster_VB}\\
 & =\prod_{\idim=1}^{\ndim}\calN_{\bmean_{\idim}}\left(\tbmean_{\idim}^{[\iVB]},\tstd_{\idim}^{[\iVB]}\IDEN_{2}\right),\nonumber \\
\ftilde_{\VB}^{[\iVB]}(\LABEL|\Data) & \propto\exp\EXPECTATION_{\ftilde_{\VB}^{[\iVB]}(\Mean|\Data)}\log\pdf(\Data,\Mean,\LABEL)\nonumber \\
 & =\prod_{\itime=1}^{\ntime}\text{\ensuremath{\Mul}}_{\bLabel_{\itime}}\left(\tbweight_{\itime}^{[\iVB]}\right),\nonumber 
\end{align}
Replacing $\LABEL$ in (\ref{eq:cluster_mean}) with $\tbWeight^{[\iVB]}=\EXPECTATION_{\ftilde_{\VB}^{[\iVB]}(\LABEL|\Data)}(\LABEL)$,
we then have: 
\begin{align}
\tbmean_{\idim}^{[\iVB]} & =\bcmean_{\idim}(\tbWeight^{[\iVB-1]}),\ \tstd_{\idim}^{[\iVB]}=\overline{\std}_{\idim}(\tbWeight^{[\iVB-1]}),\nonumber \\
\tweight_{\idim,\itime}^{[\iVB]} & \propto\frac{\calN_{\xbold_{\itime}}(\tbmean_{\idim}^{[\iVB]},\IDEN_{2})}{\exp((\tstd_{\idim}^{[\iVB]})^{2})},\label{eq:kVB}
\end{align}
where $\tbWeight^{[\iVB]}\TRIANGLEQ\setu{\tbweight}{\ntime}{[\iVB]}$
and $\sum_{\idim=1}^{\ndim}\tweight_{\idim,\itime}^{[\iVB]}=1$, $\forall\seti{\itime}{\ntime}$.
The forms $\bcmean_{\idim}$ and $\overline{\std}_{\idim}$ are given
in (\ref{eq:cluster_mean}).

From (\ref{eq:cluster_VB}), let $\fzeta_{\VB}^{[\iVB]}$ denote the
normalizing constant of $\ftilde_{\VB}^{[\iVB]}(\Mean|\Data)=\frac{1}{\fzeta_{\VB}^{[\iVB]}}\exp\EXPECTATION_{\ftilde_{\VB}^{[\iVB-1]}(\LABEL|\Data)}\log\frac{\pdf(\Data,\Mean,\LABEL)}{\ftilde_{\VB}^{[\iVB-1]}(\LABEL|\Data)}$,
similarly to (\ref{eq:CVB_posterior}). We then have:
\begin{equation}
\ELBO_{\VB}^{[\iVB]}=\log\fzeta_{\VB}^{[\iVB]}=\log\frac{1}{\fzeta\ndim^{\ntime}}\prod_{\idim=1}^{\ndim}\frac{\uweight_{\idim}(\tbWeight^{[\iVB]})}{\prod_{\itime=1}^{\ntime}\tweight_{\idim,\itime}^{[\iVB]\tweight_{\idim,\itime}^{[\iVB]}}},\label{eq:ELBO_VB}
\end{equation}
where $\uweight_{\idim}(\tbWeight^{[\iVB]})$ is given in (\ref{eq:cluster_mean}),
with $\LABEL$ being replaced by $\tbWeight^{[\iVB]}$. The convergence
of $\ELBO_{\VB}^{[\iVB]}$, as mentioned in (\ref{eq:ELBO}), can
be used as a stopping rule. 
\begin{rem}
From (\ref{eq:cluster_VB}-\ref{eq:kVB}), we can see that the VB
algorithm combines two EM algorithms (\ref{eq:EM1_EM2}-\ref{eq:kEM2})
together and takes all moments of clustering data into account. 
\end{rem}

\subsubsection{CVB approximation}

Let us now derive CVB approximation $\ftilde_{\CVB}^{[1]}=\ftilde_{2|1}^{[0]}\ftilde_{1}^{[1]}=\ftilde_{1|2}^{[1]}\ftilde_{2}^{[1]}$
for true posterior distribution $\pdf(\PARA|\Data)=\pdf(\Mean,\LABEL|\Data)$,
with $\PARA=[\Mean,\LABEL]$, via (\ref{eq:CVB_kform}), (\ref{eq:CVB_posterior}).
Firstly, let us note that, the denominator $\pdf(\Mean,\Data)$ of
$\pdf(\LABEL|\Mean,\Data)$ in (\ref{eq:cluster_reverse}) is a mixture
of $\ndim^{\ntime}$ Gaussian components, which is not factorable
over its marginals on $\bmean_{\idim}$, $\seti{\idim}{\ndim}$. Hence,
a direct application of CVB algorithm (\ref{eq:CVB_posterior}) with
$\bpara_{1}=\LABEL$ and $\bpara_{2}=\Mean$ would not yield a closed
form for $\ftilde_{CVB}(\Mean|\Data)$, when the total number $\ndim$
of clusters is not small.
\begin{itemize}
\item CVB's ternary partition:
\end{itemize}
For a tractable form of $\ftilde_{CVB}(\Mean|\Data)$, let us now
define two different binary partitions $\bpara_{1}$ and $\bpara_{2}$
for $\PARA=[\Mean,\LABEL]$ at each CVB's iteration, as explained
in subsection \ref{subsec:iterative_CVA}: 
\begin{align}
\pdf(\PARA|\Data) & =\underset{\pdf_{2|1}}{\underbrace{\pdf(\Mean|\LABEL,\Data)}}\underset{\pdf_{1}}{\underbrace{\pdf(\LABEL|\Data)}}=\underset{\pdf_{1|2}}{\underbrace{\pdf(\LABEL_{\backslash\ipick}|\Mean,\Data)}}\underset{\pdf_{2}}{\underbrace{\pdf(\Mean,\bLabel_{\ipick}|\Data)},}\nonumber \\
\ftilde(\PARA|\Data) & =\underset{\ftilde_{2|1}}{\underbrace{\ftilde(\Mean|\bLabel_{\ipick},\Data)}}\underset{\ftilde_{1}}{\underbrace{\ftilde(\LABEL|\Data)}}=\underset{\ftilde_{1|2}}{\underbrace{\ftilde(\LABEL_{\backslash\ipick}|\bLabel_{\ipick},\Data)}}\underset{\ftilde_{2}}{\underbrace{\ftilde(\Mean,\bLabel_{\ipick}|\Data)},}\label{eq:cluster_CVB}
\end{align}
for any node $\seti{\ipick}{\ntime}.$ Note that, the true conditional
$\pdf_{1|2}\TRIANGLEQ\pdf(\LABEL_{\backslash\ipick}|\Mean,\bLabel_{\ipick},\Data)=\pdf(\LABEL_{\backslash\ipick}|\Mean,\Data)$
in (\ref{eq:cluster_CVB}) does not depends on $\bLabel_{\ipick}$,
since $\pdf(\LABEL|\Mean,\Data)$ in (\ref{eq:cluster_reverse}) is
conditionally independent, i.e. $\pdf(\LABEL|\Mean,\Data)=\prod_{\itime=1}^{\ntime}\pdf(\bLabel_{\itime}|\Mean,\Data)$,
as illustrated in Fig. \ref{fig:cluster}.

Hence, given a ternary partition $\PARA=[\LABEL_{\backslash\ipick},\bLabel_{\ipick},\Mean]$
for each node $\ipick$ in (\ref{eq:cluster_CVB}), we have set $\bpara_{1}=\LABEL=[\LABEL_{\backslash\ipick},\bLabel_{\ipick}]$
and $\bpara_{2}=\Mean$ in the forward form, but $\bpara_{1}=\LABEL_{\backslash\ipick}$
and $\bpara_{2}=[\bLabel_{\ipick},\Mean]$ in reverse form in (\ref{eq:cluster_CVB}).
The equality in CVB form $\ftilde_{\CVB}^{[1]}=\ftilde_{2|1}^{[0]}\ftilde_{1}^{[1]}=\ftilde_{1|2}^{[1]}\ftilde_{2}^{[1]}$
is still valid, since we still have the same joint parameters $\PARA=[\Mean,\LABEL]$
on both sides. 
\begin{itemize}
\item CVB's initialization:
\end{itemize}
Let us consider the left form in (\ref{eq:cluster_CVB}) first. For
tractability, the initial CVB $\ftilde_{2|1}^{[0]}\TRIANGLEQ\ftilde^{[0]}(\Mean|\bLabel_{\ipick},\Data)$
will be set as a restricted form of the true conditional $\pdf_{2|1}\TRIANGLEQ\pdf(\Mean|\LABEL,\Data)$
in (\ref{eq:cluster_posteriors}), as follows: 
\begin{equation}
\ftilde_{2|1}^{[0]}=\prod_{\idim=1}^{\ndim}\ftilde^{[0]}(\bmean_{\idim}|\bLabel_{\ipick})=\prod_{\idim=1}^{\ndim}\prod_{\istate=1}^{\ndim}\calN_{\bmean_{\idim}}^{\Label_{\istate,\ipick}}\left(\tbmean_{\idim,\istate,\ipick}^{[0]},\tstd_{\idim,\istate,\ipick}^{[0]}\IDEN_{2}\right)\label{eq:cluster_CVB_f21a}
\end{equation}
where $\tbmean_{\idim,\istate,\ipick}^{[0]}\in\REAL^{2}$ and $\tstd_{\idim,\istate,\ipick}^{[0]}>0$
are initial means and variances of $\ftilde^{[0]}(\bmean_{\idim}|\bLabel_{\ipick})=\prod_{\istate=1}^{\ndim}\ftilde^{[0]}(\bmean_{\idim}|\Label_{\istate,\ipick})$. 

$ $
\begin{itemize}
\item CVB's iteration (forward step):
\end{itemize}
Let us now apply CVB algorithm (\ref{eq:CVB_posterior}) to (\ref{eq:cluster_CVB})
and approximate $\pdf_{1}\TRIANGLEQ\pdf(\LABEL|\Data)$ via $\ftilde_{2|1}^{[0]}$
in (\ref{eq:cluster_CVB_f21a}), as follows: 

\begin{align}
\ftilde_{1}^{[1]}\TRIANGLEQ & \ftilde^{[1]}(\LABEL|\Data)=\frac{1}{\fzeta_{1}^{[1]}}\exp\EXPECTATION_{\ftilde_{2|1}^{[0]}}\log\frac{\pdf(\Data,\Mean,\LABEL)}{\ftilde_{2|1}^{[0]}}\label{eq:cluster_CVB_f1}\\
 & =\frac{1}{\fzeta_{1}^{[1]}\fzeta\ndim^{\ntime}}\prod_{\idim=1}^{\ndim}\prod_{\istate=1}^{\ndim}(\tukappa_{\idim,\istate,\ipick}^{[1]}\prod_{\itime=1}^{\ntime}(\tuweight_{\idim,\istate,\itime,\ipick}^{[1]})^{\Label_{\idim,\itime}})^{\Label_{\istate,\ipick}},\nonumber 
\end{align}
in which $\pdf(\Data,\Mean,\LABEL)$ is given in (\ref{eq:cluster_Joint}-\ref{eq:cluster_MODEL})
and, hence: 
\begin{equation}
\tukappa_{\idim,\istate,\ipick}^{[1]}=2\pi e(\tstd_{\idim,\istate,\ipick}^{[0]})^{2},\ \tuweight_{\idim,\istate,\itime,\ipick}^{[1]}\TRIANGLEQ\frac{\calN_{\xbold_{\itime}}(\tbmean_{\idim,\istate,\ipick}^{[0]},\IDEN_{2})}{\exp((\tstd_{\idim,\istate,\ipick}^{[0]})^{2})},\label{eq:CVB_f1_weight}
\end{equation}
since we have $\exp\EXPECTATION_{\calN_{\bmean}(\tbmean,\tstd\IDEN_{2})}\log\frac{\calN_{\xbold_{\itime}}(\bmean,\IDEN_{2})}{\calN_{\bmean}(\tbmean,\tstd\IDEN_{2})}=\frac{\calN_{\xbold_{\itime}}(\tbmean,\IDEN_{2})\exp(-\tstd^{2})}{1/(2\pi\tstd^{2}e)}$
in general, as shown in (\ref{eq:CEF_approx}). Comparing the true
updated weights $\uweight_{\idim}(\LABEL)$ of $\pdf_{1}$ in (\ref{eq:cluster_mean})
with the approximated weights $\tuweight_{\idim,\istate,\itime,\ipick}^{[1]}$
of $\ftilde_{1}^{[1]}$ in (\ref{eq:CVB_f1_weight}), we can see that
CVB algorithm has approximated the intractable forms $\{\bcmean_{\idim}(\LABEL),\overline{\std}_{\idim}^{2}(\LABEL)\}$
with total $\ndim^{\ntime}$ elements by a factorized set of $\ntime$
tractable forms $\{\tbmean_{\idim,\istate,\ipick}^{[0]},\tstd_{\idim,\istate,\ipick}^{[0]}\}$
with only $\ndim^{2}\ntime$ elements. 

Comparing (\ref{eq:cluster_CVB}) with (\ref{eq:cluster_CVB_f1}),
we can identify the form of $\ftilde_{1|2}^{[1]}$ in (\ref{eq:cluster_CVB}),
as follows: 
\begin{align}
\ftilde_{1|2}^{[1]}\TRIANGLEQ\ftilde^{[1]}(\LABEL_{\backslash\ipick}|\bLabel_{\ipick},\Data) & =\prod_{\itime\neq\ipick}\text{\ensuremath{\Mul}}_{\bLabel_{\itime}}\left(\tTransition_{\itime,\ipick}^{[1]}\bLabel_{\ipick}\right)\label{eq:cluster_CVB_f12}
\end{align}
where $\tTransition_{\itime,\ipick}^{[1]}$ is a left stochastic matrix,
whose $\{\istate,\idim\}$-element is the updated transition probability
from $\Label_{\istate,\ipick}$ to $\Label_{\idim,\itime}$: $\ttransition_{\idim,\istate}^{[1]}(\itime,\ipick)\TRIANGLEQ\frac{\tuweight_{\idim,\istate,\itime,\ipick}^{[1]}}{\sum_{\idim=1}^{\ndim}\tuweight_{\idim,\istate,\itime,\ipick}^{[1]}}$,
$\forall\itime\neq\ipick$. For later use, let us assign $\tTransition_{\ipick,\ipick}^{[\iVB]}\TRIANGLEQ\IDEN_{\ndim}$,
with $\IDEN_{\ndim}$ denoting $\ndim\times\ndim$ identity matrix,
when $\itime=\ipick$, at any iteration $\iVB$.
\begin{itemize}
\item CVB's iteration (reverse step):
\end{itemize}
Let us apply CVB (\ref{eq:CVB_posterior}) to (\ref{eq:cluster_CVB})
again and approximate $\pdf_{2}\TRIANGLEQ\pdf(\Mean,\bLabel_{\ipick}|\Data)$
by $\ftilde_{2}^{[2]}$ in $\ftilde_{\CVB}^{[2]}=\ftilde_{2|1}^{[2]}\ftilde_{1}^{[2]}=\ftilde_{1|2}^{[1]}\ftilde_{2}^{[2]}$,
via $\ftilde_{1|2}^{[1]}$ in (\ref{eq:cluster_CVB_f12}), as follows:

\begin{align}
\ftilde_{2}^{[2]}\TRIANGLEQ & \ftilde^{[2]}(\Mean,\bLabel_{\ipick}|\Data)=\frac{1}{\fzeta_{2}^{[2]}}\exp\EXPECTATION_{\ftilde_{1|2}^{[1]}}\log\frac{\pdf(\Data,\Mean,\LABEL)}{\ftilde_{1|2}^{[1]}}\nonumber \\
= & \ftilde^{[2]}(\Mean|\bLabel_{\ipick}.\Data)\text{\ensuremath{\ftilde^{[2]}}(\ensuremath{\bLabel_{\ipick}}|\ensuremath{\Data})}\label{eq:cluster_CVB_f2}
\end{align}
in which, similar to (\ref{eq:cluster_posteriors}-\ref{eq:cluster_mean}),
we have:
\begin{align}
\ftilde_{2|1}^{[2]}\TRIANGLEQ\ftilde^{[2]}(\Mean|\bLabel_{\ipick}.\Data) & =\prod_{\idim=1}^{\ndim}\prod_{\istate=1}^{\ndim}\calN_{\bmean_{\idim}}^{\Label_{\istate,\ipick}}\left(\tbmean_{\idim,\istate,\ipick}^{[1]},\tstd_{\idim,\istate,\ipick}^{[1]}\IDEN_{2}\right)\nonumber \\
\ftilde^{[2]}(\bLabel_{\ipick}|\Data) & =\frac{1}{\fzeta_{2}^{[2]}\fzeta\ndim^{\ntime}}\prod_{\idim=1}^{\ndim}\prod_{\istate=1}^{\ndim}(\tuweight_{\idim,\istate,\ipick}^{[1]})^{\Label_{\istate,\ipick}}\label{eq:cluster_CVB_f21b}
\end{align}
Note that, as shown in (\ref{eq:CEF_approx}), we have replaced $\Label_{\idim,\itime}$
in (\ref{eq:cluster_mean}) by $\tLabel_{\idim,\itime}(\bLabel_{\ipick})\TRIANGLEQ\EXPECTATION_{\ftilde_{1|2}^{[1]}}(\Label_{\idim,\itime})=\sum_{\istate=1}^{\nstate}\ttransition_{\idim,\istate}(\itime,\ipick)\Label_{\istate,\ipick}$
in (\ref{eq:cluster_CVB_f21b}) and, hence:
\[
\tbmean_{\idim,\istate,\ipick}^{[1]}\TRIANGLEQ\frac{\sum_{\itime=1}^{\ntime}\ttransition_{\idim,\istate}^{[1]}(\itime,\ipick)\xbold_{\itime}}{\sum_{\itime=1}^{\ntime}\ttransition_{\idim,\istate}^{[1]}(\itime,\ipick)},\ \tstd_{\idim,\istate,\ipick}^{[1]}\TRIANGLEQ\frac{1}{\sqrt{\sum_{\itime=1}^{\ntime}\ttransition_{\idim,\istate}^{[1]}(\itime,\ipick)}},
\]
\begin{equation}
\tuweight_{\idim,\istate,\ipick}^{[1]}=\frac{2\pi(\tstd_{\idim,\istate,\ipick}^{[1]})^{2}\prod_{\itime=1}^{\ntime}\calN_{\xbold_{\itime}}^{\ttransition_{\idim,\istate}^{[1]}(\itime,\ipick)}(\tbmean_{\idim,\istate,\ipick}^{[1]},\IDEN_{2})}{\prod_{\itime\neq\ipick}\ttransition_{\idim,\istate}^{[1]}(\itime,\ipick)^{\ttransition_{\idim,\istate}^{[1]}(\itime,\ipick)}},\label{eq:cluster_CVB_f21c}
\end{equation}
in which, by convention, $\tbmean_{\idim,\istate,\ipick}^{[1]}=\tbmean_{\idim,\istate,\ipick}^{[0]}$,
$\tstd_{\idim,\istate,\ipick}^{[1]}=\tstd_{\idim,\istate,\ipick}^{[0]}$
are kept unchanged and $\tuweight_{\idim,\istate,\ipick}^{[1]}=1$
if $\sum_{\itime=1}^{\ntime}\ttransition_{\idim,\istate}^{[1]}(\itime,\ipick)=0$.

It is feasible to recognize that $\ftilde^{[2]}(\bLabel_{\ipick}|\Data)$
in (\ref{eq:cluster_CVB_f21b}) is actually a Multinomial distribution:
$\ftilde^{[2]}(\bLabel_{\ipick}|\Data)=\Mul_{\bLabel_{\ipick}}(\tbweight_{\ipick}^{[2]})$,
in which $\tbweight_{\ipick}^{[2]}\TRIANGLEQ\setf{\tweight}{\ndim}{\ipick}{[2]}^{\transpose}$
and $\sum_{\istate=1}^{\ndim}\tweight_{\istate,\ipick}^{[2]}=1$,
as follows: 
\begin{equation}
\tweight_{\istate,\ipick}^{[2]}\TRIANGLEQ\frac{\prod_{\idim=1}^{\ndim}\tuweight_{\idim,\istate,\ipick}^{[1]}}{\sum_{\istate=1}^{\nstate}\prod_{\idim=1}^{\ndim}\tuweight_{\idim,\istate,\ipick}^{[1]}},\ \fzeta_{2}^{[2]}=\frac{\sum_{\istate=1}^{\nstate}\prod_{\idim=1}^{\ndim}\tuweight_{\idim,\istate,\ipick}^{[1]}}{\fzeta\ndim^{\ntime}}\label{eq:cluster_CVBj}
\end{equation}

\begin{itemize}
\item CVB's form at convergence:
\end{itemize}
From (\ref{eq:cluster_CVB_f21a}) and (\ref{eq:cluster_CVB_f21b}),
we can see that $\ftilde_{2|1}^{[\iVB]}$ can be updated iteratively
from $\ftilde_{2|1}^{[\iVB-2]},$ given that only one CVB marginal
is updated per iteration $\iVB$. The iterative CVB then converges
when the $\ELBO^{[\iVB]}\TRIANGLEQ\log\fzeta_{2}^{[\iVB]}$, given
in (\ref{eq:cluster_CVBj}), converges at $\iVB=\nVB$, as shown in
(\ref{eq:ELBO}). 

Then, for any chosen $\seti{\ipick}{\nstate}$, the marginals in converged
CVB $\ftilde_{\ipick}^{[\nVB]}$ can be derived from $\ftilde^{[\nVB]}(\bLabel_{\ipick}|\Data)=\Mul_{\bLabel_{\ipick}}(\tbweight_{\ipick}^{[\nVB]})$
in (\ref{eq:cluster_CVBj}), as follows:
\begin{align}
\ftilde_{\ipick}^{[\nVB]}(\Mean|\Data) & \TRIANGLEQ\sum_{\bLabel_{\ipick}}\ftilde^{[\nVB]}(\Mean|\bLabel_{\ipick}.\Data)\ftilde^{[\nVB]}(\bLabel_{\ipick}|\Data),\label{eq:cluster_CVB_converged}\\
\ftilde_{\ipick}^{[\nVB]}(\LABEL|\Data) & \TRIANGLEQ\prod_{\itime\neq\ipick}\ftilde^{[\nVB]}(\bLabel_{\itime}|\bLabel_{\ipick}.\Data)\ftilde^{[\nVB]}(\bLabel_{\ipick}|\Data),\nonumber 
\end{align}
in which $\ftilde_{\ipick}^{[\nVB]}(\Mean|\Data)$ is a mixture of
$\nstate$ Gaussian components:
\begin{align*}
\ftilde_{\ipick}^{[\nVB]}(\Mean|\Data) & =\sum_{\istate=1}^{\ndim}\tweight_{\istate,\ipick}^{[\nVB]}\prod_{\idim=1}^{\ndim}\calN_{\bmean_{\idim}}(\tbmean_{\idim,\istate,\ipick}^{[\nVB]},\tstd_{\idim,\istate,\ipick}^{[\nVB]}\IDEN_{2}),\\
\ftilde_{\ipick}^{[\nVB]}(\bLabel_{\itime}|\Data) & =\sum_{\LABEL_{\backslash\itime}}\ftilde_{\ipick}^{[\nVB]}(\LABEL|\Data)=\Mul_{\bLabel_{\itime}}(\tbqweight_{\itime}^{[\nVB]}(\ipick)),
\end{align*}
with $\tbqweight_{\itime}^{[\nVB]}(\ipick)\TRIANGLEQ[\tqweight_{1,\itime}^{[\nVB]}(\ipick),\ldots,\tqweight_{\ndim,\itime}^{[\nVB]}(\ipick)]^{\transpose}=\tTransition_{\itime,\ipick}^{[\nVB]}\tbweight_{\ipick}^{[\nVB]}$,
$\forall\seti{\itime}{\ntime}$. The approximated posterior estimates
for cluster's means and labels in this case are, respectively: 
\begin{align}
\hMean(\ipick) & \TRIANGLEQ\EXPECTATION_{\ftilde_{\ipick}^{[\nVB]}(\Mean|\Data)}(\Mean)=\sum_{\istate=1}^{\ndim}\tweight_{\istate,\ipick}^{[\nVB]}\tMean_{\istate,\ipick}^{[\nVB]},\nonumber \\
\widehat{\bLabel_{\itime}}(\ipick) & \TRIANGLEQ\arg\max_{\bLabel_{\itime}}\ftilde_{\ipick}^{[\nVB]}(\bLabel_{\itime}|\Data)=\element_{\widehat{\idim_{\itime}}(\ipick)},\label{eq:cluster_CVB_hat}
\end{align}
where $\tMean_{\istate,\ipick}^{[\nVB]}\TRIANGLEQ[\tbmean_{1,\istate,\ipick}^{[\nVB]},\ldots,\tbmean_{\ndim,\istate,\ipick}^{[\nVB]}]$
and $\widehat{\idim_{\itime}}(\ipick)\TRIANGLEQ\argmax_{\idim}\tqweight_{\idim,\itime}^{[\nVB]}(\ipick),$
$\forall\seti{\itime}{\ntime}$.

\begin{figure*}
\begin{centering}
\includegraphics[width=0.4\linewidth]{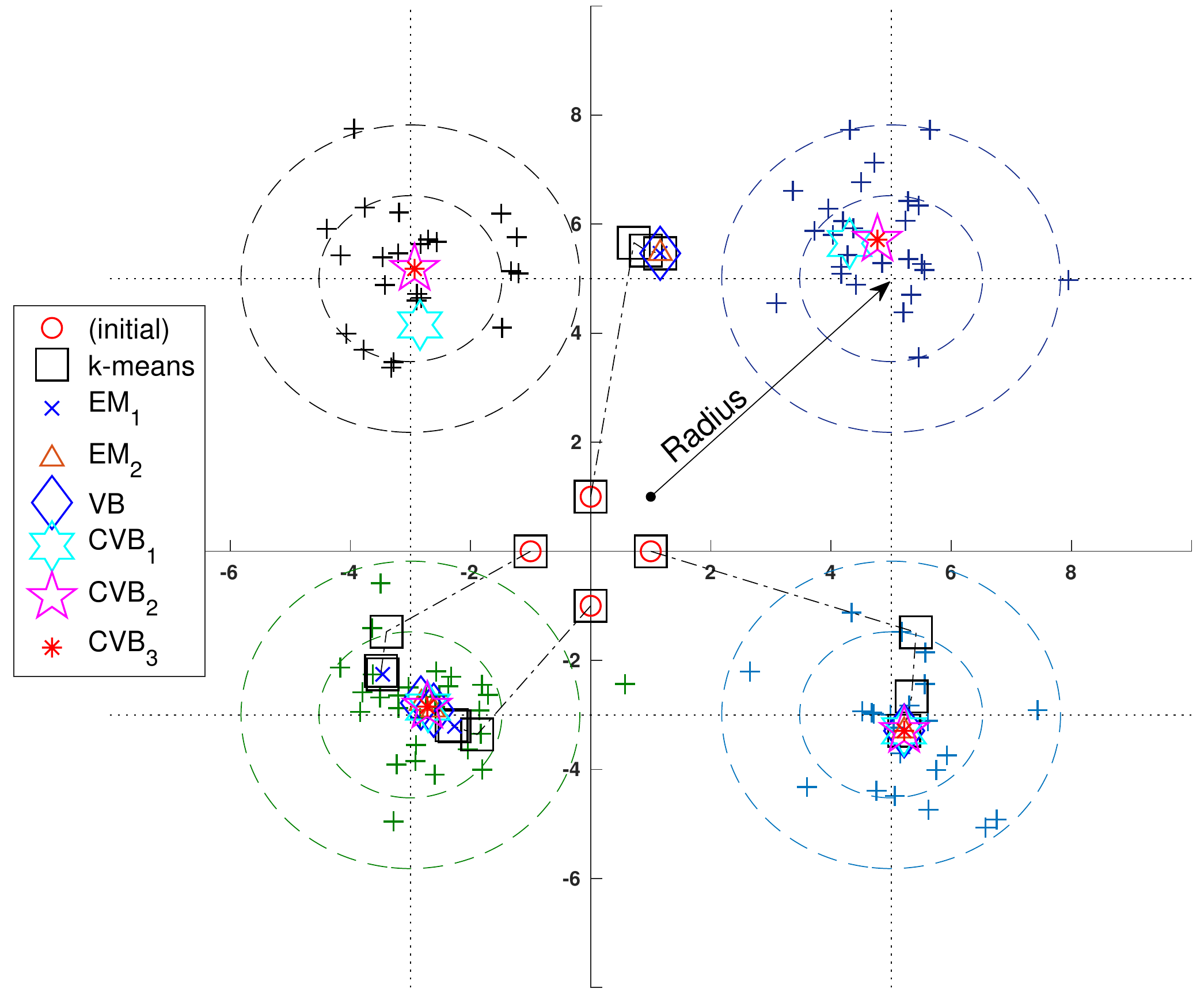}\includegraphics[width=0.4\linewidth]{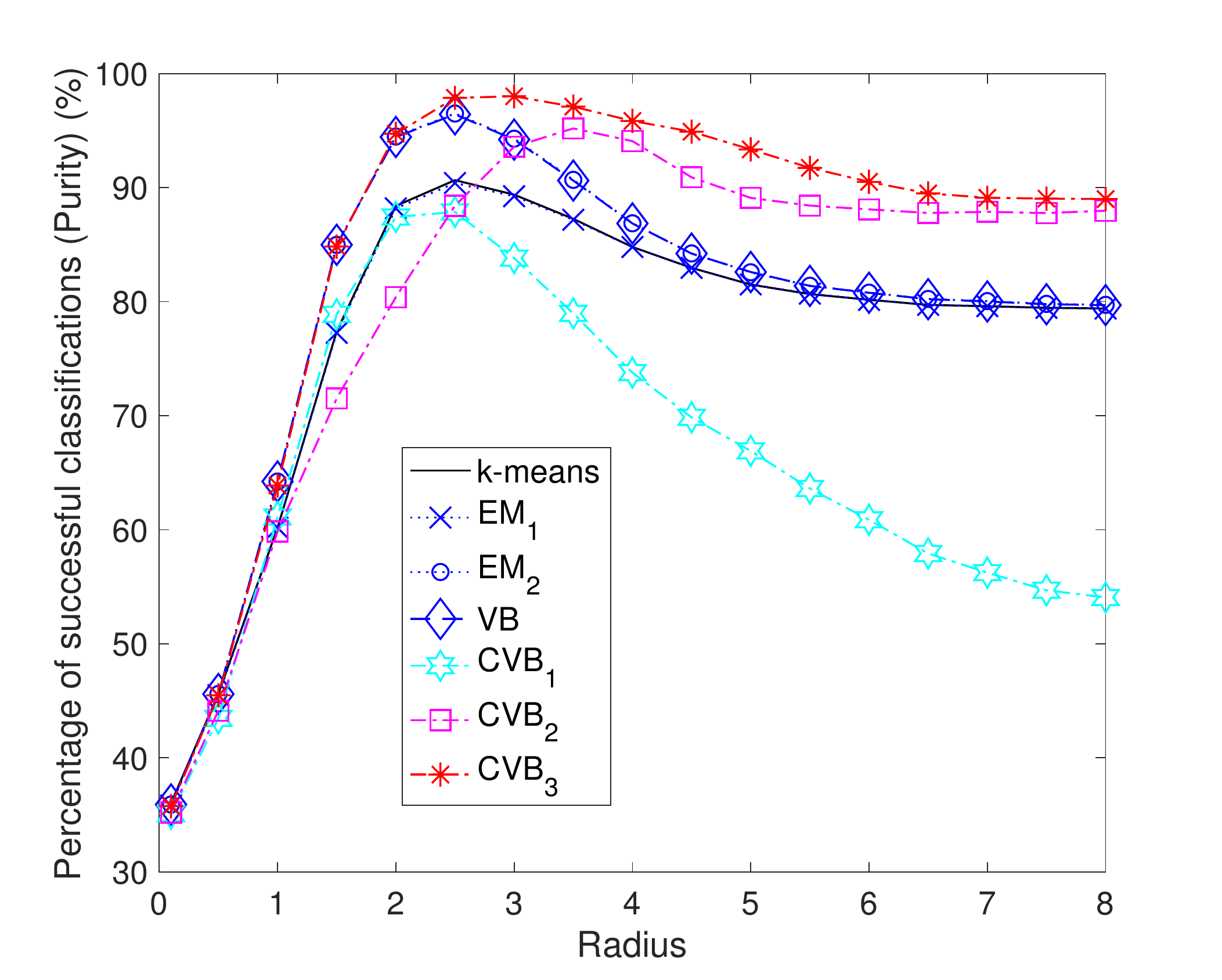}
\par\end{centering}
\centering{}\includegraphics[width=0.4\linewidth]{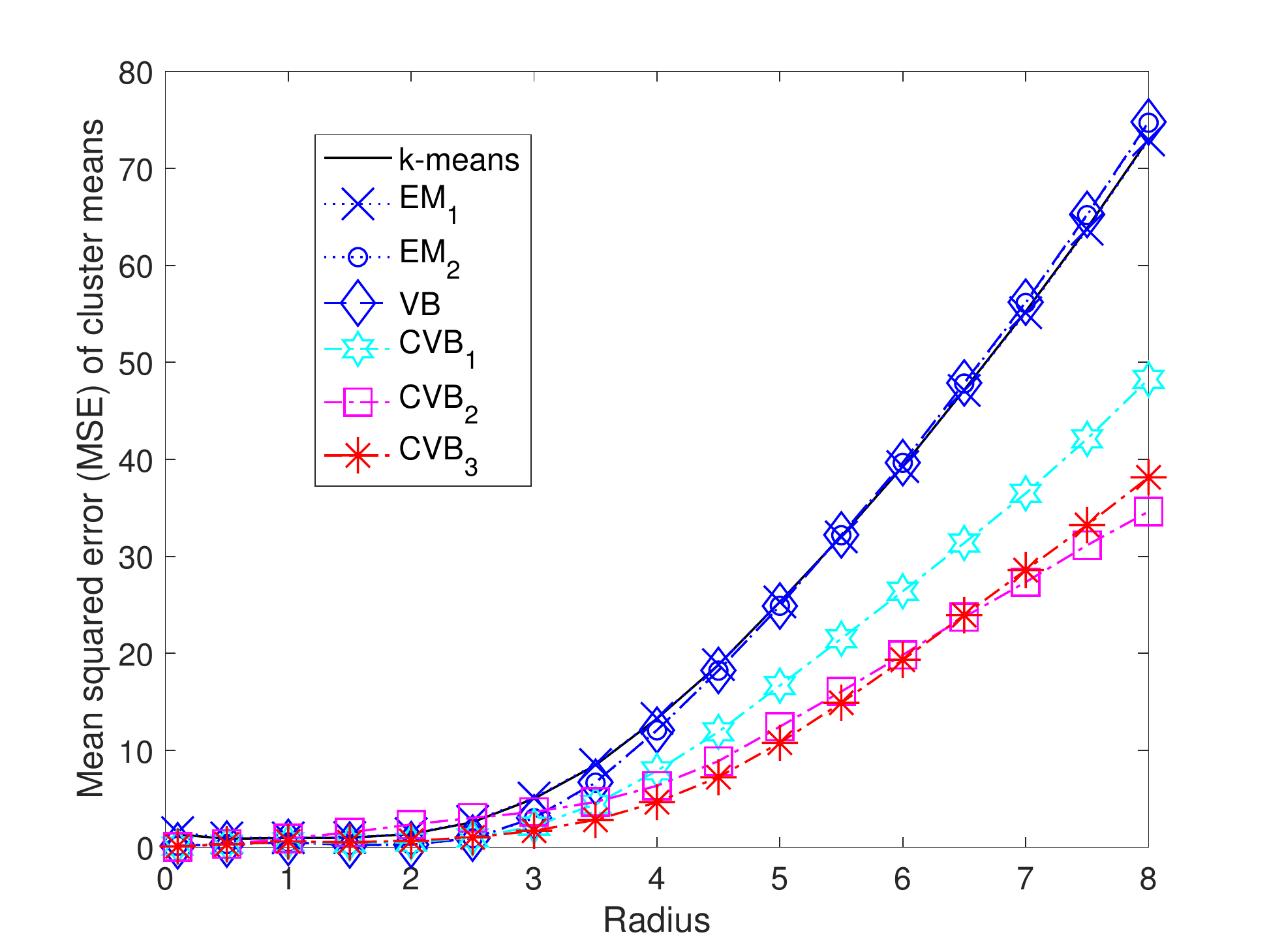}\includegraphics[width=0.4\linewidth]{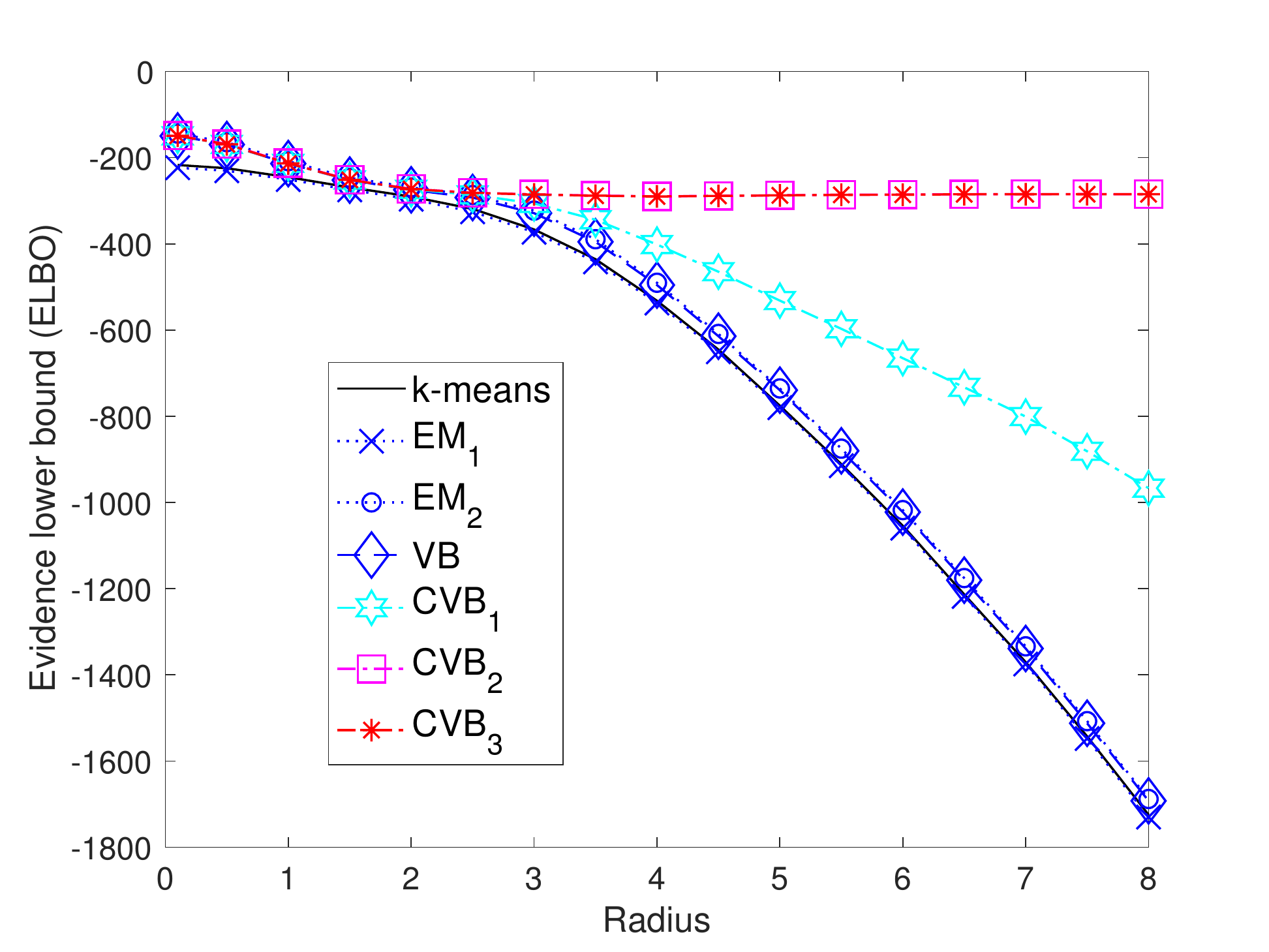}\caption{\label{fig:Cluster_CVB}CVB and mean-field approximations for $\protect\ndim=4$
bivariate independent Normal clusters $\protect\calN(\protect\bmean,\protect\IDEN_{2})$,
with mean vectors $\protect\bmean$ located diagonally and equally
at radius $\protect\Radius$ from the offset point $[1,1]^{\protect\transpose}$.
The upper left panel shows the convergent results of approximated
mean vectors for one Monte Carlo run in the case $\protect\Radius=4$,
with true mean vectors located at intersections of four dotted lines.
The dashed circles represent contours of true Normal distributions.
The plus signs $+$ are $\protect\ntime=100$ random data, generated
with equal probability from each Normal cluster. The four smallest
circles are the same initial guesses of true mean vectors for all
algorithms. The dash-dot line illustrates the k-means algorithm, from
initial to convergent points. The other panels show the Purity, MSE
and ELBO values at convergence with varying radius. The higher Purity,
the higher percentage of correct classification of data. The higher
$\protect\ELBO$ at each radius, the lower $\protect\KL$ divergence
and, hence, the better approximation for that case of radius, as shown
in (\ref{eq:ELBO}-\ref{eq:KLD_ELBO}) and illustrated in Fig. \ref{fig:fCVA},
\ref{fig:fEM}. The number of Monte Carlo runs for each radius is
$10^{4}$.}
\end{figure*}

\begin{itemize}
\item Augmented CVB approximation:
\end{itemize}
As shown above, each value $\seti{\ipick}{\ntime}$ yields a different
network structure for CVB approximation, as mentioned in section \ref{sec:Bayesian-network}.
Let us consider here three simple ways to make use of these $\ntime$
CVB's structures.

The first and heuristic way, namely $\CVB_{1}$ scheme, is to choose
$\text{\ensuremath{\widehat{\bLabel_{\ipick}}}}(\ipick)$ in (\ref{eq:cluster_CVB_hat}),
as the estimate for $\ipick$-th label, because the CVB's ternary
structure is more focused on $\bLabel_{\ipick}$ at each $\seti{\ipick}{\ntime}$,
as shown in Fig. \ref{fig:cluster}. Since every $\ipick$-th structure
is equally important in this way, we can pick the empirical average
$\hMean\TRIANGLEQ\sum_{\ipick=1}^{\ntime}\frac{1}{\ntime}\hMean(\ipick)$
as estimate for cluster's means. 

The second way, namely $\CVB_{2}$ scheme, is to pick $\ipick$ such
that $\KL_{\ftilde_{\ipick}^{[\nVB]}||f}$ at convergence is minimized,
as mentioned in (\ref{eq:KL_i}) . From (\ref{eq:ELBO}-\ref{eq:KLD_ELBO}),
we then have $\widehat{\ipick}\TRIANGLEQ\argmin_{\ipick}\KL_{\ftilde_{\ipick}^{[\nVB]}||f}=\argmax_{\ipick}\ELBO^{[\nVB]}(\ipick)=\argmax_{\ipick}\log\fzeta_{2}^{[\nVB]}(\ipick)$,
with $\fzeta_{2}^{[\nVB]}(\ipick)$ given in (\ref{eq:cluster_CVBj}).
Then, $\widehat{\bLabel_{\itime}}(\widehat{\ipick})$ and $\hMean(\widehat{\ipick})$
in (\ref{eq:cluster_CVB_hat}) will be used as estimates for categorical
label $\bLabel_{\itime}$ and cluster means, respectively, $\forall\seti{\itime}{\ntime}$.

The third way, namely $\CVB_{3}$ scheme, is to apply the augmented
approach for CVB, given in (\ref{eq:optimal_weight}). Then, from
(\ref{eq:mixture_moments}) and (\ref{eq:cluster_CVB_hat}), the augmented
CVB's estimates for cluster's means and labels in this case are $\hMean^{*}\TRIANGLEQ\sum_{\ipick=1}^{\ntime}\qweight_{\ipick}^{*}\hMean(\ipick)$
and $\widehat{\bLabel_{\itime}}^{*}=\element_{\widehat{\idim_{\itime}}^{*}},$
respectively, with: 
\begin{align}
\widehat{\idim_{\itime}}^{*} & \TRIANGLEQ\argmax_{\idim}\sum_{\ipick=1}^{\ntime}\qweight_{\ipick}^{*}\tqweight_{\idim,\itime}^{[\nVB]}(\ipick),\nonumber \\
\qweight_{\ipick}^{*} & =\frac{\exp(-\KL_{\ftilde_{\ipick}^{[\nVB]}||f})}{\sum_{\ipick=1}^{\ntime}\exp(-\KL_{\ftilde_{\ipick}^{[\nVB]}||f})}=\frac{\fzeta_{2}^{[\nVB]}(\ipick)}{\sum_{\ipick=1}^{\ntime}\fzeta_{2}^{[\nVB]}(\ipick)},\label{eq:CVB_final}
\end{align}
in which $\qweight_{\ipick}^{*}$ is found via (\ref{eq:optimal_weight})
and $\KL_{\ftilde_{\ipick}^{[\nVB]}||f}=-\ELBO^{[\nVB]}(\ipick)+\log\pdf(\Data),$
as shown in (\ref{eq:KLD_ELBO}) $\forall\seti{\ipick}{\ntime}$.

Although we can compute all moments of augmented CVB $\ftilde_{0}^{[\nVB]}=\sum_{\ipick=1}^{\ntime}\qweight_{\ipick}^{*}\ftilde_{\ipick}^{[\nVB]}$via
(\ref{eq:mixture_moments}) and (\ref{eq:CVB_final}), it is difficult
to evaluate $\KL_{\ftilde_{0}^{[\nVB]}||f}$ and its $\ELBO$ value
directly, as mentioned in subsection \ref{subsec:Augmented-CVB}.
Hence, for comparison with $\CVB_{2}$ scheme in simulations, let
us instead compute heuristic values $\sum_{\ipick=1}^{\ntime}\frac{1}{\ntime}\ELBO^{[\nVB]}(\ipick)$
and $\sum_{\ipick=1}^{\ntime}\qweight_{\ipick}^{*}\ELBO^{[\nVB]}(\ipick)$
at convergence for $\CVB_{1}$ and $\CVB_{3}$ schemes, respectively,
with $\ELBO^{[\nVB]}(\ipick)\TRIANGLEQ\log\fzeta_{2}^{[\nVB]}(\ipick)$
given in (\ref{eq:cluster_CVBj}).
\begin{rem}
\label{rem:init-CVB}Note that, the CVB $\ftilde_{\ipick}^{[\nVB]}$
still belongs to a conditional structure class of node $\ipick$ at
convergence, even if the initialization $\{\tbmean_{\idim,\istate,\ipick}^{[0]},\tstd_{\idim,\istate,\ipick}^{[0]}\}$
of CVB is exactly the same as that of VB. Indeed, in below simulations,
even though initially we set $\tbmean_{\idim,\istate,\ipick}^{[0]}=\tbmean_{\idim}^{[0]}$,
$\tstd_{\idim,\istate,\ipick}^{[0]}=\tstd_{\idim}^{[0]}$, $\forall\istate,\ipick$
and, hence, $\ftilde_{2|1}^{[0]}=\ftilde^{[0]}(\Mean|\bLabel_{\ipick},\Data)$
in (\ref{eq:cluster_CVB_f21a}) independent of $\bLabel_{\ipick}$,
the conditional $\ftilde^{[\iVB]}(\Mean|\bLabel_{\ipick},\Data)$
in (\ref{eq:cluster_CVB_f21b}-\ref{eq:cluster_CVB_f21c}) depends
on $\bLabel_{\ipick}$ again in subsequent iterations, as already
explained in subsection \ref{subsec:iterative_CVA} for this case
of ternary partition. 
\end{rem}

\subsubsection{Simulation's results}

Since k-means algorithm (\ref{eq:k-means}) works best for independent
Normal clusters, let us illustrate the superior performance of CVB
to mean-field approximations even in this case. For this purpose,
a set of $\ndim=4$ bivariate independent Normal clusters $\calN(\bmean_{\idim},\IDEN_{2})$
are generated randomly, with true means $\Mean=\Mean_{0}\Radius+\left[\begin{array}{c}
1\\
1
\end{array}\right]$ and $\Mean_{0}\TRIANGLEQ\left[\begin{array}{cccc}
-1 & 1 & 1 & -1\\
1 & 1 & -1 & -1
\end{array}\right]$. At each time $\itime$, a cluster is then chosen with equal probabilities
$\weight_{\idim}=\frac{1}{\ndim}$, $\seti{\idim}{\ndim}$ in order
to generate the data $\xbold_{\itime}\in\REAL^{2}$, $\seti{\itime}{\ntime},$
with $\ntime=100$, as shown in Fig. \ref{fig:Cluster_CVB}. The varying
radius $\Radius$ then controls the inter-distance between clusters.
In order to quantify the algorithm's performance, let us compute the
Purity and mean squared error (MSE) for estimates $\hbLabel_{\itime}$,
$\hMean$ of categorical labels $\bLabel_{\itime}$ and mean vectors
$\Mean$, respectively. The Purity, which is a common measure for
percentage of successful label's classification \cite{purity:cite:06},
is calculated as follows: $\text{Purity}=\sum_{\idim=1}^{\ndim}\frac{1}{\ntime}\max_{\istate}\sum_{\itime=1}^{\ntime}\delta[\hLabel_{\idim,\itime}=\Label_{\istate,\itime}]$
in each Monte Carlo run. The higher $\text{Purity}\in[0,1]$, the
better estimate for labels. The MSE in each Monte Carlo run is calculated
as follows: $\text{MSE}=\frac{1}{\ndim}\min_{\phi\in\Phi}||\phi(\hMean)-\Mean||^{2}$,
where $\Phi$ is all $\ndim!$ possible permutations of $\ndim$ estimated
cluster means in $\hMean\in\REAL^{2\times\ndim}$.

For comparison at convergence, the initialization $\tMean^{[0]}=\Mean_{0}$
and $\tstd^{[0]}=1$ are the same for all algorithms. The k-means
(\ref{eq:k-means}) and $\EM_{1}$ algorithms (\ref{eq:kEM1}) will
converge at iteration $\nVB$ if there is no update for categorical
labels, i.e. $\hLABEL^{[\nVB]}=\hLABEL^{[\nVB-1]}\Leftrightarrow\ELBO^{[\nVB]}=\ELBO^{[\nVB-1]}$
in this case. The other algorithms are called converged at iteration
$\nVB$ if $0\leq\ELBO^{[\nVB]}-\ELBO^{[\nVB-1]}\leq0.01$. The averaged
values of $\nVB$ over all cases in Fig. \ref{fig:Cluster_CVB} are
$[16.4,16.4,27.2,27.4,27.8]\pm[5.0,5.1,10.4,10.4,7.8]$ for k-means,
$\EM_{1}$, $\EM_{2}$,$\VB$ and $\CVB$ algorithms. Only one approximated
marginal is updated per iteration.

We can see that both performance and number of iterations of k-means
and $\EM_{1}$ algorithms are almost identical to each other, since
they use the same approach with point estimates for categorical labels.
Although the $\EM_{1}$ (\ref{eq:kEM1}) takes one extra data-driven
step, in comparison with k-means, by using the total number of classified
labels in each cluster as an indicator for credibility, the $\EM_{1}$
is virtually the same as k-means in estimate's accuracy. Likewise,
since the point estimates of labels are data-driven and use hard decision
approach, the k-means and $\EM_{1}$ yield lower accuracy than other
methods, which are model-driven and use soft decision approach. 

The $\EM_{2}$ (\ref{eq:kEM2}) and $\VB$ (\ref{eq:kVB}) also have
almost identical performance and number of iterations, even though
$\EM_{2}$ does not update the cluster mean's credibility via total
number of classified labels like VB does. Hence, like the case of
$\EM_{1}$ versus k-means, this extra step of data-driven update seems
insignificant in terms of estimate's accuracy. Nevertheless, since
both $\EM_{2}$ and $\VB$ use the model's probability of each label
as weighted credibility and make soft decision at each iteration,
their performance is significantly better than k-means and $\EM_{1}$
in the range of radius $\Radius\in[2,4]$. Hence, the model-driven
update step seems to exploit more information from the true model
than the data-driven update step, when the clusters are close to each
other. 

For a large radius $\Radius>4$, there is not much difference between
soft and hard decisions for these standard Normal clusters, since
the tail of Normal distribution is very small in these cases. Hence,
given the same initialization at origin, the performances of all mean-field
approximations like k-means, $\EM_{1}$, $\EM_{2}$ and VB are very
close to each other when the inter-distance between clusters is high.
Also, since the computation of soft decision in VB and $\EM_{2}$
requires almost double number of iterations, compared with hard decision
approaches like k-means and $\EM_{1}$, the k-means is more advantageous
in this case, owing to its low computational complexity.

The CVB algorithms are the slowest methods overall. Since the CVB
in (\ref{eq:CVB_final}) requires nearly the same number of iterations
as VB for each structure $\seti{\ipick}{\ntime}$, as illustrated
in Fig. \ref{fig:cluster}, the CVB's complexity is at least $\ntime$
times slower than VB method, where $\ntime$ is the number of data.
In practice, we may not have to update all $\ntime$ CVB's potential
structures, since there might be some good candidates out of exponentially
growing number of potential structures. In this paper, however, let
us consider the case of $\ntime$ structures in order to illustrate
the superior performance of augmented CVB form in $\CVB_{3}$ (\ref{eq:CVB_final}),
in comparison with VB, heuristic $\CVB_{1}$ and hit-or-miss $\CVB_{2}$
approaches. 

The heuristic $\CVB_{1}$, which takes uniform average for mean vectors
over all $\ntime$ potential structures, returns a lower MSE than
mean-field approximations in all cases. This result seems reasonable,
since cluster means are common parameters of all potential CVB structures
in Fig. \ref{fig:cluster}. In contrast, $\CVB_{1}$ returns label's
estimate $\hbLabel_{\ipick}$ via $\ipick$-th structure only, without
considering label's estimates from other CVB's structures. Hence,
the label's Purity of $\CVB_{1}$ is only on par with that of mean-field
approximations for short radius $\Radius\leq2$ and deteriorates over
longer radius $\Radius>2$. As illustrated in Fig. \ref{fig:Gauss},
CVB might be the worst approximation if the CVB's structure is too
different from true posterior structure. In this case, a single $\ipick$-th
structure seems to be a bad CVB candidate for estimating label $\bLabel_{\ipick}$
at time $\seti{\ipick}{\ntime}.$ 

The hit-or-miss $\CVB_{2}$, which picks the single best structure
$\widehat{\ipick}$ in terms of $\KL$ divergence, yields the worst
performance in the range $\Radius\in[1,2.5]$, while in other cases,
it is the second-best method. The structure $\widehat{\ipick}$, as
illustrated in Fig. \ref{fig:cluster}, concentrates on the $\widehat{\ipick}$-th
label. Hence, the classification's accuracy of $\CVB_{2}$ depends
on whether the hard decision on  $\widehat{\ipick}$-th label serves
as a good reference for other labels, as illustrated in Fig. \ref{fig:Gauss}.
For this reason, $\CVB_{2}$ may be able to achieve globally optimal
approximation, but it may also be worse than mean-field approximations.
When $\Radius<3$, which is less than three standard deviation of
a standard Normal cluster, the clusters data are likely overlapped
with each other. Within this range, the hard decision of $\CVB_{2}$
on $\widehat{\ipick}$ destroys the correlated information between
clusters and, hence, becomes worse than other methods. For $\Radius\geq3$,
the $\CVB_{2}$ becomes better, which indicates that the classification's
accuracy now relies more on the most significantly correlated structure
between labels.

Generalizing both schemes $\CVB_{1}$ and $\CVB_{2}$, $\CVB_{3}$
(\ref{eq:CVB_final}) can return the optimal weights for the mixture
of $\ntime$ potential structures and achieve the minimum upper bound
of $\KL$ divergence (\ref{eq:augmented_weight}), as illustrated
in Fig. \ref{fig:mixtureKLD}. Hence, the $\CVB_{3}$ yields the best
performance in Fig. \ref{fig:Cluster_CVB}. When $\Radius<3$, the
$\CVB_{3}$ is on par with VB approximation, since the probabilities
computed via Normal model are high enough for making soft decisions
in VB. When $\Radius>3$, however, VB has to rely on hard decisions
like k-means, since the standard Normal probabilities are too low.
The $\CVB_{3}$, in contrast, automatically move the mixture's weights
closer to hard decision on the best structures like $\CVB_{2}$. 

Note that, although the computed ELBO values for $\CVB_{2}$ in Fig.
\ref{fig:Cluster_CVB} are correct, the computed ELBO values for $\CVB_{1}$
and $\CVB_{3}$ are merely heuristic and not correct values, since
their ELBO values are hard to compute in this case. Nonetheless, from
their performance in Purity and MSE, we may speculate that the true
$\ELBO$ values of $\CVB_{1}$ and $\CVB_{3}$ are lower and higher
than those of $\CVB_{2}$, respectively. Equivalently, in terms of
KL divergence, the $\CVB_{3}$ seems to be the best posterior approximation
for this independent Normal cluster model, followed by $\CVB_{2}$,
$\CVB_{1}$ and mean-field approximations, which yield almost identical
$\ELBO$ values. 

Intuitively, as shown in the case of $\Radius=4$ in the upper left
panel of Fig. \ref{fig:Cluster_CVB}, the mean-field approximations
like VB, $\EM$ and k-means seems not to recognize the correlations
between data of the same clusters, but focus more on the inter-distance
between clusters as a whole. The $\CVB$ approximations, in contrast,
exploit the correlations between each label $\bLabel_{\ipick}$ to
all other labels, as shown in Fig. \ref{fig:cluster}. Although the
heuristic $\CVB_{1}$ becomes worse when $\Radius$ increases, the
$\CVB_{2}$ and $\CVB_{3}$ are still able to pick the best correlated
structures to represent the data. When inter-distance of cluster is
much higher than cluster's variance, these two CVB methods stabilize
and accurately classify $90\%$ of total data in average. The successful
rate is only about $80\%$ for all other state-of-the-art mean-field
approximations. 

\section{Conclusion \label{sec:Conclusion}}

In this paper, the independent constraint of mean-field approximations
like VB, EM and k-means algorithms has been shown to be a special
case of a broader conditional constraint class, namely copula. By
Sklar's theorem, which guarantees the existence of copula for any
joint distribution, a copula Variational Bayes (CVB) algorithm is
then designed in order to minimize the Kullback-Leibler (KL) divergence
from the true joint distribution to an approximated copula class.
The iterative CVB can converge to the true probability distribution
when their copula structures are close to each other. From perspective
of generalized Bregman divergence in information geometry, the CVB
algorithm and its special cases in mean-field approximations have
been shown to iteratively project the true probability distribution
to a conditional constraint class until convergence at a local minimum
KL divergence. 

For a global approximation of a generic probabilistic network, the
CVB is then further extended to the so-called augmented CVB form.
This global CVB network can be seen as an optimally weighted hierarchical
mixture of many local CVB approximations with simpler network structures.
By this way, the locally optimal approximation in mean-field methods
can be extended to be globally optimal in copula class for the first
time. This global property was then illustrated via simulations of
correlated bivariate Gaussian distribution and standard Normal clustering,
in which the CVB's performance was shown to be far superior to VB,
EM and k-means algorithms in terms of percentage of accurate classifications,
mean squared error (MSE) and KL divergence. Despite being canonical,
these popular Gaussian models illustrated the potential applications
of CVB to machine learning and Bayesian network. The application of
copula's design in statistics and a faster computational flow for
augmented CVB network may be regarded as two out of many promising
approaches for improving CVB approximation in future works.

\appendices{}

\section{Bayesian minimum-risk estimation \label{subsec:Bayesian-minimum-risk}}

Let us briefly review the importance of posterior distributions in
practice, via minimum-risk property of Bayesian estimation method.
Without loss of generalization, let us assume that the unknown parameter
$\para$ in our model is continuous. In practice, the aim is often
to return estimated value $\hat{\para}\TRIANGLEQ\hat{\para}(\xbold)$,
as a function of noisy data $\xbold$, with least mean squared error
$\text{MSE}(\hat{\para},\para)\TRIANGLEQ\EXPECTATION_{f(\xbold,\para)}||\hat{\para}(\xbold)-\para||^{2}$,
where $||\cdot||$ is $\Lnorm_{2}$-normed operator. Then, by basic
chain rule of probability $f(\xbold,\para)=f(\para|\xbold)f(\xbold)$,
we have \cite{Bayes:BOOK:Bernado,VH:PhDThesis:14}: 
\begin{align}
\hat{\para} & \TRIANGLEQ\argmin_{\tilde{\para}}\text{MSE}(\tilde{\para},\para)\nonumber \\
 & =\argmin_{\tilde{\para}}\EXPECTATION_{f(\para|\xbold)}||\tilde{\para}(\xbold)-\para||^{2}\label{eq:MSE}\\
 & =\EXPECTATION_{f(\para|\xbold)}(\para),\nonumber 
\end{align}
which shows that the posterior mean $\hat{\para}=\EXPECTATION_{f(\para|\xbold)}(\para)$
is the least MSE estimate. Note that, the result (\ref{eq:MSE}) is
also a special case of Bregman variance theorem (\ref{eq:Bregman_mean})
when applied to Euclidean distance (\ref{eq:Euclidean}). In general,
we may replace the $\Lnorm_{2}$-norm in (\ref{eq:MSE}) by other
normed functions. For example, it is well-known that the best estimators
for the least total variation norm $\Lnorm_{1}$ and the zero-one
loss $\Lnorm_{\infty}$ are the median and mode of the posterior $f(\para|\xbold)$,
respectively \cite{Bayes:BOOK:Bernado,VH:PhDThesis:14}. 

\section*{Acknowledgement}

I am always grateful to Dr. Anthony Quinn for his guidance on Bayesian
methodology. He is the best Ph.D. supervisor that I could hope for.

\bibliographystyle{IEEEtran}
\bibliography{CVB_bibtex}

\begin{IEEEbiography}[{\includegraphics[width=1\textwidth]{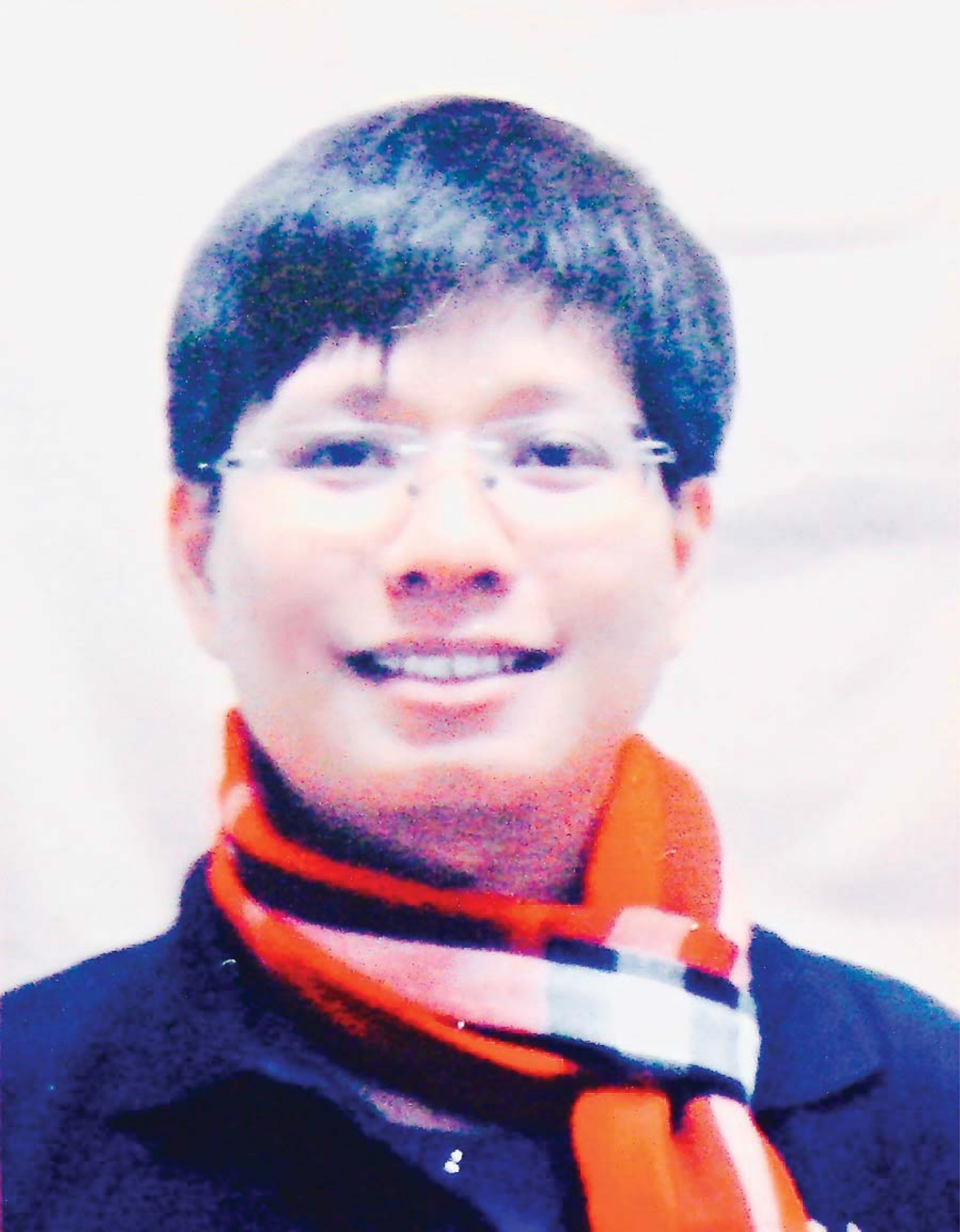}}]{Viet Hung Tran}
 received the B.Eng. degree from Hochiminh city University of Technology,
Vietnam, in 2008, the master`s degree from ENS Cachan, Paris, France,
in 2009, and the Ph.D. degree from the Trinity College Dublin, Ireland,
in 2014. From 2014 to 2016, he held a post-doctoral position with
Telecom ParisTech. He is currently a Research Fellow at University
of Surrey, London suburb, U.K. His research interest is optimal algorithms
for Bayesian learning network and information theory. He was awarded
the best mathematical paper prize at IEEE Irish Signals and Systems
Conference, 2011. 
\end{IEEEbiography}

\end{document}